%% file: main.tex
\documentclass[a4paper,11pt]{article}
\usepackage{jinstpub} 
\usepackage{lineno}
\usepackage{amsmath}
\usepackage{graphicx}
\usepackage{xspace}
\usepackage{placeins}
\usepackage{comment}
\RequirePackage[version=3]{mhchem}

\usepackage[rawfloats=true]{floatrow}

\title{\boldmath Upgrades of the ATLAS Zero Degree Calorimeter System for Run 3 at the Large Hadron Collider}

\author[g]{G.~Avoni}
\author[g]{M.~Bruschi}
\author[d]{G.~Canale}
\author[a]{Z.~Citron}
\author[d,e]{B.A.~Cole}
\author[f]{E.~Dahle}
\author[d]{B.~Dziedzic}
\author[a]{B.~Glik}
\author[f]{M.~Grosse-Perdekamp}
\author[e]{Y.~Guo}
\author[h]{K.~Korcyl}
\author[f,m]{M.~Hoppesch}
\author[f,m]{M.~Housenga}
\author[f]{C.~Lantz}
\author[f]{Y.~Liu}
\author[l,m,f]{R.~Longo}
\author[f]{S.~Lund}
\author[f]{D.~MacLean}
\author[g]{S.~Meneghini}
\author[i]{M.~Milovanovic}
\author[b]{G.~Mladenovic}
\author[e]{S.~Mohapatra}
\author[f]{F.~Mohammed Rafee}
\author[a]{Y.~Moyal}
\author[g]{C.~Sbarra}
\author[g]{A.~Sbrizzi}
\author[e]{B.~Seidlitz}
\author[a]{S.~Shenkar}
\author[c]{P.~Steinberg} 
\author[a]{L.~Sudit} 
\author[d]{D. Valuch}
\author[f]{K.~Young}

\affiliation[a]{Ben-Gurion University of the Negev, Dept. of Physics, Beer-Sheva 84105, Israel}
\affiliation[b]{University of Belgrade - Faculty of Mechanical Engineering
Kraljice Marije 16
11000 Belgrade, Serbia}
\affiliation[c]{Brookhaven National Laboratory, Upton, NY 11973, USA}
\affiliation[d]{CERN, Geneva; Switzerland.}
\affiliation[e]{Columbia University Nevis Laboratories, 136 South Broadway, Irvington, NY, USA}
\affiliation[f]{University of Illinois at Urbana-Champaign, Dept.\ of Physics, Urbana, IL 61801-3080, USA}
\affiliation[g]{INFN, Sezione di Bologna, Viale Berti Pichat 6/2, 40127 Bologna, Italy}
\affiliation[h]{Henryk Niewodniczański Institute of Nuclear Physics (IFJ), ul. Radzikowskiego 152, 31-342 Kraków, Krakow, Poland.}
\affiliation[i]{II. Physikalisches Institut, Justus-Liebig-Universität Gießen
Heinrich-Buff-Ring 16
DE - 35392 Gießen}
\affiliation[l]{Università di Torino, Dipartimento di Fisica, via Pietro Giuria 1, 10125, Torino (TO), Italy}
\affiliation[m]{INFN, Sezione di Torino, via Pietro Giuria 1, 10125, Torino (TO), Italy}

\emailAdd{cole@nevis.columbia.edu}

\abstract{
Experimental studies of ultra-relativistic heavy ion collisions at the Large Hadron Collider (LHC) depend crucially on Zero Degree Calorimeters (ZDCs) that measure neutrons produced at near-beam rapidity in nucleus-nucleus collisions. In hadronic nuclear collisions these neutrons are mainly spectator neutrons, those that do not scatter from opposing nucleons during the collision.
As a result, the ZDCs provide a vital probe of heavy ion collision geometry. The ZDCs are also essential in the study of ultra-peripheral collisions that are initiated by photons associated with the electric fields of one or both nuclei. 
Coherent photon emission typically leaves the photon emitter intact, making the observation of no ZDC signal, on one or both sides, a tag of such processes.
The ATLAS ZDCs, built prior to Run~1 were substantially upgraded for LHC Run~3. The primary upgrades included replacement of the quartz Cherenkov radiator with $\text{H}_2$-doped fused silica rods; installation of fast air-core signal cables between the ZDC and the ATLAS USA15 cavern; new LED-based calibration system; and new electronics implemented for readout and fully-digital triggering.
The ZDCs were also augmented with new ``Reaction Plane Detectors'' (RPDs) designed to measure the transverse centroid of multi-neutron showers to allow event-by-event reconstruction of the directed-flow plane in nuclear collisions. The Run~3 ZDC detectors, including the RPDs, are described in detail with emphasis on aspects that are new for Run~3.
}


\begin{document}
\maketitle
\flushbottom
\input{macros.tex}

\section{Introduction}
\label{sec:intro}
\input{intro.tex}

\section{Detector configuration and environment}
\label{sec:ZDCover}
\input{overview.tex}

\section{Detector implementation details}
\label{sec:detimpl}
\input{detector_details.tex}

\section{Electronics and Trigger}
\label{sec:FEEtrig}
\input{readout.tex}

\subsection{Trigger design and implementation}
\input{trigger.tex}
\input{monitoring.tex}

\section{Detector working points}
\label{sec:workpoints}
\input{OperatingPoints.tex}

\section{Detector calibration }
\label{sec:calib}
\subsection{LED calibration system}
\label{sec:LED}
\input{LED.tex}

\subsection{FEE calibration system}
\label{sec:pulser}
\input{pulser.tex}
\subsection{Data-driven ZDC energy calibration}
\label{sec:OneNeutcalib}

\input{calib.tex}

\subsection{RPD position calibration}
\input{rpdcalib.tex}

\section{Monte Carlo simulations}
\label{sec:MC}
\input{MC.tex}

\FloatBarrier
\section{Summary and future upgrades}
\label{sec:summary}
\input{summary.tex}

\acknowledgments
\input{acknowledgements.tex}

\clearpage
\appendix
\input{optimization}

\bibliographystyle{JHEP}
\bibliography{biblio.bib}

\end{document}

%% file: macros.tex
\renewcommand{\AA}{\mbox{A+A}\xspace}
\newcommand{\snn}{\ensuremath{\sqrt{s_{\text{NN}}}}\xspace}
\newcommand{\Nspec}{\ensuremath{N_{\text{spec}}}\xspace}
\newcommand{\Em}{EM}
\newcommand{\Had}[1]{HAD#1\xspace}
\newcommand{\pxp}{$p$+$p$\xspace}

\newcommand{\xnxn}{\mbox{XnXn}\xspace}
\newcommand{\znxn}{\mbox{0nXn}\xspace}
\newcommand{\znzn}{\mbox{0n0n}\xspace}
\newcommand{\ZDCA}{{\tt ZDC\_A}\xspace}
\newcommand{\ZDCC}{{\tt ZDC\_C}\xspace}
\newcommand{\ZDCAC}{{\tt ZDC\_A\_C}\xspace}
\newcommand{\adcmax}{\ensuremath{\text{ADC}^{\text{max}}_{\text{sub}}}\xspace}
\newcommand{\mum}{\ensuremath{\mu \text{m}}\xspace}
\newcommand{\muA}{\ensuremath{\mu \text{A}}\xspace}
\newcommand{\ohm}{\ensuremath{\Omega}\xspace}

\newcommand{\Slink}{\mbox{S-Link}\xspace}

\newcommand{\BC}{BCX\xspace}
\newcommand{\BCs}{BCXs\xspace}
\newcommand{\aircore}{air core\xspace}
\newcommand{\Aircore}{Air core\xspace}

\newcommand{\Geant}{GEANT4\xspace}

\newcommand{\onen}{\Nn{1}}
\newcommand{\Nn}[1]{\ensuremath{#1n}\xspace}

\newcommand{\sumadcf}{\ensuremath{\sum\text{ADC}/4}\xspace}

\newcommand*{\Npart}{\ensuremath{N_{\text{part}}}\xspace}
\newcommand*{\PbPb}{\ce{Pb}+\ce{Pb}\xspace}
\newcommand*{\sqn}{\ensuremath{\sqrt{s_{_\text{NN}}}}\xspace}
\newcommand{\pp}{\ensuremath{pp}\xspace}

\newcommand*{\pPb}{\ensuremath{p}+\ce{Pb}\xspace}
\newcommand*{\electronvolt}{eV}

\newcommand*{\TeV}{\ensuremath{\text{T\electronvolt}}\xspace}
\newcommand*{\GeV}{\ensuremath{\text{G\electronvolt}}\xspace}
\newcommand*{\MeV}{\ensuremath{\text{M\electronvolt}}\xspace}

%% file: intro.tex
\subsection{Zero degree calorimeters in high-energy nuclear collisions}
Zero degree calorimeters (ZDCs) have been a key component of
experiments studying high-energy heavy ion collisions
at a variety of energies.  Their use began in fixed target experiments at the Brookhaven Alternating Gradient Synchrotron \cite{Abbott:1990ka,PhysRevC.45.819,Mikhaylov:2015juf}
and the CERN Super Proton Synchrotron \cite{ARNALDI19981} with nucleon-nucleon center of mass energies ranging from  $\snn \sim 5 - 20$~GeV, and have continued at collider experiments at the Relativistic Heavy Ion Collider ($\snn = 7-200$ GeV) \cite{ADLER2001488} and the Large Hadron Collider (LHC) ($2.76-5.36$ TeV) \cite{ALICE:2004cyf,Suranyi:2021ssd,Jenni:1009649}. 
ZDCs are used to measure ``spectator''
nucleons in high-energy nuclear collisions: those nucleons that do not interact hadronically during the crossing of the two
nuclei.  Instead, they are freed by the impulsive shock of the collision and   
propagate in approximately the same direction as -- {\it i.e.} at zero degrees with respect to -- the beam and with the
same energy per nucleon. 
The remaining nucleons -- those that ``participate'' in the collision -- typically produce particles as a result
of inelastic scatterings and the strongly-interacting matter created by
this particle production is typically the focus of experiments
studying high-energy nuclear collisions. 
For example, at the LHC, the
four major experiments, including ATLAS, study the nature of the quark-gluon plasma (QGP) created in ultra-relativistic \PbPb\ collisions. 
The initial temperature, and transverse size of the plasma are almost
completely determined by the number of participants (\Npart).
Since the number of spectator nucleons (\Nspec) is essentially $2A-\Npart$, measurements of the energy flow at zero degrees are able to provide valuable experimental control over the initial conditions of the QGP.

The nuclear overlap, determined by the nucleus-nucleus impact parameter ($b$, the transverse distance between the nuclear barycenters), determines 
$\Npart$ -- and, thus, the number of spectator nucleons.  At high energies, nucleons in the nucleus can be treated in the eikonal approximation, so as essentially frozen in place at the moment of impact.  In this approach, the interactions between nucleons in the colliding nuclei become essentially semi-classical, with each inelastic interaction determined by the nucleon-nucleon total inelastic cross section \cite{Miller:2007ri}. 
This means that, on general grounds, \Npart increases and \Nspec decreases
with decreasing impact parameter. 
In ``central'' \AA\ collisions
where $b$ is much smaller than the radius of the colliding nuclei, $b
\ll R$, $\Nspec \rightarrow 0$ while $\Npart \rightarrow 2A$. At the
opposite extreme, in the most ``peripheral'' collisions where $b \gtrsim 2R$
and the nuclei only graze each other, $\Nspec \rightarrow 2A$ and
$\Npart \rightarrow 0$. However, in these collisions, the incident
nuclei remain mostly intact with the result that the spectator
nucleons are mostly bound in remnant nuclear fragments.  

At intermediate impact parameters, $b \sim R$, the collision can deliver 
transverse momentum to {\it both} nuclei with the result 
that the spectator nucleons have net momentum in the ``directed flow'' \cite{Bozek:2010bi} plane, but in opposite directions for each nucleus.
A measurement of the azimuthal orientation of this plane,
typically represented by the angle $\Psi_1$, allows the study of the
collective directed flow response of the QGP by correlating the
azimuthal angle dependence of particle production near mid-rapidity
with the $\Psi_1$ angle. 
The ZDCs provide access to this parameter via event-by-event measurements 
of the centroid of the many-neutron showers in both ZDCs, 
which directly reflect the direction of the transverse kick.

At RHIC and the LHC, charged spectators and beam-rapidity
nuclear fragments are bent by the insertion magnets in the intersection regions (IRs) 
on trajectories well away from zero degrees. 
Thus, the ZDCs detect only neutral particles, primarily spectator neutrons. 
This implies that, in contrast to the naive
expectation stated above, the energy observed in the ZDC {\it decreases} in more peripheral collisions, with large event-to-event fluctuations.
In the opposite extreme, \textit{i.e.} more central collisions where the incident nuclei are
completely disintegrated, coalescence of neutrons and protons into
deuterons or other light fragments can also reduce the already small number of neutrons
reaching the ZDCs.

\subsection{Zero degree calorimeters and ultra-peripheral collisions}
At the ultra-relativistic energies reached at RHIC and the LHC, 
the mainly transverse electromagnetic
fields of the fully ionized nuclei produce a high flux of nearly-real
photons. For photon wavelengths longer than the nuclear radius,
$\lambda > R$, the photons are emitted coherently by the entire
nucleus with the result that the photon flux scales as $Z^2$, where
$Z$ is the atomic number of the nucleus. For large-$Z$ ions like lead
(Pb), the photon fluxes are sufficiently large that nuclear collisions
at the LHC have large cross-sections for photon-initiated
processes. In symmetric nuclear collisions with impact parameters
larger than twice the nuclear radius, $b > 2R$, hadronic collisions
between the nuclei are strongly suppressed and photon-initiated
processes dominate. These are frequently referred to as
``ultra-peripheral collisions'' (UPC) \cite{Baltz:2007kq} and they have become an
important part of the LHC physics program \cite{Contreras:2015dqa,Klein:2020fmr}.

ZDCs are valuable for identifying and triggering on UPCs due to the
fact that the coherent emission of photons from a nucleus typically
leaves it intact. 
For example, in a photonuclear ($\gamma+A$) process, 
the ZDC in the direction of the
photon-emitting nucleus usually observes no energy while the
ZDC in the direction of the struck nucleus has a high
probability to register spectator nucleons, or even 
evaporation neutrons produced by the
de-excitation of the compound nuclear state formed in the
interaction. These one-sided neutron topologies are usually referred-to by 
the label \znxn. Conversely, photon-photon collisions typically leave
both nuclei intact, so no neutrons are observed in either direction -- a
topology referred-to as \znzn. Finally, events with neutrons
in both directions -- \xnxn topologies -- result from
hadronic interactions, as well as mutual Coulomb dissociation.

Long-range Coulomb interactions between the scattering nuclei can
stimulate the excitation of giant dipole or higher-order resonances in photon-emitting nuclei. 
The nucleus then decays via low-energy neutron emission in its own rest frame, where the resulting
neutrons are typically within the acceptance of the ZDCs. 
On one hand,
these dissociated nucleons can confuse the interpretation of
measurements by causing migration between different neutron
topologies. On the other hand, the strong impact parameter dependence
of the Coulomb excitation process means that the absence, presence or
number of neutrons can be used as a ``knob'' to study the impact
parameter dependence of UPC processes, as long as the nature of the
process can be separately determined, e.g. by using particle production signatures in the central rapidity region.

The above-described Coulomb excitation processes can occur in
\PbPb\ collisions in which no other EM or hadronic exchange
occurs \cite{Pshenichnov:2011zz}. The cross-section for such ``dissociative'' processes to
produce neutrons in the ZDC is large \cite{PhysRevLett.109.252302}, in excess of 200~b for the excitation of a single nucleus and$~\sim 400$~b for producing at least one neutron in either direction. 
While the number of neutrons
produced in Coulomb excitation processes is typically smaller on
average than that in hadronic \PbPb\ collisions, the dissociative
cross-section is larger than the total hadronic cross-section by a
factor of~50. Thus, the rate for neutrons arriving at the ZDC and the
total radiation dose delivered to the detectors is dominated by
dissociative processes. 

\subsection{Requirements on ZDC performance}
The complications in the ZDC response in peripheral collisions due to
the loss of spectator neutrons to charged nuclear fragments has, so far,
prevented the incorporation of ZDC data into ATLAS evaluations of
centrality in \PbPb\ collisions. However, although centrality is in general challenging in \pPb collisions, in these collisions the ZDC {\em has} been used to directly characterize centrality \cite{ATLAS:2022iyq}. The ZDCs are crucial to
the ATLAS heavy-ion program for measurements in UPC collisions,
for minimum-bias triggering in \PbPb\ collisions, and, more recently
for studying ultra-central \PbPb\ collisions \cite{ATLAS:2024jvf}. The ZDCs are also used
together with measurements of transverse energy in the forward
calorimeters to suppress events containing multiple \PbPb\ collisions
in the same crossing (in-time pileup).
An additional physics goal for the ZDCs in Run~3 is to provide event-by-event detection of the directed-flow event plane angle. This can be accomplished by measuring the direction of the net deflection of the spectator neutrons which, in turn, requires a detector capable of measuring the average transverse position of multi-neutron showers in the ZDC.

For UPC physics and for minimum-bias triggering, it is essential
that the ZDCs provide clean separation between events having no
neutrons and those with at least one neutron. This separation is
needed both in offline analysis and in the trigger and
necessarily requires good control over the energy scale, at least for
small numbers of neutrons. Fortunately, \PbPb\ collisions provide a
significant rate for events having a single neutron in one calorimeter
and, with sufficient energy resolution, the one-neutron (\onen) peak
can be easily identified and used to accurately establish the
calorimetric absolute energy scale. As will be discussed below, the ATLAS ZDC single-neutron energy
resolution is dominated by intrinsic momentum fluctuations of the bound neutrons and  fluctuations in longitudinal containment of
the multi-TeV hadronic showers. Nonetheless, the typical resolution is
better than 20\%, more than adequate to identify the \onen peak and
use it to determine the ZDC energy scale. 

There is a clear advantage in using the ATLAS ZDCs for luminosity measurement during heavy-ion data-taking. 
One of the most important reasons is that the alternatives in ATLAS to the primary luminosity measurement provided by the LUCID detector \cite{Lumi} are not effective in \PbPb collisions due to the relatively low (relative to \pp) collision rates. In addition, with a trigger threshold set to select at least one multiple-TeV neutron, there is little background to a ZDC-based luminosity measurement from noise or  ``afterglow'' \cite{afterglow}. Furthermore, the large  dissociative cross-sections mean that van~der~Meer calibrations   \cite{Grafstrom:2015foa} using ZDC-measured luminosities are not susceptible to statistical fluctuations even at large beam separations.

The ZDCs must be able to operate in a high-radiation environment
resulting from the large numbers of beam-energy neutrons that are
produced in heavy ion collisions. More details on the radiation levels
that the ZDCs must accommodate will be provided in
Section~\ref{sec:RadiationEnvironment}. However, the radiation doses
delivered to the vicinity of the ZDC during high-luminosity \pp\ operation
are such that even radiation hard detectors would be severely
damaged. Thus, the ZDCs are typically only installed during
nucleus-nucleus and proton-nucleus operation. However, the ZDCs  were
used during two low-$\mu$ \pp runs in Run~3, namely a special run at
13.6~TeV for the LHCf experiment \cite{LHCf:2008lfy} and a $\sqrt{s} =
5.36$~TeV run in 2024 intended to provide reference data for heavy ion
measurements. 

\subsection{The ATLAS ZDCs in LHC runs 1 and 2}
\label{sec:introrun12}
Two ZDCs were constructed for the ATLAS experiment prior to the start
of Run~1 operation of the LHC, and those detectors, with the upgrades described in this paper are still in operation during Run~3.  The ZDCs are, primarily,
tungsten-sampling calorimeters with quartz Cherenkov radiator rods as
the active elements. Both calorimeters are divided into four
longitudinal sections (modules) that are each read out by a single
photomultiplier tube (PMT). For Run~1 the second module in both calorimeters was
equipped with transverse segmentation provided by rods running
lengthwise through the module. In one calorimeter, transverse sampling
was also instrumented in the first module to allow measurement of the
transverse positions of individual photons for detection of neutral
pions. More details on the design and geometry of the calorimeters is
provided in Sec~\ref{sec:calor} below.

In Runs 1 and 2, signals from the ZDC PMTs were transmitted to the ATLAS USA15 cavern over 300~m CC50 cables, where they were split and amplified
using custom electronics modules. The resulting signals were digitized
using ``pre-processor modules'' (PPMs) originally designed for the
ATLAS Level-1 Calorimeter (L1Calo) trigger system \cite{L1Calo} and the resulting
data were transmitted to the ATLAS data acquisition system using L1Calo
readout driver (ROD) modules. The PPMs provided 16 independent
channels with 10-bit flash ADCs (FADCs) designed to sample at 40~MHz, though
they could also be operated at 80~MHz. Because of the limited dynamic
range of the PPM ADCs, the electronics modules produced separate copies of
the input signals, with a factor of 10 amplification, that were
read out independently to provide effective dual-gain
digitization. Because the 80~MHz mode of the PPMs was not originally
foreseen, the custom electronics built for the ZDCs also produced
two copies of the input signals delayed by 12.5~ns, again with gains
differing by a factor of 10. These two copies were also independently
digitized to provide an effective 80~MHz sampling. 
Level-1 triggers were provided by the ZDC by performing an analog sum
of the four analog PMT signals from each calorimeter and then
discriminating the resulting signal, ostensibly well below the
sum amplitude corresponding to single-neutrons.

The originally constructed ZDCs had a number of issues and limitations
that significantly affected their use by ATLAS. One of the primary problems
was associated with the poor radiation hardness of the
quartz radiator. Exposure of the ZDCs to over $1~\text{fb}^{-1}$ of
\pp\ collisions during 7~TeV operation of the LHC in 2011
severely damaged the quartz radiator rods and the longitudinal rods
that provided the transverse segmentation. As a result, the
performance of the ZDCs for the 2011~2.76~TeV \PbPb\ run was sufficiently poor that
the data were unusable. The quartz was replaced prior to
the 2013~\pPb\ run and needed replacement again after the 2016~\pPb\ data-taking.

Another significant limitation of the initially constructed ZDCs
resulted from the high currents in the dynode chain of the Hamamatsu R6091 
PMTs used to read out the calorimeter modules (originally implemented in 
Hamamatsu H6559 assemblies). 
In the operation of the ZDCs during the 2015 \PbPb\ run, substantial (as large as a factor of 2)
variation of the single-neutron amplitude with instantaneous luminosity was
observed. This gain variation was attributed to
luminosity-dependent changes in the dynode potentials resulting from large
($\gg 100~\mu\text{A}$) currents in the dynode chain. To solve this
problem, prior to the 2016~\pPb\ run, the original PMTs were replaced
with Hamamatsu R6091 PMTs equipped with boosters on the last three dynode
stages (H6559 MODB assemblies). The addition of the dynode boosters largely removed the luminosity-dependent variations in PMT gain and the ZDCs have been operated using the boosters during the remainder of Run~2 and all of Run~3.

Due to a combination of the long CC50 cables used to propagate the PMT signals to the ATLAS electronics room, and intrinsic
performance of the custom electronics, the ZDC signals arriving at the
PPMs had a typical rise-time of 10~ns and a fall-time of 25~ns. While
these rise and fall times were reasonably matched to the sampling
rate of the PPMs and were nominally adequate for Run~1 operation, they
became problematic during Run~2, as the LHC reduced the bunch spacing
during heavy ion operation first to 100~ns and then 75~ns in the 2018
\PbPb\ run. The smaller separation between colliding bunches, combined
with the large dynamic range of the signals in the ZDC -- an event with
$> 50$  neutrons in a single ZDC could be followed by an event with a
single neutron -- led to events with significant out-of-time (OOT) pileup
where large tails from pulses generated by a preceding collision would
underlie and often obscure small pulses from an event of interest. The effects from OOT pileup
were exacerbated by energy deposits in the ZDC between filled bunch
crossings, likely due to presence of ghost charge \cite{ghostq} in the machine. 
The OOT effects were mostly mitigated in offline analysis by fitting
the FADC data to a sum of a pulse and a background term describing the
tail of the preceding pulse, but they nonetheless made the signal 
reconstruction more difficult and
reduced sensitivity to events with few neutrons. Worse, the OOT
pileup caused more severe problems with the Level-1 trigger due to the
fact that signals from large energy deposits might remain above the
discriminator threshold long enough to prevent the discriminator
re-firing on an event of interest. This problem was partially
mitigated by spitting the signal from the analog sums and
differentiating one copy in each calorimeter using a passive
high-pass filter. The resulting signals were then separately discriminated and
logically ORed with the output of the discriminator acting directly on
the analog sums. This method made it possible to preserve 90\% trigger
efficiency for single neutrons in the 2018 \PbPb\ run, but for the
high-luminosity 8~TeV portion of the 2016 \pPb\ run, the ZDC Level-1 trigger
had minimal efficiency for fewer than two neutrons.

\subsection{Run 3 upgrades to ATLAS ZDCs}
\label{sec:introrun3}
Given the difficulties and performance limitations of the originally
constructed ZDCs, which are only partially described above, ATLAS
undertook a program to improve the detector, the readout chain, and the trigger
during the second long shutdown of the LHC in preparation for
Run~3. The main updates were:
\begin{itemize}
\item The quartz Cherenkov radiator rods were replaced
with rods of fused silica doped to provide significantly reduced
sensitivity to radiation damage\cite{Yang:2022xfv}.
\item The cables transmitting ZDC signals
to USA15 were replaced with low-dispersion coaxial \aircore cables.
\item New front-end and readout
electronics were implemented based on the ``LUCROD'' boards designed
for the ATLAS LUCID2 detector~\cite{LUCID2}.
\item Fully-digital triggering was implemented in the LUCRODs.
\item Each calorimeter was
instrumented with a ``reaction plane detector'' (RPD) that
provides transverse imaging of the showers in the ZDC.
\item A new LED calibration system that operates continuously during data-taking was implemented.
\end{itemize} 
This paper describes these important upgrades to the ATLAS ZDCs and documents the
configuration of the ZDCs for the 2023 and 2024 \PbPb\ operation of
the LHC. 

The remainder of the paper is structured as follows:
Section~\ref{sec:ZDCover} provides an overview of the detector
configurations during the two years; Section~\ref{sec:detimpl}
describes the implementation of the calorimeter and RPD detectors;
Section~\ref{sec:FEEtrig} describes
the LUCROD boards for use as ZDC front-end electronics (FEE) and the
implementation of ZDC Level-1 triggers. 
Section~\ref{sec:workpoints} provides details on the configuration of the detector working points during the data taking. 
Section~\ref{sec:calib} describes the calibration of the ZDCs, both systems used to measure the response of the PMTs and the FEE and data-drive techniques used to establish an absolute energy scale. Section~\ref{sec:MC} describes procedures used for Monte Carlo simulation of the detector and presents results that illustrate the development of
hadronic showers in the ZDC and the energy deposition in the calorimeter modules. Section~\ref{sec:summary} provides a summary and a brief review of future upgrades of the ZDCs planned for the HL-LHC.

%% file: overview.tex

\subsection{Target Absorbers for Neutrals}
The Target Absorbers for Neutrals (TANs) are passive components of the
LHC installed to absorb the flux of far-forward neutral particles
produced at the high luminosity interaction regions of the
accelerator. They also contain the transition where the two separate
beam pipes of the LHC merge into a single pipe that leads to the ATLAS
interaction point (IP). The TAN absorbers are composed of an inner copper absorber (1100 mm
wide $\times$ 3500~mm long),
an outer steel shielding (comprised of five assemblies) and a front
marble shielding. Each TAN contains a slot 1000~mm long, by 96~mm wide
and 605~mm deep located between the two beam pipes to house
forward detectors like the Beam RAte of Neutrals (BRAN)~\cite{MATIS2017114}, the LHCf detector~\cite{LHCf:2008lfy} as well as the
ZDC and RPD.

\begin{figure}[!b]    
\centering 
\includegraphics[width=0.98\textwidth]{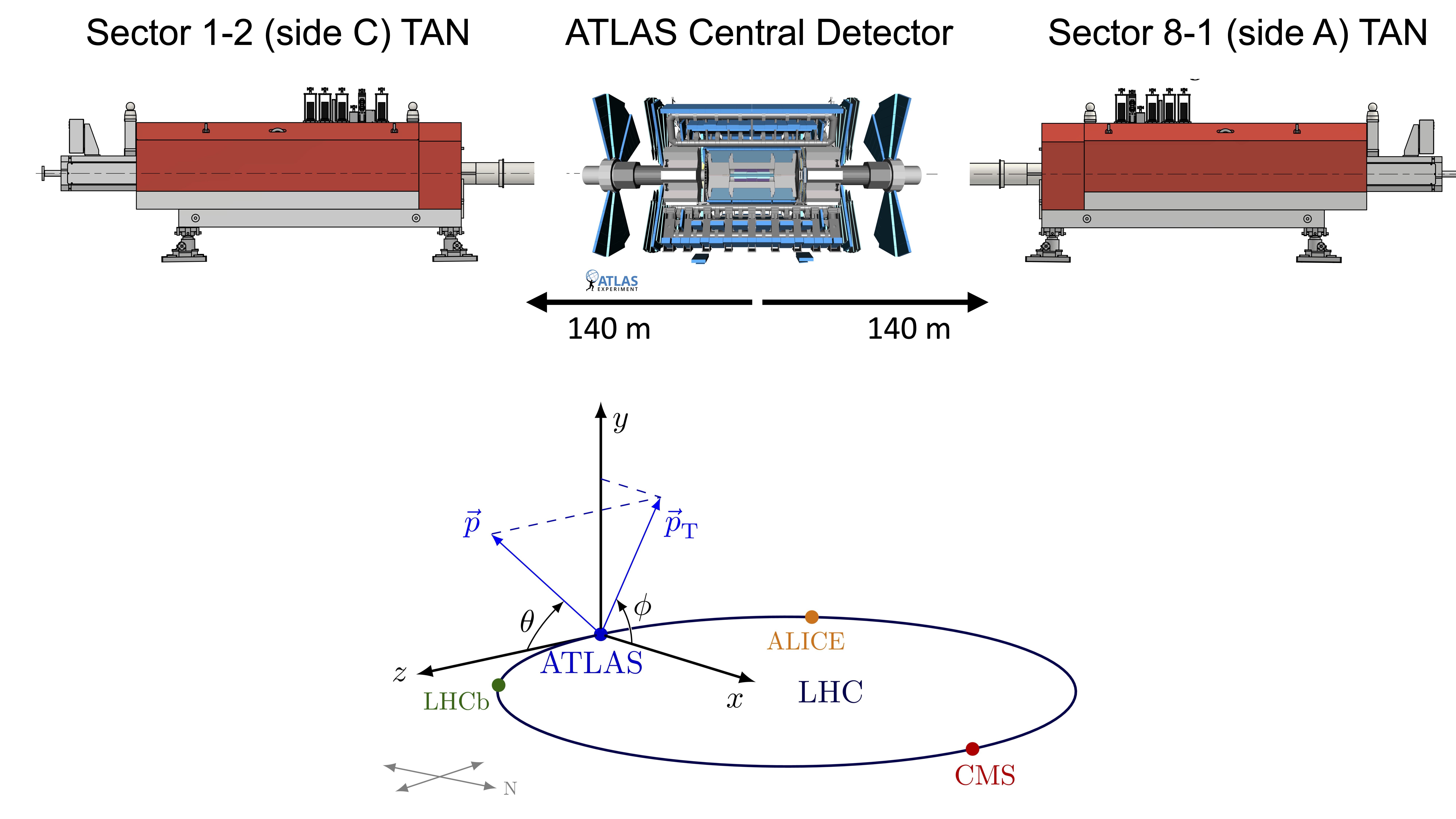} 
\caption{Top: A not-to-scale demonstration of the placement of the LHC TANs and the ZDCs relative to the ATLAS detector. Bottom: the ATLAS coordinate system. The blue arrows indicate possible momentum vectors in three dimensions ($\vec{p}$) or in the $x-y$ plane ($\vec{p}_T$) to illustrate the polar and azimuthal angles, respectively.
}
\label{fig:TANgeom}
\end{figure}
The top portion of Figure~\ref{fig:TANgeom} shows a view of the two TANs on the sector
1-2 (left) and 8-1 (right) sides of ATLAS. According to ATLAS
convention, the TAN on the Sector~12 side of ATLAS is labeled ``C'' and
that on the Sector~81 side of ATLAS is labeled ``A''.
The bottom portion of Figure~\ref{fig:TANgeom} specifies the ATLAS right-handed coordinate system \footnote{ATLAS uses a right-handed coordinate system with its origin at the nominal interaction point (IP)
in the centre of the detector and the \(z\)-axis along the beam pipe.
The \(x\)-axis points from the IP to the centre of the LHC ring,
and the \(y\)-axis points upwards.
Polar coordinates \((r,\phi)\) are used in the transverse plane, 
\(\phi\) being the azimuthal angle around the \(z\)-axis.
The pseudorapidity is defined in terms of the polar angle \(\theta\) as \(\eta = -\ln \tan(\theta/2)\) and is equal to the rapidity
$ y = \frac{1}{2} \ln \left( \frac{E + p_z c}{E - p_z c} \right) $ in the relativistic limit.
Angular distance is measured in units of \(\Delta R \equiv \sqrt{(\Delta y)^{2} + (\Delta\phi)^{2}}\).
}, with the $x$ axis pointing
horizontally towards the center of the LHC ring, the $y$ axis pointing
vertically upward, and the $z$ axis pointing tangent to the beams in
the counter-clockwise direction as viewed from above.  Based on this convention, the C-side and A-side TANs are located at $z =+140$~m and $z=$-140~m, respectively.

For standard high-luminosity \pp\ operation of the LHC, the TAN slot is filled with
nine $99 \times 94 \times 605 ~\text{mm}^3$ copper bars and the BRAN
detector, positioned between the third and the fourth bar from the
IP. The copper bars provide the stopping power needed to prevent
the energy deposition from very forward neutral particles from causing quenches
of the twin aperture superconducting beam separation dipoles (D2). Prior to heavy-ion data taking, the copper bars are extracted using a crane permanently installed over the TAN and replaced by the ZDC and RPD. Following the completion of a  heavy ion run and prior to the start of high-luminosity \pp\ operation of the LHC the ZDC modules and RPDs are removed and the Cu bars restored. 

\begin{figure}[!tb]    
\centering
\includegraphics[width=0.95\textwidth]{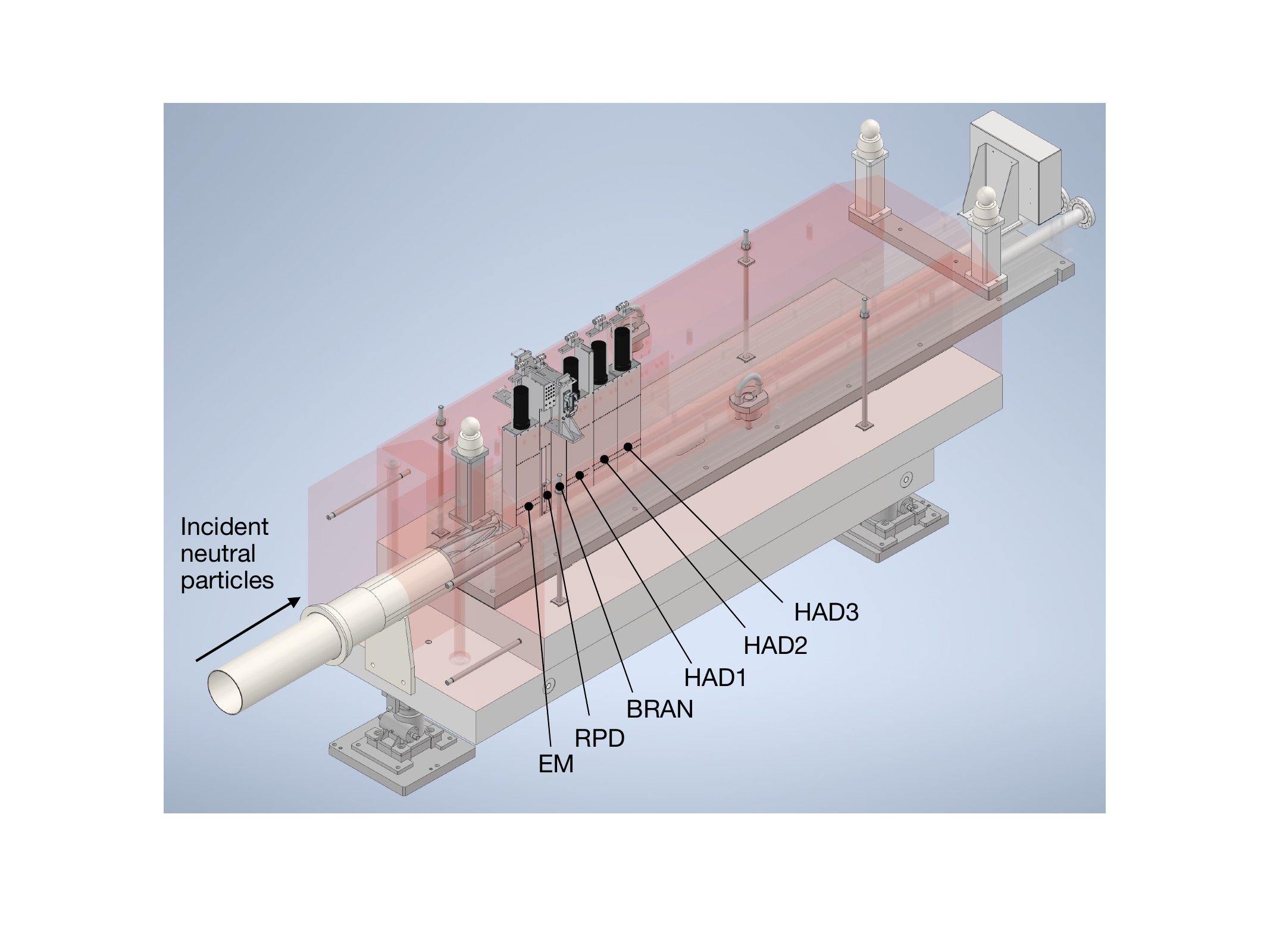}

\caption{Diagram showing the TAN on the A side of the ATLAS detector
  along with the components of the ZDC, as installed for the 2023
  \PbPb\ data-taking period. The  arrow indicates the
  direction of impinging neutral particles produced by
  \PbPb\ collisions. The TAN is rendered as semi-transparent to illustrate
  its internal structure and to allow the active portions of the ZDC
  to be seen.  }
\label{fig:TANA2023}
\end{figure}

Figure~\ref{fig:TANA2023} shows the structure of the A-side TAN. The single beam pipe from the Point 1 IR enters the TAN on the
left and exits as two separate pipe on the right. Neutral particles
produced in collisions in the center of the ATLAS detector are
incident from the left; after passing through the chamber where the beam pipes split, they encounter the ZDC.

\subsection{Run 3 ZDC configurations}
\label{sec:zdc_overview}
The ATLAS ZDC system consists of two similar (but non-identical)
detectors, each installed in a TAN slot.
Each detector consists of four calorimeter
sections, henceforth referred to as ``modules'', and a reaction plane detector (RPD). 
The LHC BRAN detectors are also positioned in the TAN, although their placement relative to the ZDC
modules differs between 2023 and 2024 (and subsequent years of Run~3 ZDC
operation) as described below. 

Each ZDC calorimeter module contains about 1.1 interaction lengths and 29 radiation
lengths of tungsten absorber. Since the first module on each side absorbs essentially
all incoming electromagnetic radiation, these modules are referred-to as electromagnetic or ``EM'' modules. The other three modules are required to achieve near-full longitudinal containment of
multi-TeV hadronic showers and are called ``HAD1'', ``HAD2'' and
``HAD3''. The structure of the different calorimeter
modules is identical, except for the HAD1 modules, and the C-side EM
module. These modules have empty space behind the absorber
for the routing of the longitudinal fibers described in Section~\ref{sec:introrun12}. Although these
fibers were severely damaged by radiation in Run~1, and are no longer used, 
they remain in the detector because removing them and re-constructing the modules was deemed to have substantial risk.

As noted above, the ZDC shares space in each TAN with the BRANs, which provide luminosity information to the LHC operators.
During Runs~1 and 2, the BRAN was an ionization detector, composed of a ceramic casing with copper electrodes. 
It occupied a fixed position within each TAN, 
such that the ZDC EM modules were placed in front of the BRANs, 
and the three ZDC HAD modules were positioned behind them. 
During Run~3, a new version of the BRAN detectors was installed,
with a thick copper absorber and fused silica radiator.
The new design provided half an interaction length of
material in the middle of the ZDC, very close to the shower maximum. 
During the 2023 \PbPb\ run, the nominal placement of the BRAN detectors was
maintained, and their impact on the ZDCs was directly observed in
significantly worse than expected resolution for the 1-3 neutron
peaks. Subsequent \Geant \cite{Agostinelli:2002hh} MC simulations that included the modified
BRANs verified that fluctuations in the energy deposition in the BRANs
contributed significantly to the degradation of the ZDC energy resolution. 
To solve this problem, a decision was made jointly
with the LHC BRAN team, prior to the 2024 \PbPb\ run,  
to move the BRAN detectors behind the ZDC 
modules for subsequent heavy-ion data-taking.

The Reaction Plane Detector (RPD) is a position-sensitive, calorimetrically-thin detector designed to
measure the centroid of multi-neutron showers in the plane
transverse to the beam direction. The RPD implements a staggered
``pan flute``
arrangement of vertical fibers designed to provide an effective four-fold
segmentation in each of the vertical and horizontal directions. 
A more detailed description of the RPD is given in Section~\ref{sec:RPD}.  
In 2023, the originally envisioned module arrangement was followed where an 
RPD was installed behind each ZDC EM module. 
However, the different structure of the A and C EM modules,
due to the additional space required for the longitudinal fibers,
led to an additional 10~cm air gap between the end of the tungsten 
absorber in the EM module and the front face of the RPD. 
This gap was found to have a large impact on
the showers observed by the RPD, with the total number of particles
seen on the C side reduced by about a factor of two, with a noticeably
larger transverse spread of the showers. 
Simulations using \Geant demonstrated that the observed behavior results from the
angular divergence of the large electromagnetic component of the
showers in the ZDC.  To improve the performance of the RPD on the C
side and restore symmetry between the two sides, the modules in the C 
calorimeter were rearranged during the 2024 \PbPb\ run, by exchanging the 
positions of the EM and HAD3 modules.  

\begin{figure}[!tb]    
\centering 
\includegraphics[width=0.95\textwidth]{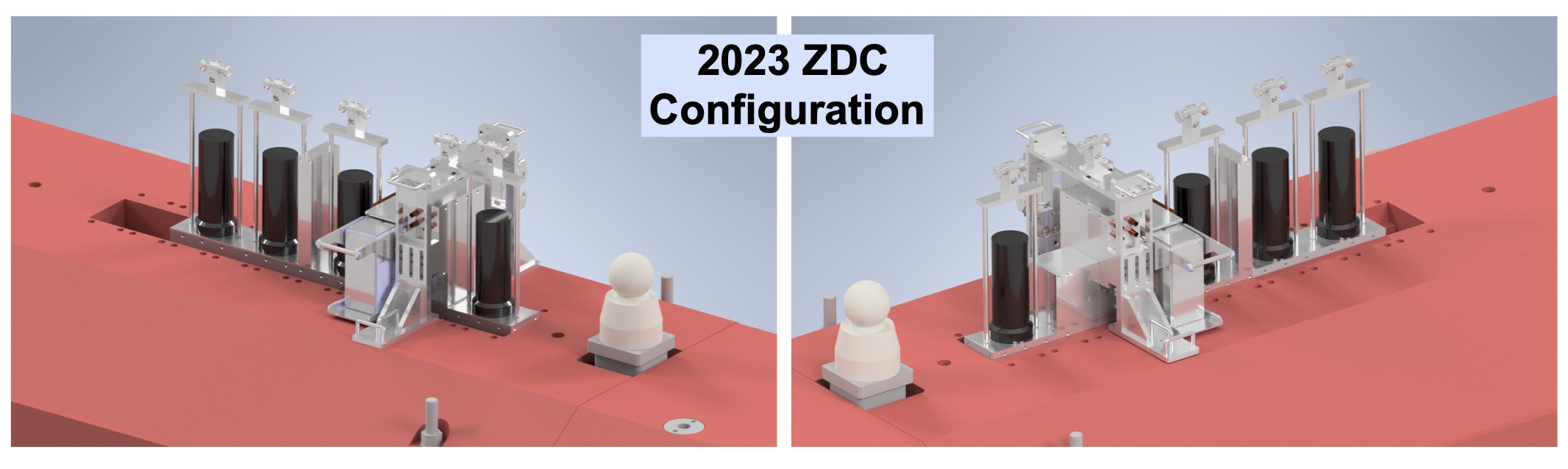} 
\caption{The ZDC and BRAN configuration in the C (left) and A (right) side TANs during the 2023 Heavy Ion Run. 
}
\label{fig:ZDCschematic2023}
\end{figure}

\begin{figure}[!tb]    
\centering 
\includegraphics[width=0.95\textwidth]{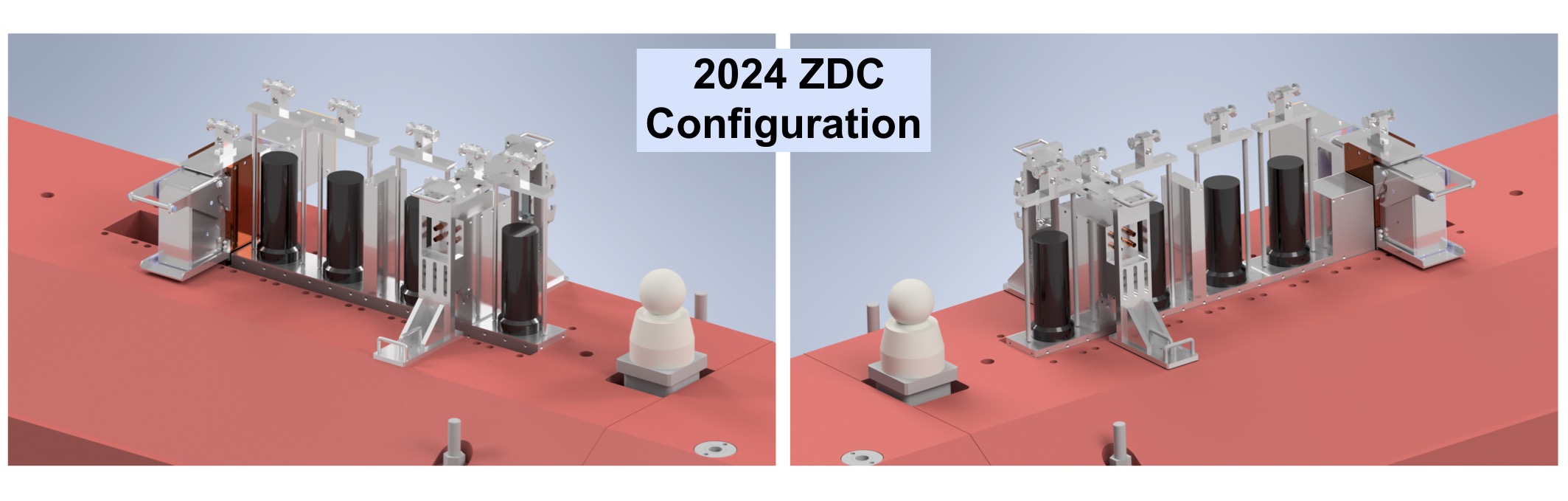} 
\caption{The ZDC and BRAN configuration in the C (left) and A (right) side TANs during the 2024 Heavy Ion Run. 
}
\label{fig:ZDCschematic2024}
\end{figure}

To illustrate the effect of the above-described rearrangements, 
schematics of the ATLAS ZDCs as installed during the 2023 and
2024 \PbPb\ runs are shown in Figure~\ref{fig:ZDCschematic2023} and
Figure~\ref{fig:ZDCschematic2024}, respectively. The schematics show
the order of the components in each detector along with their external
structure and placement. 

\subsection{Pb+Pb beam conditions}
\label{sec:PbPbcond}
In 2023 and 2024 the LHC provided \PbPb\ collisions at a 
center-of-mass energy per-nucleon of 5.36~TeV.
During both years, the LHC operated with a nominal $\beta^* = 50$~cm in ATLAS,
yielding an emittance of about 1.5~$\mu\text{m}$. The typical number of ions
per bunch increased between 2023 and 2024, reaching a maximum of about
$1.6\times10^8$ at the start of fills. 
The typical number of hadronic collisions per crossing was about, $\mu_h = 0.004$
at the start of \PbPb\ fills. 
Based on measured cross section at similar beam energies, the
corresponding number of EM interactions  per crossing producing neutrons in at
least one ZDC side was $\approx 0.2$. During both
2023 and 2024, the LHC operated with trains having 50~ns bunch spacing
such that, at ``full machine,'' 960 and 1032 bunches, respectively,
collided in ATLAS. In 2023, at the start of fills, the luminosity in
ATLAS was leveled at $3.4\times 10^{27}~\text{cm}^{-2}~\text{s}^{-1}$  
while in 2024 the leveling target was $6.4\times 10^{27}~\text{cm}^{-2}~\text{s}^{-1}$.

For typical LHC operations, the beams have a small crossing angle to reduce
parasitic collisions as well as beam-beam interactions.  
During 2023 (2024) \PbPb\ operation, the Pb beams in ATLAS had a vertical
crossing angle of +170 (+150)~$\mu\text{Rad}$. 
In both years the vertical position of the RPD (see Section~\ref{sec:RPD}) 
was set so that
a 150~$\mu\text{Rad}$ crossing angle would place the outgoing neutrons at the
vertical center of the active RPD detector area. 

In both years, the
RPDs were intended to be horizontally centered on the projected beam
direction. However, in 2023 a net horizontal displacement of the showers 
was observed with opposite direction on the A and C sides, suggestive of a 
$\approx 15~\mu\text{Rad}$ horizontal tilt
of the beam axis.  The presence of such a tilt was also implied by
measurements from the Point~1 Diode ORbit and OScillation (DOROS) Beam Position Monitors (BPMs)\cite{Gasior:2017ana} in both 2023 and 2024. 
During commissioning for the 2024~\PbPb\ operation of the LHC, a scan was
performed in ATLAS in which the crossing angle was kept fixed but the
horizontal angles of both beams were changed in steps of
25~$\mu\text{Rad}$. Monitoring plots of the RPD shower centroids made
during the scan and a subsequent rapid offline analyses -- the results
of which are presented
in Sec~\ref{sec:rpd_calib} -- confirmed the presence of the horizontal tilt in
ATLAS. Adjustments were then made by the LHC to remove this tilt such that
the Pb beams were aligned with
the nominal beam direction in the 2024 \PbPb\ run to better than
$5~\mu\text{Rad}$.

\subsection{Radiation environment}
\label{sec:RadiationEnvironment}
The energy deposition induced in the TAN by forward-going neutral particles
from collisions at the ATLAS interaction point are quite large, and can be
characterized via FLUKA simulations (see \cite{Yang:2022xfv} and
references therein). 
Simulations of the radiation dose in Pb+Pb collisions at 5.36~TeV were performed using an approximate geometry based on the CMS ZDC design, which has a similar material distribution along the beam direction. The results found the peak dose to be 0.7~MGy/nb$^{-1}$ (0.05~MGy/nb$^{-1}$) in the ZDC (RPD). 
In 2023 and 2024 a total of about 3.8~nb$^{-1}$ of Pb+Pb collisions at 5.36~TeV was delivered to ATLAS, corresponding to $\sim$2.66~MGy (0.19~MGy) deposited at peak in the ZDC (RPD). 
Based on radiation-hardness studies of the fused silica used in the Run 3 ZDC, no significant degradation of the radiator transmittance is expected in the range of wavelengths accepted by the borosilicate windows on the PMTs until 5~MGy -- the maximum dose for which data are available in the literature \cite{Yang:2022xfv}. Using this 5~MGy value as a target maximum accumulated dose for the detectors, the ZDCs have been exposed to half this dose during 2023 and 2024 operation. The total dose from upcoming runs in 2025 and 2026 is expected to be less than that from 2023 and 2024 and thus safely within the target 5~MGy. This assumes no high-luminosity \pPb\ run is scheduled. If such data-taking occurs, the accumulated dose to the ZDCs may exceed the range for which the potential impact on the fibers is known. 

The intense flux of forward neutral particles impinging on the TAN during high-luminosity \pp\ operation causes substantial activation of the TAN. Even with a few days of cool-down typically provided by machine-development periods scheduled prior to the start of LHC technical stops, high activation levels  at the TAN surface and in the Cu bars could potentially deliver significant radiation dose to the workers installing the ZDCs. For example, 
during the technical stop preceding the ZDC installation in 2023 (2024), the peak dose rate measured on the copper bars extracted from the TAN was 18 (50)~mSv/h. 
For this reason, the ZDC calorimeter modules and the RPD are installed using a fully remote-controllable mini-crane system that is located above the TAN and operated by the CERN transport team. 
The use of the mini-crane limits the personnel exposure only to the detector connection time, an operation that can be accomplished in typically 10-15 minutes.

\begin{figure}[!tb]    
\centering 
\includegraphics[width=0.48\textwidth]{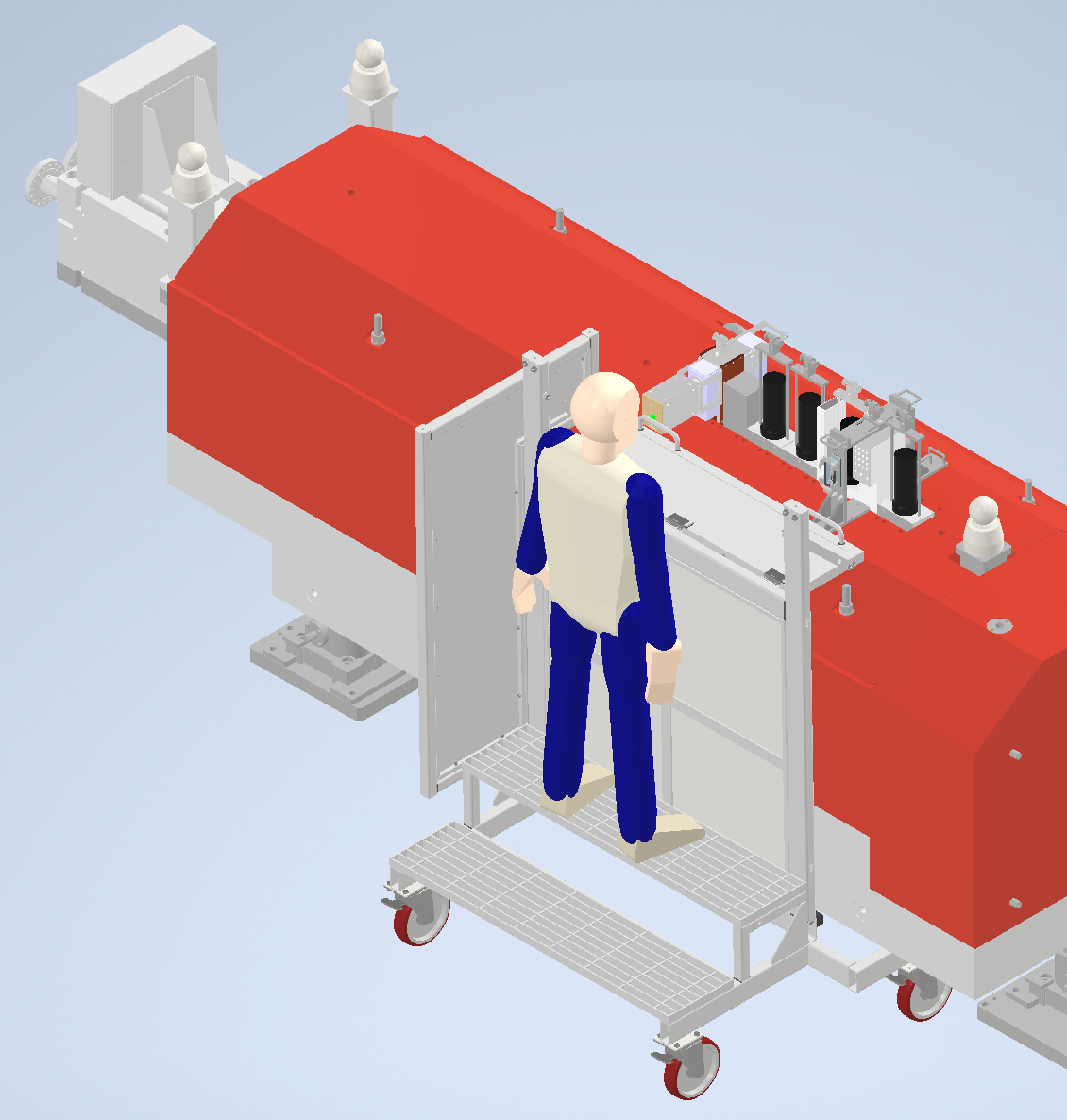} 
\includegraphics[width=0.43\textwidth]{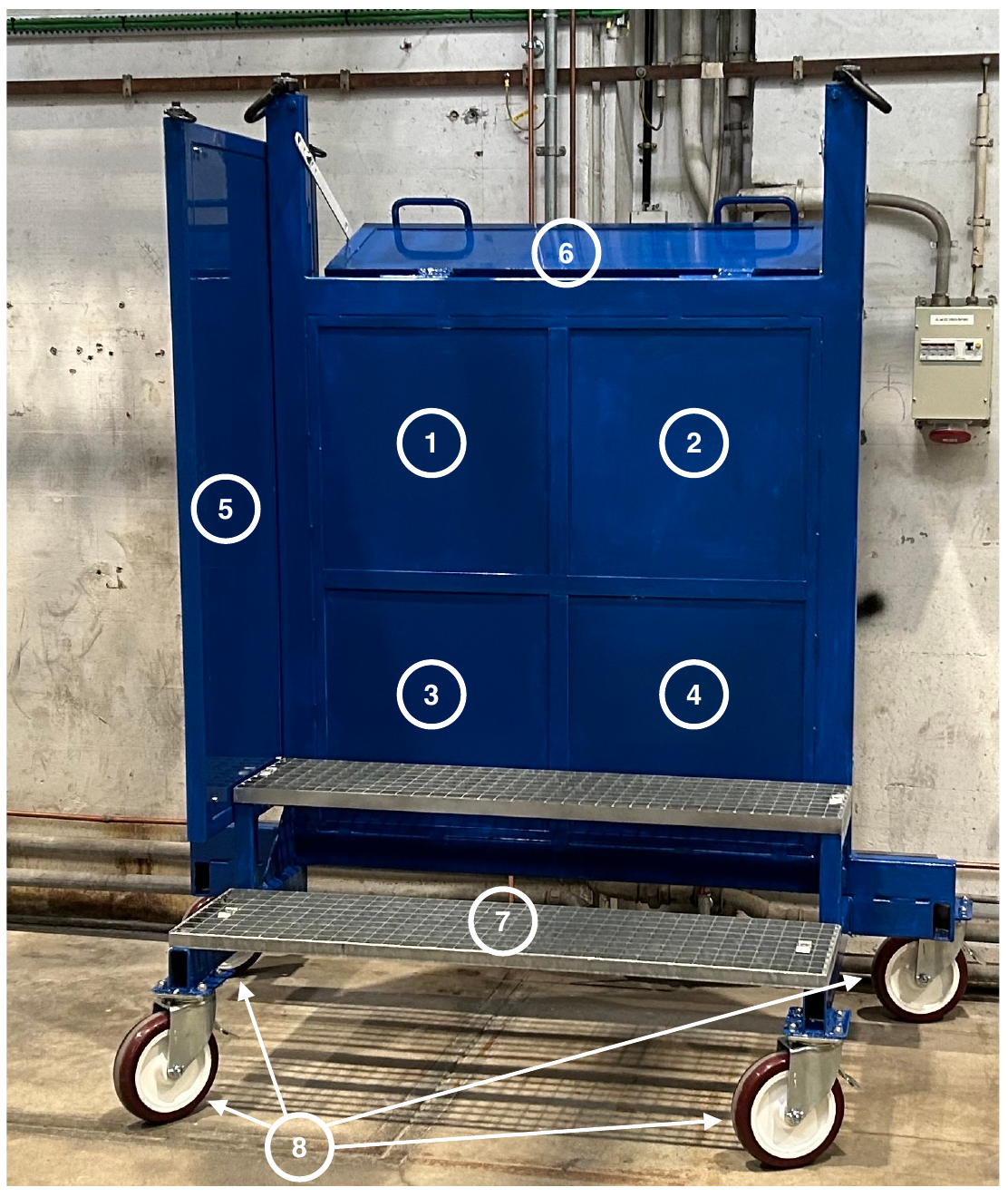} 
\caption{Left panel: a CAD 3D model showing the shielding nearby the TAN, as used during detector installation. Right panel: picture of the ZDC shielding, fully assembled and painted. The numbering on the picture indicates different components: front shield panels (1), (2), (3) and (4), side shielding panel (5), balcony shielding (6), built-in steps (7) and heavy-duty locking swivel casters (8). 
}
\label{fig:shielding}
\end{figure}

To moderate the radiation to personnel involved in the connection of the detector, new lead-based radiation shielding was designed and constructed to be used for the ZDC installation. A three-dimensional drawing showing the shielding placed near the TAN  is shown in the left panel of Figure~\ref{fig:shielding}. The shielding is comprised of 5 stacked layers of 2~mm lead sheet. The lead is safely encapsulated between steel plates, which are mounted on a steel frame. 

The top part of the shielding is connected to the main body via hinges, and can be folded down to rest on top of the TAN. The resulting balcony  provides additional shielding to the chest area for personnel working on the detector. The shielding also includes a side panel that protects against radiation originating in LHC collimators upstream of the TAN. This panel is detachable and can be moved by portable crane to either side of the frame to adapt to the different directions of the collimators on the A and C sides of ATLAS. 
Steps provide the user a better vantage point for working on the ZDC. The frame easily rolls on heavy-duty locking swivel casters.
A photograph of the shielding as fully constructed is shown in the right panel of  Figure~\ref{fig:shielding} with labels indicating major components. 
An assessment of the effectiveness of the shielding was provided by measurements carried out by the CERN Radiation Protection team. The reduction in dose delivered to operators was found to be about a factor of three. 

%% file: detector_details.tex
\input{calor.tex}

\subsection{RPD}
\label{sec:RPD}
\input{RPD.tex}

\subsection{High voltage generation and distribution}
\label{sec:HV}
\input{HV.tex}

%% file: calor.tex
\subsection{ZDC}
\label{sec:calor}

The ZDC modules are sampling calorimeter sections composed of alternating layers of tungsten absorber and close-packed fused silica Cherenkov radiator rods. 
In each module, Cherenkov photons
produced by particles traversing the rods with velocities $\beta \gtrsim 0.7$ are transmitted via total internal reflection to the top end of each rod.  
At that point, the light exits the rods and is directed by a reflective
air light guide into a photomultiplier tube. 

\begin{figure}
\includegraphics[width=0.95\textwidth]{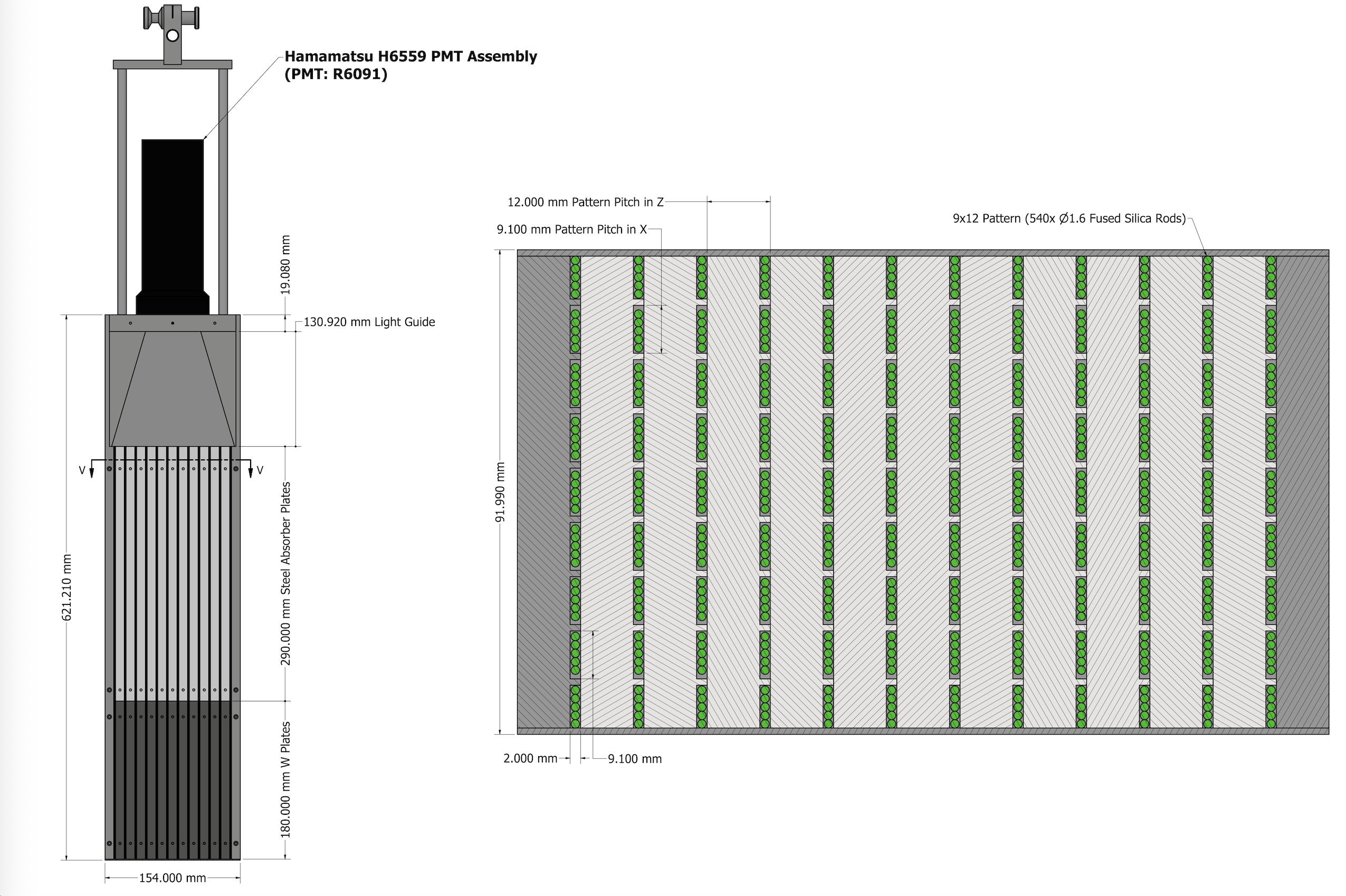}
\caption{Schematic showing the structure of a typical ZDC calorimeter module (one having no longitudinal fibers). The left diagram shows a side view of the internal layout of a module. The components from bottom to top are: tungsten plates, steel plates, the light guide, and the PMT. The right diagram shows a top view of the internal structure with the tungsten plates indicated by the light-gray hatched areas and the cross-section of the fibers represented by the green circles. The dark gray hatched area represents the steel case of the detector. }
\label{fig:ZDCModules}
\end{figure}

\begin{figure}
\includegraphics[width=0.36\textwidth]{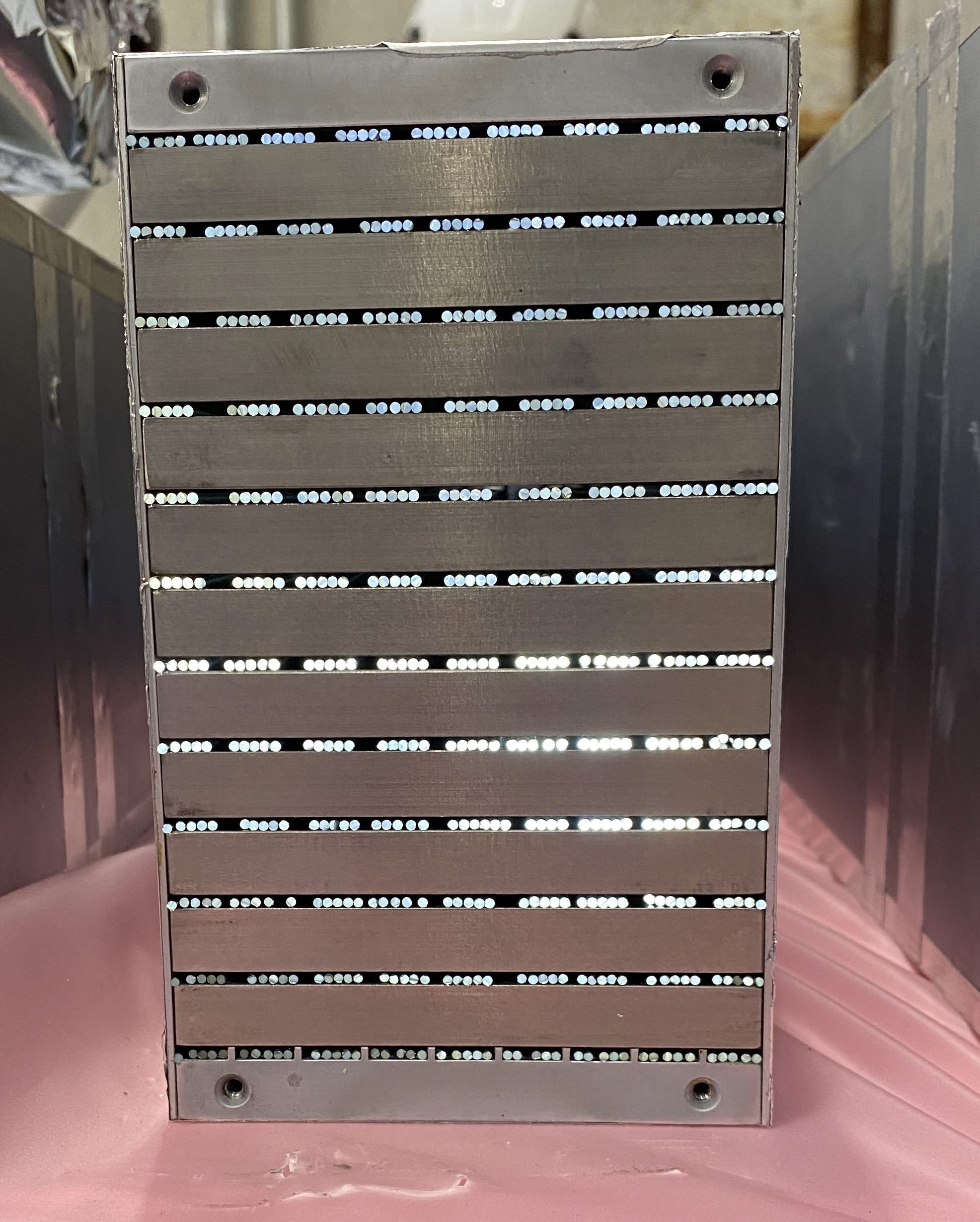}
\includegraphics[width=0.59\textwidth]{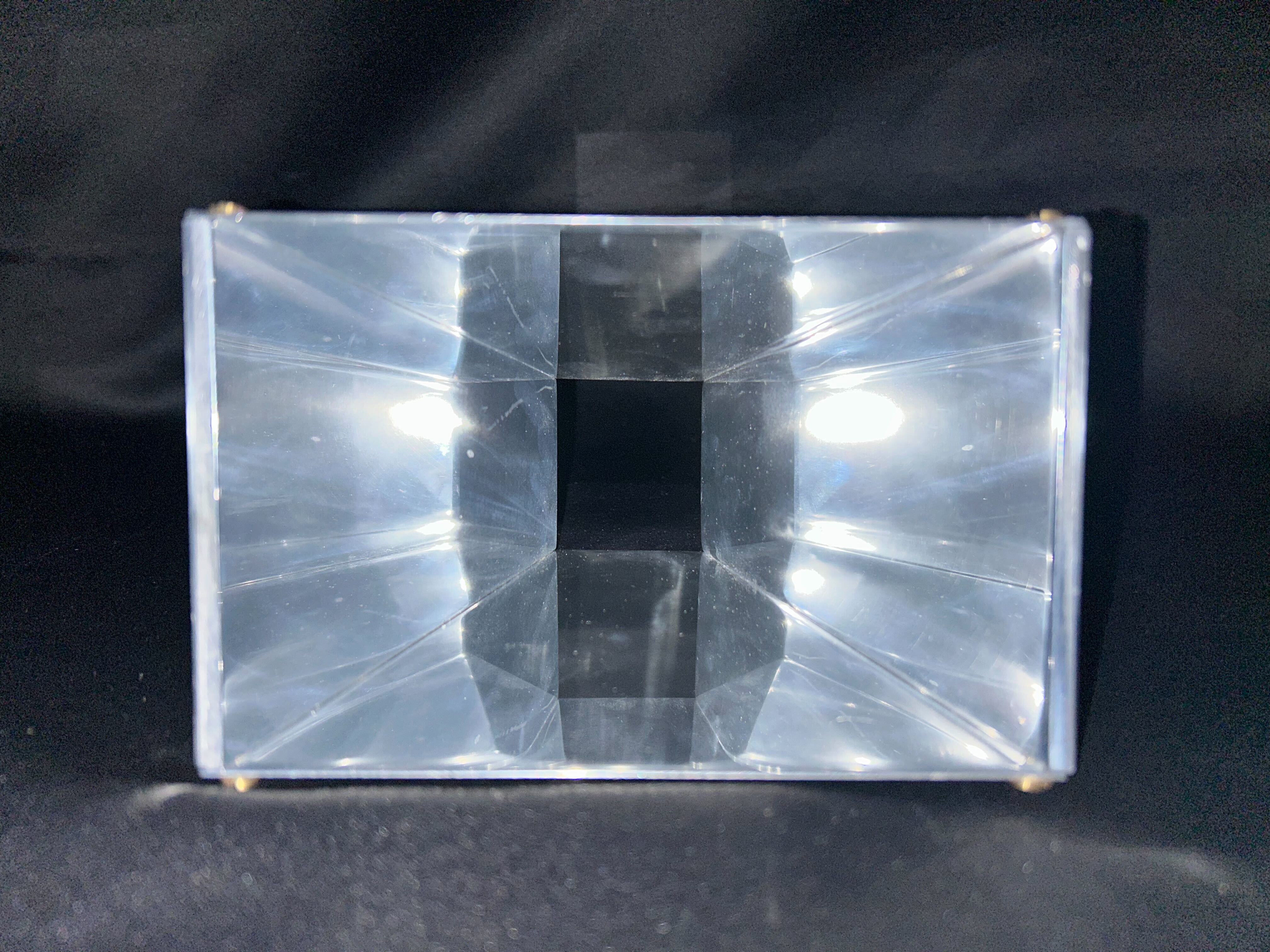}
\caption{Left: bottom view of a fully loaded ZDC module, featuring polished radiation-hard fused silica rods. The rods are illuminated by directing light through the ZDC’s trapezoidal light guide.
Right: A ZDC trapezoidal reflective light guide captured in a dark environment. The camera flash, positioned near the center of the light-guide hole, highlights its reflective properties. }
\label{fig:ZDCModule_real}
\end{figure}

A schematic diagram showing the
structure of a typical hadronic module -- i.e. without the Run~1 transverse
segmentation -- is shown in Figure~\ref{fig:ZDCModules}. 
The left diagram shows a side view of the internal structure of the module while the right diagram shows a top view, with the PMT removed. 
The active area of each module is $\approx 90$~mm wide and 180~mm high. 
Eleven W plates, 10~mm thick, located at the bottom of the module, provide the primary absorber material. These are interleaved with 12 layers of vertical fused silica rods, which are 470~mm long and 1.5~mm in diameter. 
As shown in the right side of Figure~\ref{fig:ZDCModules}, the rods are arranged lateral to the beam in nine groups of five rods each. 
Steel plates, 89.6~mm wide, 290~mm high and 10~mm thick are placed above the tungsten plates. These attenuate the mostly electromagnetic components of showers that extend above the active area of the calorimeter. 
This limits the contribution of these particles to the total light yield and reduces the radiation dose delivered to the PMTs.  
Grooves in the steel plates guide the placement of the rods, and maintain their vertical orientation. 
The trapezoidal reflector shown in the left of Figure~\ref{fig:ZDCModules} is 131~mm high, has a base that is $92\times 144~{\text{mm}}^2$ in area, and has a top of dimensions $54\times 54~{\text{mm}}^2$. 
Figure~\ref{fig:ZDCModule_real} shows both the rod arrangement in one ZDC module (left) and one mirror-finished trapezoidal light guide (right). 

The calorimeter module is fully enclosed by steel plates that provide both structural integrity and positioning for the above-described detector components. The plates are 10~mm thick at the front and back of the detector, 1.2~mm thick on the sides and bottom, and 20~mm thick on the top. The tungsten plates and steel plates are aligned and held in place with two rows each of screws that penetrate the side plates. The front and back steel plates are thinned to 5~mm thickness over the top 150~mm to provide space and support for the reflector. The top plate has a hole of radius 36~mm into which the PMT is inserted. A harness comprised of steel rods, 280~mm long and 10~mm in diameter, that are screwed directly into the top plate of the ZDC and a 15~mm thick steel plate with dedicated crane attachment is used to lift and transport the module.

The dimensions of the ZDC sensitive region, larger in the vertical direction than horizontal, have been chosen to both fill the transverse space available in the TAN, and to provide some flexibility in the 
choice of vertical crossing angle.  
The aperture for neutral particles is limited by the D1 dipoles located upstream of the ZDC. These magnets effectively restrict the ZDC acceptance to $\pm 4.5$~cm relative to the geometric center of
the absorber. 
The absorber is centered at the vertical center of the outgoing beam pipes, about 6.7~cm above the bottom surface of the TAN.
This gives a  pseudorapidity acceptance of $|\eta|>8.7$ in the horizontal and vertical directions, but $|\eta|>8.4$ along the diagonal.

As noted in Section~\ref{sec:introrun12}, 
the ZDC was instrumented during LHC Runs 1 and 2 with GE214 quartz radiator rods. This quartz is susceptible
to radiation damage through the creation of color-absorption centers within the quartz crystal structure. These color centers significantly
degrade the transmission of light over a broad range of wavelengths extending from the UV region, where a substantial fraction of the detectable Cherenkov emission occurs, up to 900~nm. To improve the radiation hardness of the ZDCs, an R\&D program
was undertaken jointly with the LHC BRAN team to
study the effects of radiation damage in fused silica \cite{Yang:2022xfv}. 
Samples of fused silica with different levels of OH and H$_{2}$ doping were placed in a BRAN prototype module and exposed to 40-50~$\text{fb}^{-1}$ of 13~\TeV \pp\ collisions during LHC beam operations in Run~2. 
Analyses of these samples demonstrated that Spectrosil 2000 with high
H$_{2}$ doping (${\sim} 1000$~ppm), would provide sufficient
radiation hardness for a radiation dose equivalent to
1-2~$\text{fb}^{-1}$ of 13~\TeV \pp\ collisions.  This dose is greater than that foreseen for the entirety of heavy ion operations during Run~3. 
Thus, as part of the Run~3 upgrade, the quartz rods in the ZDC were replaced with $\text{H}_2$-doped Spectrosil 2000 fused silica rods provided by Heraeus Quarzglas. 

\begin{figure}[!tbp]
\centering 
    \includegraphics[width=0.9\textwidth]{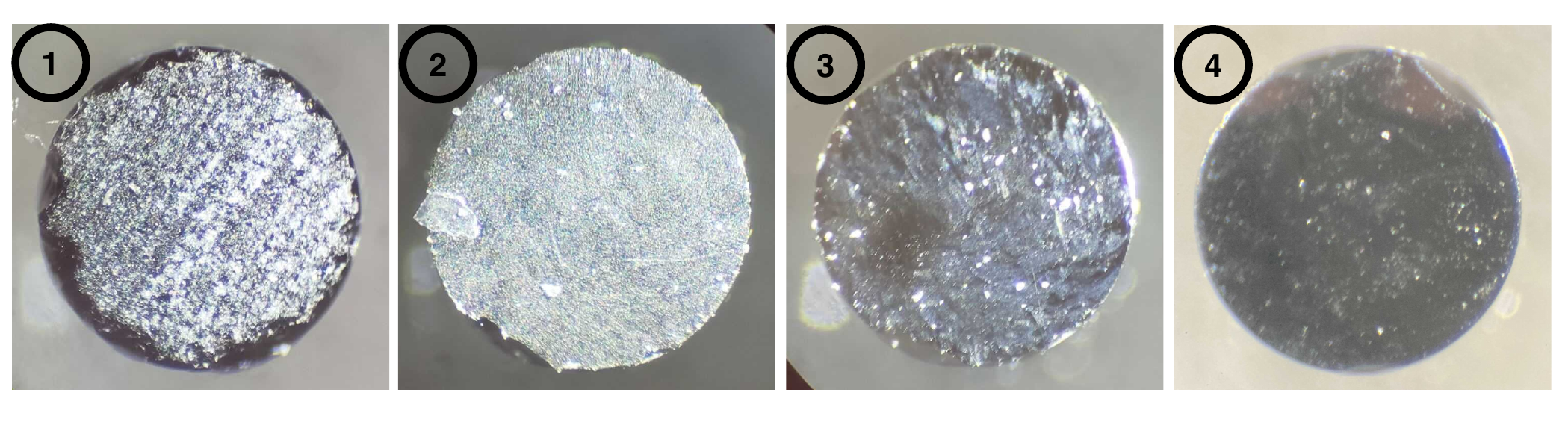}
    \includegraphics[width=0.95\textwidth]{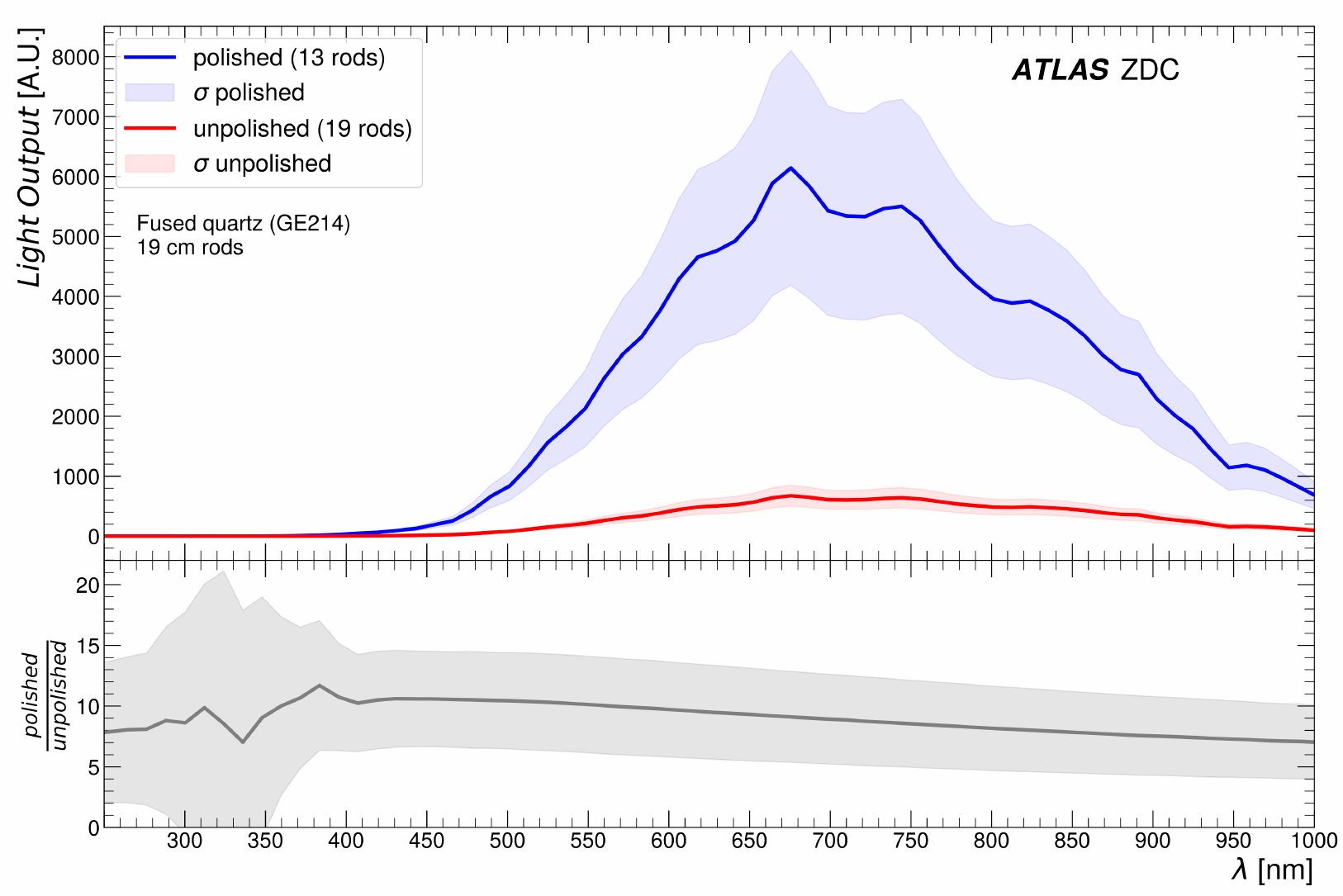}
    \caption{A comparison of light transmission through polished and unpolished rods. Top: a microscope image of the face of a rod before and after each of the three stages of polishing: 1-unpolished; 2-after 9~\mum abrasive; 3-after 3~\mum abrasive; 4-after 0.5~\mum abrasive.  The light originates from the camera direction. Bottom figure: the top panel shows the average light output in arbitrary units as a function of wavelength for polished (blue) and unpolished (red) GE214 rods and the bottom panel shows the relative improvement in transmission of polished over unpolished rods. }
    \label{fig:trans_eff}
\end{figure}

To maximize light transmission from the rods, the upper ends of the
rods were polished while the bottom ends were left unpolished. 
Custom polishing setups were located at Ben Gurion
University and the University of Illinois Urbana-Champaign
\cite{Shenkar:2023} based on Buehler ECOMET~3 polishing machines. 
These were augmented with an additional DC motor that provide a
second (independent) off-center rotation of the polishing base plate.  
An apparatus for holding a batch of 17 rods was attached to the polishing
plate, and the rods were polished using a sequence of 9, 3, and 0.5~\mum
grade abrasives. Figure \ref{fig:trans_eff} demonstrates the improvement
in light transmission following the polishing. The top figure shows microscope photographs of the evolution of the surface of a rod through the different polishing steps from unpolished (left) to fully polished (right). The bottom plot show results of a measurement of the wavelength dependence of  the transmittance of polished and unpolished GE214 
rods\footnote{The quartz rods were used instead of fused silica for the initial studies of polishing as they are much less expensive}. The lower panel in the bottom figure quantifies the improvement in the transmittance as a result of the polishing.

The ZDC modules are equipped with Hamamatsu~H6559 assemblies that utilize the R6091 PMT \cite{Hamamatsu_R6091} which contains a large, 65~mm diameter bialkali photocathode protected by a borosilicate window. This PMT has high 
quantum efficiency for wavelengths between 300 and 650~nm, with the sensitivity peaking at 420~nm. It has a 12-stage, linear-focused dynode
chain. The nominal and maximum anode high voltages for the PMT are 1500 and 2500~V, respectively. The H6559 modB assemblies provide booster connections to support the potentials on the last three dynode stages. The R6091 PMTs have rise times of about 2~ns, commensurate with the fast Cherenkov signal in the fused silica radiator, and a small transit time spread of 1.5~ns. However, they exhibit significant ($> 1$~ns) transit time variation with applied HV. 
Measurements provided by Hamamatsu of the anode luminous sensitivity for the eight PMTs used for the ZDC showed results that varied from $386~\text{A}\,\text{lm}^{-1}$ to $1380~\text{A}\,\text{lm}^{-1}$. Separate measurements performed using an LED calibration system at CERN yielded results that mostly tracked the luminous sensitivity measurements though some of the PMTs showed 10-25\% deviations in (relative) response compared to the Hamamatsu measurements.

As noted in Section~\ref{sec:introrun12}, during LHC Runs 1 and 2, signals from the ZDC calorimeter modules were transmitted to the USA15 cavern over 300~m of ``fast'' CC50 coaxial cables, characterized by a 
signal propagation speed of $0.83c$ but with
non-trivial attenuation (21~dB/100~m at $\sim$~GHz frequencies which are characteristic of the PMT pulses). The resulting broadening of the pulses created serious difficulties with OOT pileup at bunch spacings smaller than 200~ns. To solve this problem, eight new \aircore HELIFLEX HCA78-50JFN cables
were installed between the TANs and the ATLAS USA15 cavern. \Aircore cables are typically utilized in defense and professional
radio/television installations and have notably larger cross section than typical signal cables used in physics experiments. 
The inner copper conductor has 9~mm diameter and the
outer a diameter of 20.2~mm. The outer conductor is surrounded 
by a flame-retardant polyethylene jacket with metalhydroxite filling. 
The spacing between conductors is maintained by a helical polyethylene
spacer, and the cables are filled with
nitrogen gas. The mass density of the
spacer is sufficiently low that the nitrogen determines the dielectric
response of the cable. The HELIFLEX HCA78-50JFN cables have a propagation speed of $0.93c$. 
Lab tests studying the propagation of pulses, with similar rise and
fall time as those produced by the ZDC, over 200~m of this
cable showed only 2-3~dB attenuation of the pulse with
over $90$\% of the charge contained in a single 25~ns LHC
bunch crossing. These results indicated a dramatic improvement in performance over the CC50 cables used in Runs 1 and 2.
 
\begin{figure}[!tbp]
\centering 
    \includegraphics[width=0.95\textwidth]{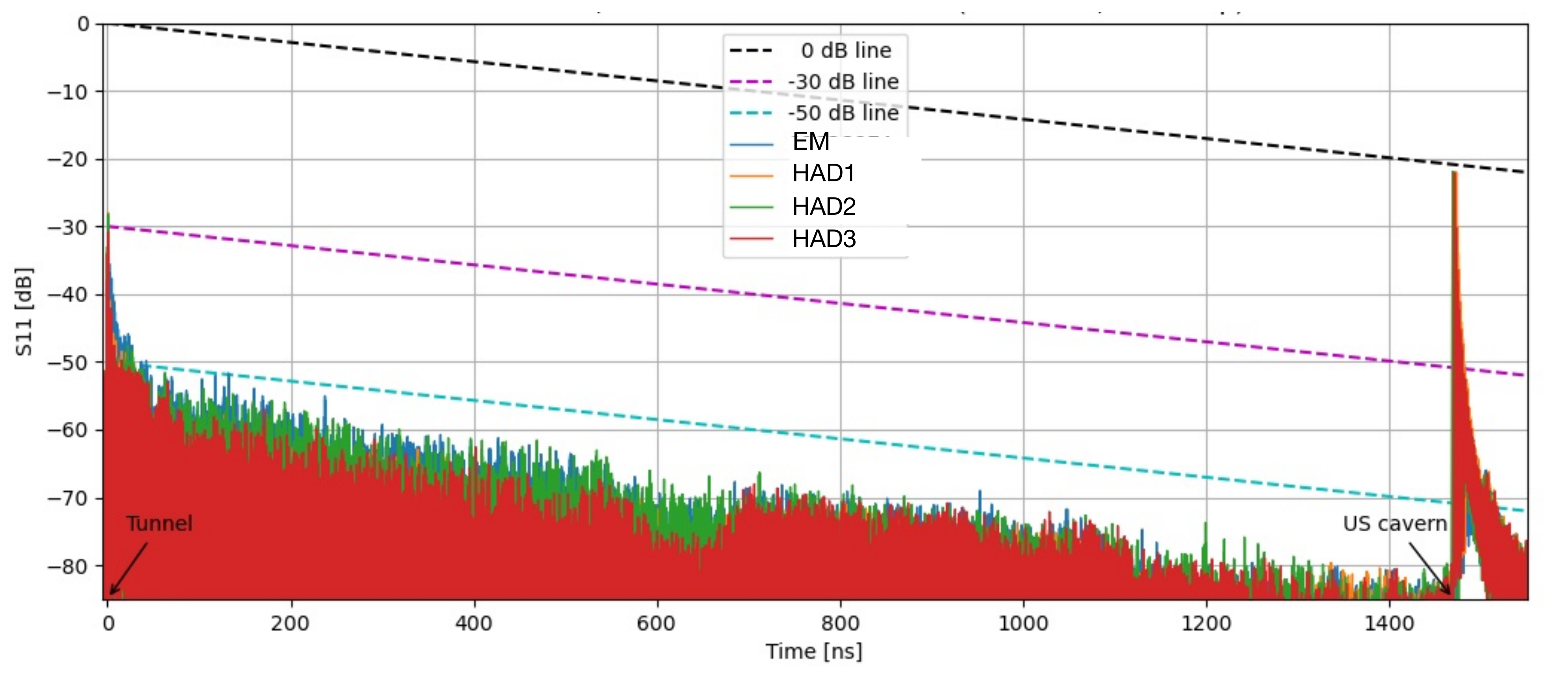}
    \includegraphics[width=0.95\textwidth]{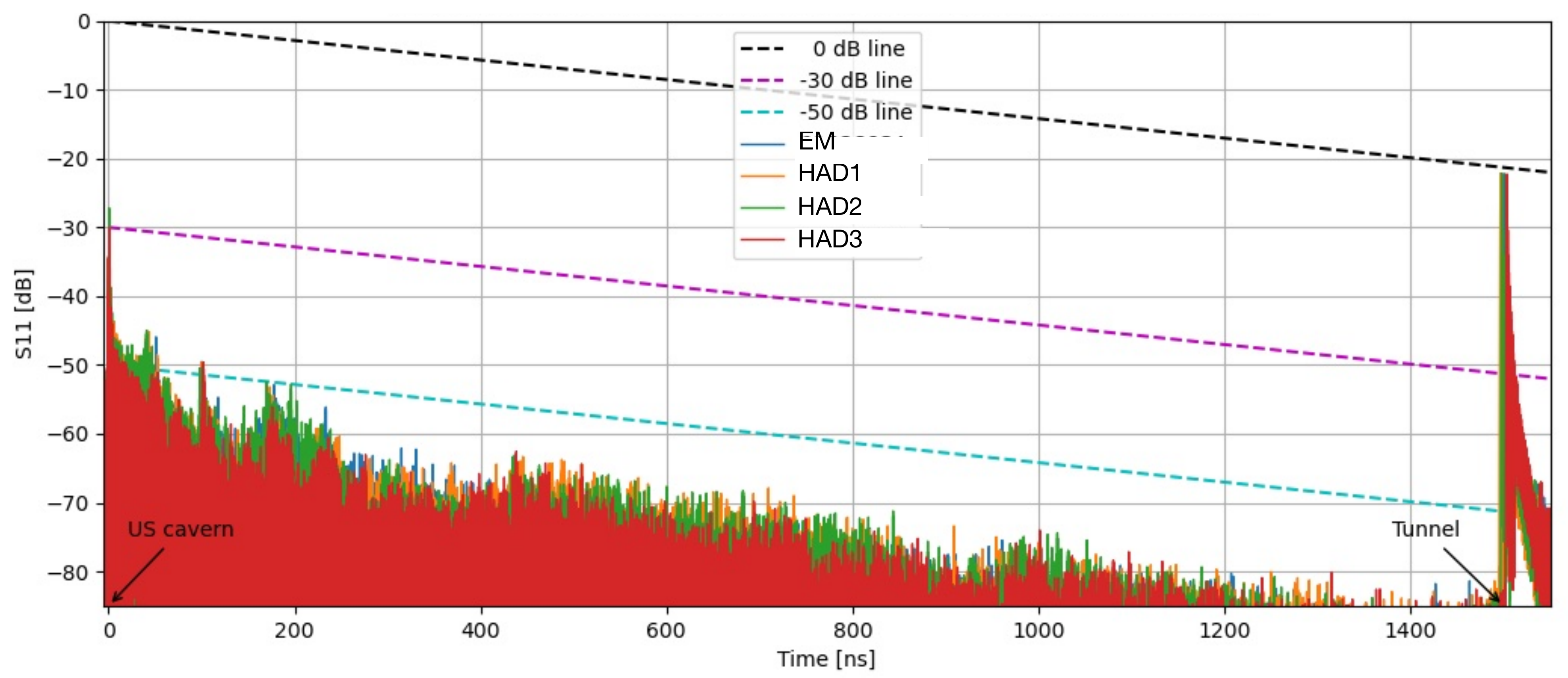}
    \caption{Results of reflectometry measurements performed on the \aircore cables installed for the ZDC in 2021. The figures show $S_{11}$ expressed in dB as a function of time for the reflected signal. Results are overlaid for the four cables on each side: top - side C, bottom - side A. The dashed lines show expected $S_{11}$ values for full reflection (black line), -30~dB reflection, and -50~dB reflection. }
    \label{fig:vnameas}
\end{figure}

\Aircore cables are notably heavier and more rigid than
typical coaxial cables and their performance is much more
sensitive to the delicate conductor geometry.  
Thus, a special installation campaign was organized 
with the CERN EN-EA department during LS2.  
The shortest possible signal path was desired, so the cables were routed through the ``survey galleries'' which provide the most direct access to USA15 without the cables having
to pass through the ATLAS cavern.
The cables were installed by a team of about 20 workers in June 2021 over four days.  Both prior-to and after installation, the cables were tested by the CERN RF team using a vector network analyzer (VNA), which performs frequency-domain reflectometry measurements over a wide range of frequencies. The resulting data are then Fourier-analyzed to yield reflected power as a function of transit time. Defects in the cables typically introduce reflections at specific time intervals associated with the location of the defect along the cable. The results of the reflectometry measurements for the eight \aircore cables are shown in Figure~\ref{fig:vnameas}. The dominant reflection corresponds to the open end of the cable with some spread arising from the Fourier analysis. Neglecting similar effects at zero time, only $\lesssim -45$~dB reflections can be seen along the cables; these have negligible impact on the propagation  of signals.  

The VNA measurements also provide a direct measure of the propagation time for signals along the cables. Because the cables take slightly different routes to the survey gallery, the cables on the C side are shorter. The propagation times, averaged over the four cables, are 736~ns on the C side and 750~ns on the A side. The cable-to-cable deviations from these averages are at most 2~ns.

%% file: RPD.tex
\begin{figure}[!tbp]
\centering 
    \includegraphics[width=0.95\textwidth]{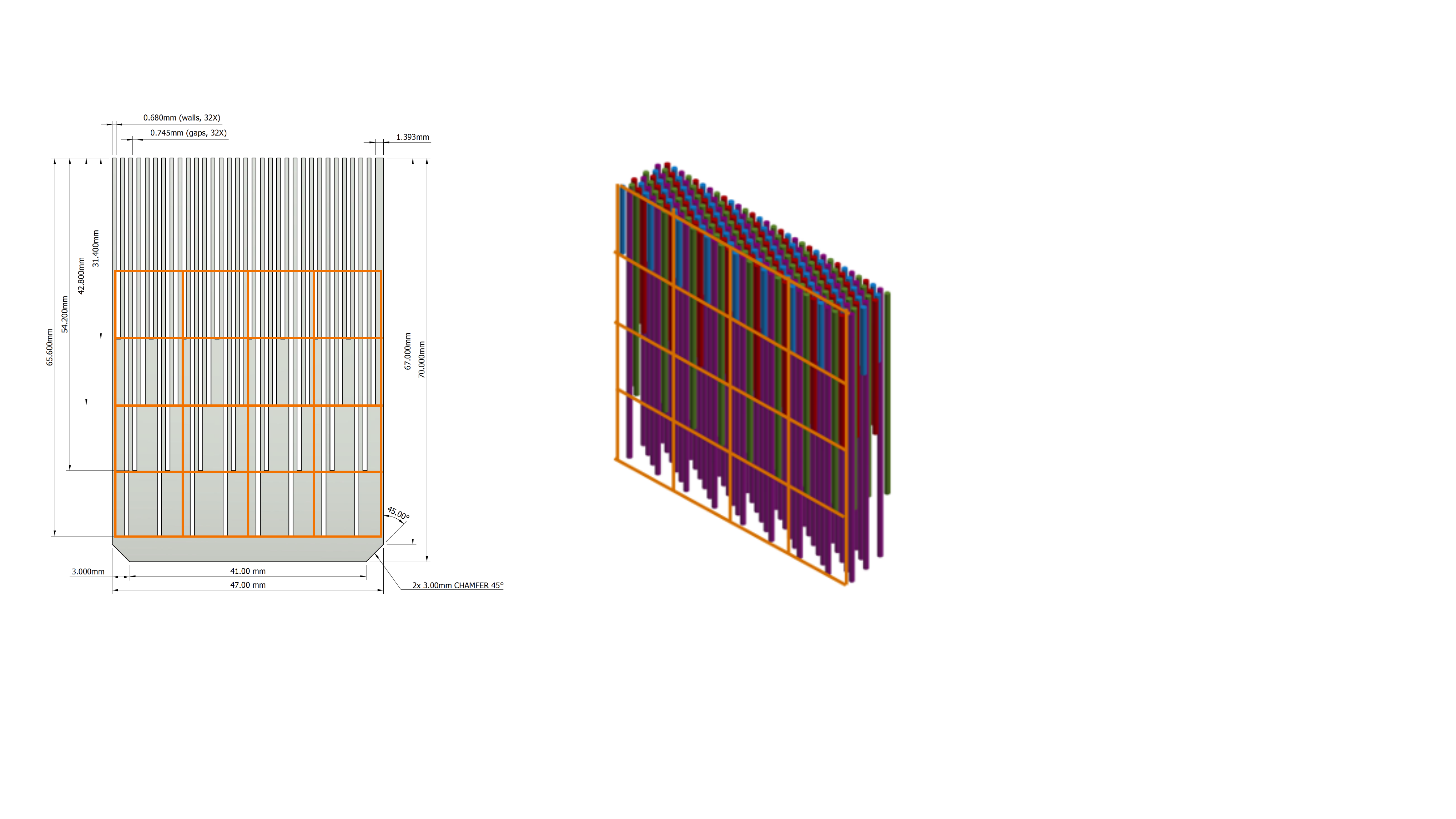}
    \caption{Left: schematic of one of the RPD harps. Right: arrangement of the fibers in RPD active area, implemented using the harps. A representation of the RPD ``virtual'' tiles is reported on each of the drawings. }
    \label{fig:RPD_harps}
\end{figure}

The RPDs are designed to image and measure the centroid of
multi-neutron showers in the ZDC. The constraints on the 
available space imposed by the TAN slot pose a challenge for measuring the vertical position of centroids, since the RPD sits in a narrow slot and thus can only be accessed from above.  
This challenging situation is addressed using a novel ``pan flute''
design originally developed for the future Run~4 ZDC \cite{ZDCHLLHC}.
In this design, 256 fused silica fibers are arranged in a $4
\times 4$ grid of virtual square tiles that cover an active area of $45.6 \times 45.6 \; \text{mm}^2$. The arrangement of the fibers is illustrated in Figure~\ref{fig:RPD_harps}. The horizontal segmentation is implemented
directly via the grouping of the fibers, while the vertical
segmentation is implemented using four different fiber lengths. The longest fibers cover the full vertical acceptance of the RPD, the next longest image the top 3/4 of active area, the next the top 1/2, and the shortest fibers the top 1/4 of the active area. 
Fibers of different layers overlap horizontally by 150~\mum on both sides, for a total of 300~\mum of overlap between fibers of consecutive rows. This pattern is chosen such that every incident charged particle of a shower passing through the active region encounters at least one fiber.

\begin{figure}[!tbp]
\centering 
    \includegraphics[width=0.95\textwidth]{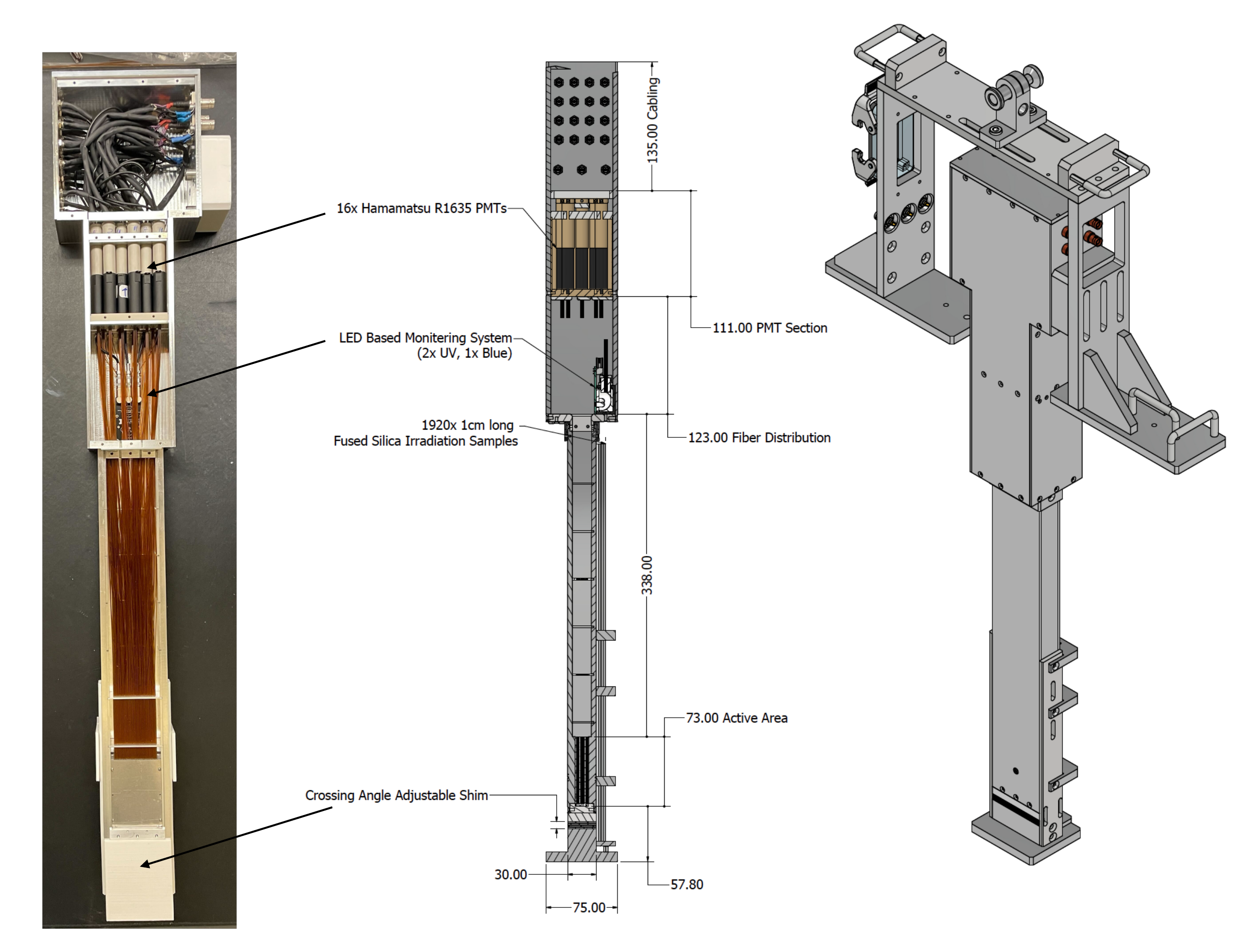}
    \caption{Left: picture of the open, fully assembled, RPD. Center: schematics showing a side view of the detector. Right: 3D view of the detector assembled with its support structure for craning and interface with service nearby the TAN.}
    \label{fig:RPD_fullDetector}
\end{figure}

The fibers are radiation-hard Polymicro FBP600660710 \cite{FBP_Datasheet}. Each fiber has a diameter of 710~$\mu$m, consisting of a 600~$\mu$m fused-silica core, a 60~$\mu$m thick doped fused-silica cladding, and a 50~$\mu$m thick polyamide buffer. The core-cladding configuration results in a numerical acceptance of 0.22, corresponding to a critical angle of about 8$^\circ$ from the vertical. The fibers extend vertically 472~mm beyond the active area to connect with the PMTs. They are secured using two sets of aluminum support plates, each featuring a unique pattern of machined fiber guide channels. 

\begin{figure}[!tb]
\centering 
    \includegraphics[width=0.95\textwidth]{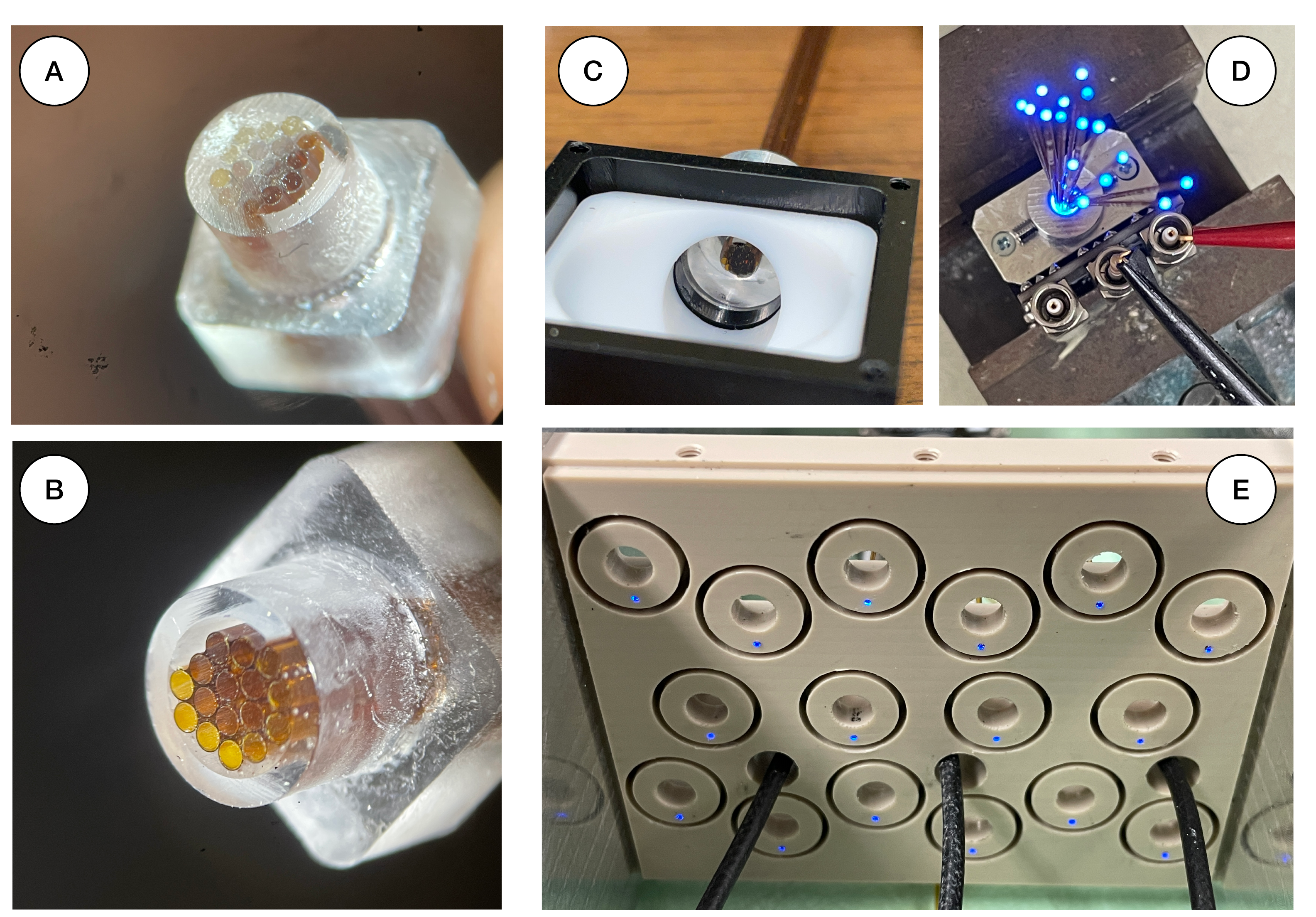}
    \caption{(A) RPD fiber plug before optical finishing of the surface. (B) RPD fiber plug after optical finishing of the surface, ready for installation in the detector. (C) Detail of the RPD LED showing the optically finished fiber plug inserted into the PTFE homogenizer box. (D) Fibers exiting the RPD LED system, illuminated using one of the 370 nm LEDs inside the homogenizer. (E) Top view of the RPD PMT interface plate. LED fibers are illuminated and visible on each of the PMT slots. The three cables entering the plate carry the LED driving signals from the top of the chassis to the LED board inserted into the homogenizer. }
    \label{fig:RPD_Details}
\end{figure}

The connection between the fibers and the PMTs is achieved via a PEEK interface plate. This plate has recesses on one side to accommodate the fibers and on the other side for the PMTs, which are equipped with mu-metal shields (see panel E of Figure~\ref{fig:RPD_Details}). To ensure a low-dispersion optical interface with the PMTs, the fibers corresponding to each channel are glued into an acrylic plug. After trimming the plugs to the desired length, the surface interfacing with the PMTs remains rough (see panel A of Figure~\ref{fig:RPD_Details}). To address this issue and maximize the uniformity and efficiency of optical transmittance, the plugs undergo a custom polishing procedure specifically developed for these components. The results of this polishing process, demonstrating the improved optical finish, are shown in panel B of Figure~\ref{fig:RPD_Details}.

The light from the 16 groups of fibers is detected using Hamamatsu R1635 PMTs \cite{Hamamatsu_R1635}, which are 10~mm in diameter and have 8 stages of linear-focused dynodes. Each PMT is equipped with a borosilicate window and an 8~mm diameter Bialkali photocathode which provides sensitivity over wavelengths from 300 to 650~nm, with a peak sensitivity at 420~nm. The HV-gain characteristic of the PMTs was provided by Hamamatsu and used to sort them into groups of four, to minimize HV-gain variations within each group.
In the detector, each group receives an independent HV supply with the result that only four independent HV lines are needed for each RPD detector. Potential degradation of the PMT windows from radiation damage is monitored using an LED calibration system, installed below the interface plate (see Figure~\ref{fig:RPD_fullDetector}). The system is described in Section~\ref{sec:led}. 

The signal and HV lines within the detector are routed using LEMO cables (HV wires) to surface-mount LEMO (SHV) connectors affixed to the RPD chassis. Each HV wire is encased in a dedicated heat-shrunk knitted mesh, that is connected to ground via the detector chassis and provides electrical shielding to avoid cross-talk between the lines during operations. 

Signals from the RPD are transmitted from the TANs to the USA15 cavern over $\sim$300~m long CB50 cables that were installed for use by the
LHCf detector\cite{LHCf:2008lfy}.
Short LEMO cables are used to connect the surface-mount LEMO connectors on the detector to a Harting Han-Modular multi-pin connector mounted on the ``wings'' support structure of the detector. This setup facilitates the transition to the LHCf CB50 cables through a single connection. The relatively poor dispersion of the CB50 cables is such that signals produced by the RPD are attenuated by
a factor of $\sim 5$ between the detector and the front-end
electronics in USA15, with a long fall time of about 35~ns. 

Proper positioning of the active area relative to the beam direction is critical to the performance of the detector. The transverse alignment is primarily defined by the upper part of the RPD body, which was designed to fit the TAN slot width with 2 mm of tolerance on either side. To aid the positioning during the installation, a T-shaped foot, covering the entire shadow of the top portion of the detector below the active area, was included in the design, see Figure~\ref{fig:RPD_fullDetector} central and right panels.

\begin{figure}[!tb]
\centering  
    \includegraphics[width=0.98\textwidth]{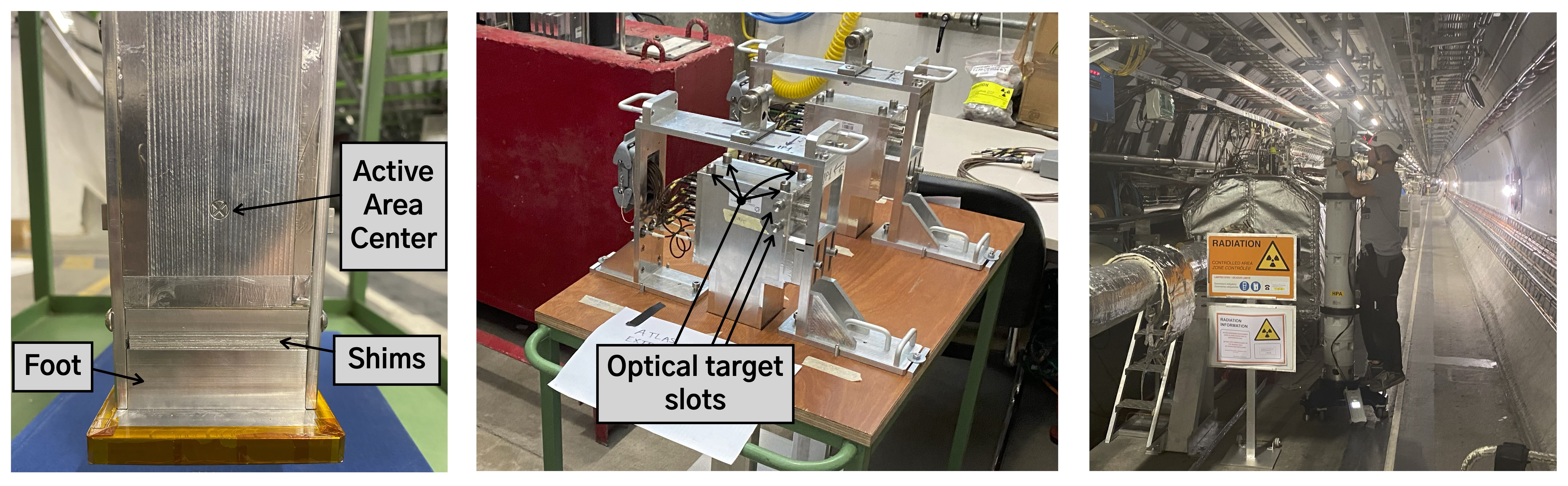}
    \caption{Left: close view of the RPD bottom region in the 2023/4 configuration. The engraved bullseye indicates the center of the RPD active area. The foot and the shims are also visible. Center: location of the optical target slots on the RPD. Right: survey of the RPD in the TAN, performed by the BE-GM team before the Heavy Ion data-taking.  }
    \label{fig:RPD_align}
\end{figure}

The vertical position of the RPD is chosen based on the vertical crossing angle set by the accelerator during heavy-ion data taking in a given year. In both 2023 and 2024, this angle was set to +150~$\mu$rad, which results in a +21.2~mm vertical shift at the RPD face. The vertical position of the RPD is adjusted by varying the foot height before installation. This adjustment is achieved using a series of threaded shims placed between the detector and the foot, allowing precise positioning of the active area ( see Figure~\ref{fig:RPD_align}, left panel). The vertical granularity is determined by the shim thickness, with each shim providing a 1~mm adjustment. The adjustable foot height can accommodate positions corresponding to crossing angles in the range +100~$\mu$rad to +200~$\mu$rad. 
During installation, the detector is craned into the slot in its fully extended configuration (+200~$\mu$rad). In this arrangement, when the foot is resting on the TAN slot floor, the wings remain suspended.
The next step involves loosening the bolts that secure the wings' position and gently lowering them until they rest on the TAN surface. The horizontal and longitudinal position of the detector in the TAN is then fixed using brackets that latch onto dowel pins inserted in dedicated holes present in the TAN slots. Once the RPD is locked in place laterally and longitudinally, the bolts on the wings are tightened again, fixing the vertical position of the RPD relative to the wings. 
A configuration for a zero vertical crossing can also be achieved by utilizing the full vertical range of the mounting rails on the wings. In this scenario, the foot is not installed, and the detector rests directly on the wings, preventing any downward movement. 

After installation and before operations begin, the CERN BE-GM group surveys the detector’s position relative to the beam axis. This is achieved using a series of optical target slots located on the top and front surfaces of the detector (see Figure~\ref{fig:RPD_align}, central panel). The 3D structure of both RPDs was mapped in the BE-GM database during a \textit{fiducialization} measurement carried out after the detector construction. This procedure must be repeated whenever the detector is mechanically opened for maintenance. The survey team uses the results of the fiducialization to project the body of the detector in the LHC reference system after measuring the optical targets -- a procedure shown in the right panel of Figure~\ref{fig:RPD_align}. The survey is conducted during the technical stops of the LHC, following the installation and connection of the detector and HV commissioning, in order to avoid any post-measurement change in detector position.

%% file: HV.tex
High voltages for the ZDC and RPD PMTs are generated using a combination of CAEN A1535N and A1833N modules mounted in a CAEN SY4527 mainframe that is located in the USA15 cavern. The high voltages are delivered to the the TANs on the two sides of ATLAS using three 37-conductor cables. Two of these are routed from USA15 to the Side-A TAN, and the other is routed to the Side~C TAN. These cables are equipped with REDEL SAG.H51 connectors at both ends. Custom passive HV patch modules are used at both ends to translate connections from the HV supply in USA15 to the multi-conductor HV cables, and to break out the connections at the TANs. The ZDC main HV and the RPD HV connections are made using standard SHV cables that run between the patch boxes located below the TANs to the detectors. The ZDC booster high voltage connections are made using three-conductor HTC-50-1-1 cables equipped with three SHV connections to the patch boxes and a Hirose RM12BPE-4S(71) connector at the PMT end.

For the ZDC PMTs, the ratios of the booster voltages to the primary voltage are nominally fixed by the H6559 voltage divider
chain, but when operating the PMTs, the booster voltages are raised relative to the nominal
settings to ensure at least 10~\muA current draw from the HV
supply during quiescent operation of the detector. This choice is based on prior ATLAS experience, especially with operation of boosters on the LUCID PMTs.

%% file: readout.tex
\subsection{LUCROD overview}
\label{sec:readout}

As noted in Section~\ref{sec:introrun3}, the full electronics chain for
the ZDC was upgraded for Run~3 using LUCROD boards originally designed
for the LUCID2 detector in ATLAS \cite{LUCID2}. The LUCRODs are 9U VME
modules that provide up to 16 independent channels of digitization and
readout. Each board has 16 LEMO inputs and 16 LEMO outputs for analog
signals, four LEMO outputs for trigger, busy, and diagnostic signals,
an optical receiver for ATLAS trigger and timing signals, and two
optical outputs. One of these optical connections implements
the \Slink protocol \cite{slink} needed to transfer ZDC data to the  ATLAS data acquisition system \cite{ATLASTDAQ}; the other is currently unused. The LUCROD boards
contain 10 field programmable gate arrays (FPGAs). Eight of these,
labeled FPGAch, are dedicated to receiving and processing input data, the other two manage the operation of the board and
handle external communication. The
programming for all FPGAs on the board are stored in flash memory
which can be written over VME or via JTAG.

Analog signals are received at the LUCRODs on 16 dedicated cards that provide 50~\ohm termination, amplification and
baseline adjustment controlled by digital-to-analog converters (DACs) located on the main LUCROD board, and conversion to
differential output using Analog Devices AD8138 drivers. The analog cards also
provide a copy of the input signal on an output LEMO connector, 
amplified and with its own baseline adjustment. 
The offset and gain for this 
copied input signal
are also controlled by dedicated DACs. 

The signals received from the analog cards are digitized on
the LUCROD main board using Analog Devices AD9434 (AD9434BCPZ-500) 12-bit flash ADCs (FADCs) operated at a 320~MHz sampling rate. These FADCs have a dynamic range of 1.5~V that is nominally matched to that of the analog input cards. However, the dynamic range of the analog signals is affected by the applied baseline offsets with the result that the analog pulses were sometimes found to saturate at amplitudes lower than 1.5~V, {\it i.e.} at FADC values below 4095. This behavior could adversely affect the signal processing in the trigger, but it was easily cured via application of a $\times 1.1-1.15$ amplification of the input signals.

The LUCROD firmware
combines pairs of inputs to provide a single readout channel associated with each FPGAch. In each pair, data from only one of the FADCs is selected for subsequent processing. This implementation provides redundancy in case of failure of one of the analog cards or FADC. It also
allows implementation of more complex processing of the data in the FPGAch
which is implemented using an ALTERA
Cyclone IV chip (EP4CE40F).  Specialized firmware to generate ZDC triggers
processes the received data in a pipeline that is described in
Section~\ref{sec:trigger}, and the data are also separately formatted and
stored in a ring buffer on the channel FPGA for retrieval on
receipt of Level-1 trigger accept. The 320~MHz sampling frequency implies that
eight 12-bit words need to be recorded for each channel 
in a single LHC bunch crossing (\BC). 
The phase of the FADC samples within the crossing can be adjusted
separately for each channel using a register implemented on the FPGAch. This FPGA can also store data for asynchronous VME readout in a ``spy'' buffer that records FADC samples from 8~\BCs starting at a programmed bunch crossing number.

Two FPGAs implemented using ALTERA Cyclone~IV chips
(EP4CGX30F) control the operation of the LUCROD and manage all
inputs and outputs for the board. 
Both FPGAs are linked to the VME bus and can
respond to VME transfers.  One of these FPGAs (FPGAv) manages communication with the channel FPGAs, provides the
final steps of trigger processing, and implements the state machine
for the operation of the board. It also receives signals from the
TTCrq \cite{Butterworth:2004nla} card which decodes the master clock and trigger data from the
ATLAS trigger system. The other FPGA (FPGAm) receives data from the FPGAch, formats it and transfers the data over \Slink to the ATLAS data acquisition system.

A 40~MHz (nominal) clock is distributed to all FPGAs using a dedicated
clock manager chip whose input can be selected to be either the ATLAS
clock or that produced by a 40~MHz oscillator on the LUCROD. Each FPGA
implements a phase-lock loop (PLL) that generates five different clocks at frequencies
of 20, 40, 160, and (2) 320~MHz. The 100~MHz clock needed by the \Slink interface is generated
from a separate on-board oscillator. For LUCID, an 80~MHz oscillator
was chosen and that choice was initially preserved for the ZDC. The
\Slink clock was then generated in the FPGAm using frequency division
by four and multiplication by ten. However, a subset of the boards
fabricated for the ZDC generated frequent \Slink down events, 
which were ultimately traced to a sudden change in phase of the synthesized 100~MHz clock. 
This problem was solved by
replacing the 80~MHz oscillator on the LUCRODs with a 100~MHz
oscillator and simplifying the associated PLL in the firmware. 

Data are received from the six ZDC LUCRODs in a dedicated ATLAS
Readout System (ROS) \cite{ATLASROSRuns23} computer equipped with a
ROBINNP PCI-express card \cite{ATLASTDAQRun2}. This card can support
up to 12 optical connections, although only half of these are used for
the ZDC. The card implements the \Slink protocol and provides 8~GByte
of on-board buffering and 1.6~GByte/s aggregate transfer rate over PCIe.

\subsection{ZDC and RPD readout configuration}
To cover the full dynamic range of signals from \PbPb\ collisions,
dual-gain readout of the ZDC was maintained in Run~3.  In 2023, the
eight signals received in USA15 from the calorimeter modules were
passively attenuated by a factor of 10 to match the 1.5~V dynamic
range of the FADCs and were read out in a single LUCROD to produce
``low gain'' data. The analog output signals from the low gain
channels, amplified by a factor of 10 were then provided to the
inputs of a second LUCROD to produce ``high gain'' readout.  
For 2024 operation, the analog signals from the PMTs were attenuated by a factor
of 10 and passed through  linear fan-in/fan-out (FI/FO) modules to allow the incorporation of injected test pulses as described below. 
Separate outputs from
the linear fan-in/fan-out 
were used for the low and high gain readout. For the
high gain channels, the additional $\times 10$ amplification was provided by the input amplifiers on the LUCROD. 

The 32 channels of the RPD detector are read out using four
LUCRODs, two for each side. The signals are received on CB50 cables
and are input directly to the LUCRODs. They are amplified using
per-channel gain factors such that the combined PMT and
LUCROD amplification gain yield the same nominal factor for all
channels (see Section~\ref{sec:RPDWP}). 

The amplifier DAC settings used for the ZDC and RPD are calibrated via measurements of the gain versus DAC setting using both measurement of noise RMS versus DAC setting and the measured amplitude of pulses injected directly into the electronics (as discussed in Section~\ref{sec:pulser}). These methods give consistent results. An example gain curve is shown in Section~\ref{sec:RPDWP}, Figure~\ref{fig:rpd_operating_points}. 

During normal data-taking, the ZDC and RPD LUCRODs are configured to
read out FADC data from three \BCs for each received L1 accept. The
phases on all channels, and the readout offset in the channel FPGA ring
buffers, are adjusted such that the pulse maximum lies nominally in
the 10th sample. An automatic procedure implemented at the start of
every data-taking run adjusts the per-channel baseline settings so that the
pedestal for each ADC is $\approx~100$. The timing of the readout of the LUCRODs is set by comparing the times of level-1 accepts received from the CTP to the timing of pulses found by the peak-finding algorithm operating in the trigger (see below). In "ramp-up" fills during heavy ion periods, the filled bunches are sufficiently separated that the this timing can be well-determined and used to set the required  pointers into the LUCROD ring buffers. These readout pointers were established during the 2022 \pp special run for LHCf and since then have not needed adjustment. Confirmation that the correct data are being read out is provided by online monitoring histograms that compare ZDC energies to, for example, the total energy measured in the forward calorimeter (See Section~\ref{sec:promptmonitoring}).

\begin{figure}[!b]
\centering 
    \includegraphics[width=0.98\textwidth]{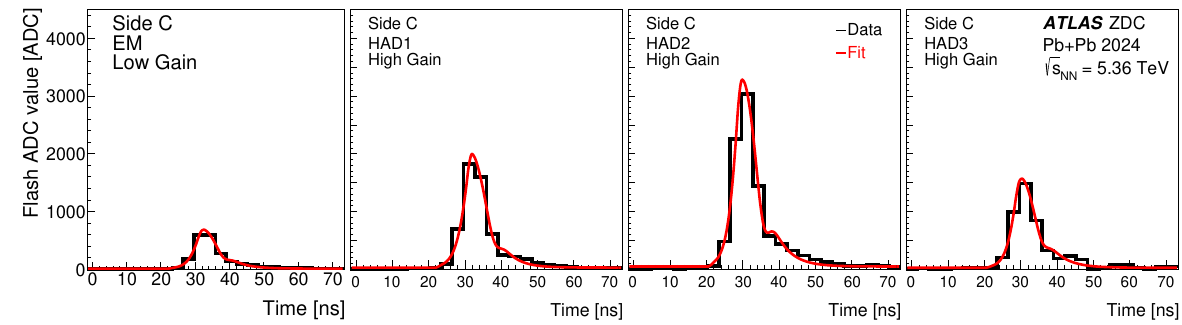}
    \includegraphics[width=0.98\textwidth]{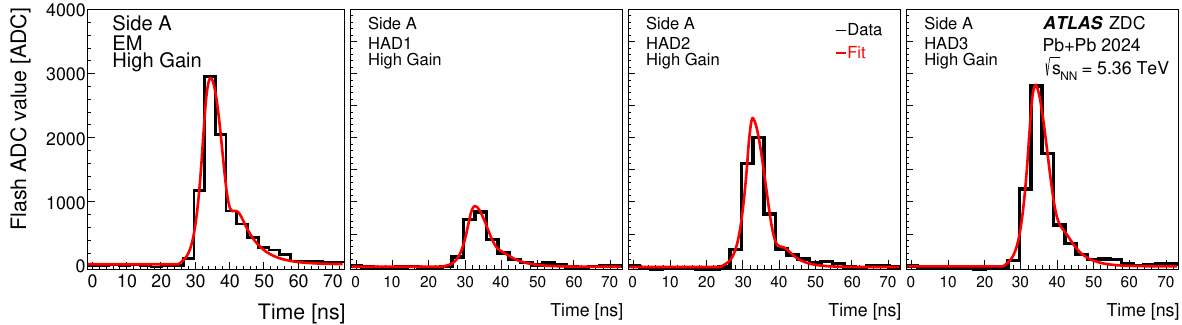}
    \caption{Example data read from the ZDC during 2024 \PbPb\ data-taking. Each panel shows 24 FADC samples plotted as a function of time with each sample corresponding to a time interval of 3.125~ns. The data are shown after subtraction of the baseline measured (typically) using the first sample in the event. The top row shows data for side C and bottom for side A. For the data on side C, the pulse in the EM module overflows the high gain FADC range so the low gain data -- that would be used in analysis -- are shown. In all of the panels, results of a fit of the data performed in the offline analysis is also shown.
    }
    \label{fig:FADC_example}
\end{figure}

\begin{figure}
    \centering
    \includegraphics[width=0.99\linewidth]{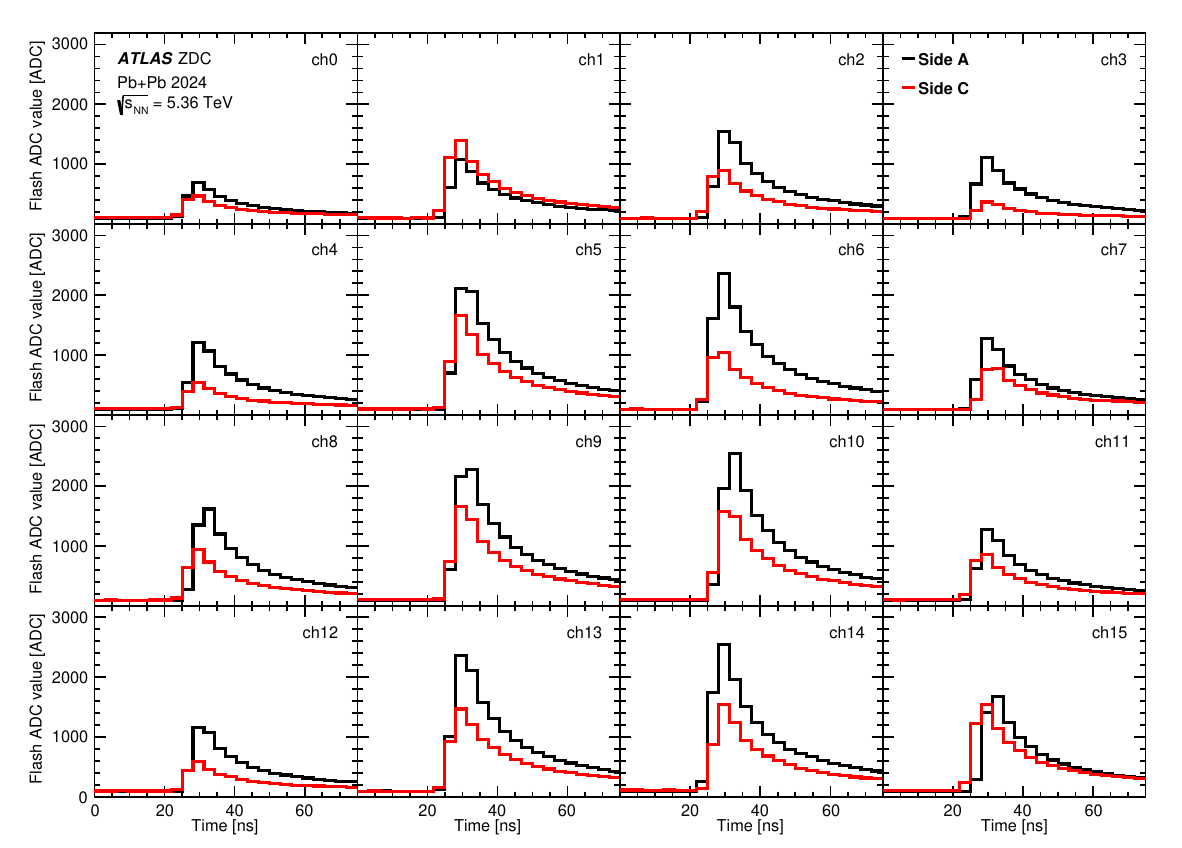}%
    \caption{Example data read from the RPD during the 2024 Pb+Pb data-taking. Each panel shows 24 FADC samples as a function of time for both the detector, displayed in black (side A) and red (side C). }
    \label{fig:FADC_RPD}
\end{figure}

A demonstration of a typical event read from the ZDC is shown in Figure~\ref{fig:FADC_example}. The FADC samples are plotted versus time from the start of the readout window with each sample corresponding to 3.125~ns. Also shown on the plots are the results of fits to the pulses performed by the ZDC reconstruction software, which is not described in this paper. The data for side~C, shown in the top row, have a calibrated total energy of 23~TeV corresponding to 8-9 neutrons. The FADC overflows in the EMC module for that event so the low gain data are shown. The data for side A, shown in the bottom row, correspond to a total calibrated energy of 16~TeV or about 6 neutrons. An example of data read out from the RPD in shown in Figure~\ref{fig:FADC_RPD}  where the FADC data in  both sides of the detector are shown for one 35-neutron event taken from the 2024 Pb+Pb data stream. 

\begin{figure}[!tbp]
\centering 
    \includegraphics[width=0.98\textwidth]{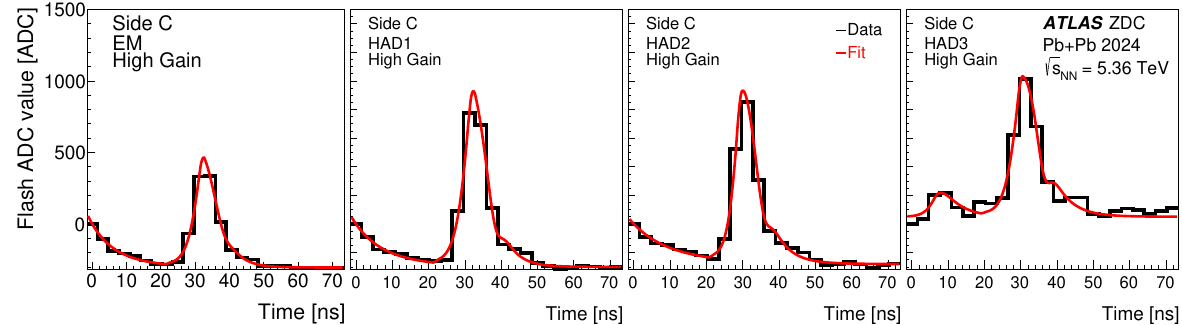}
    \includegraphics[width=0.98\textwidth]{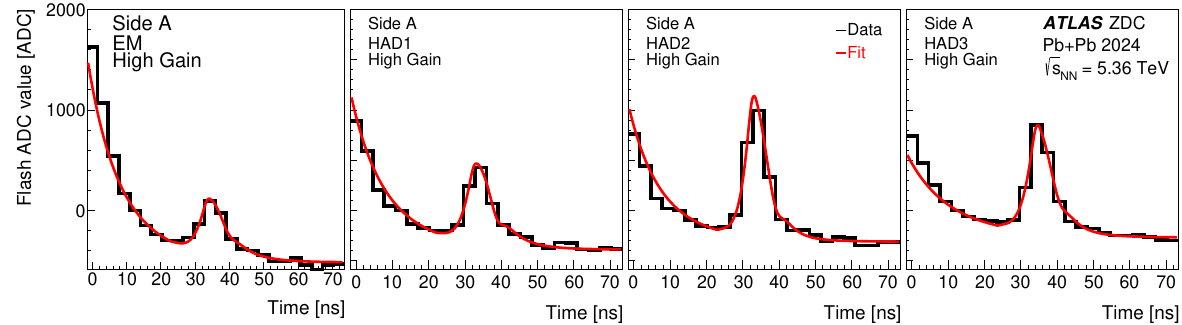}
    \caption{Example data read from the ZDC during 2024 \PbPb\ data-taking used to demonstrate the role of OOT pileup. Each panel shows 24 FADC samples plotted as a function of time with each sample corresponding to a time interval of 3.125~ns. The data are shown after subtraction of the baseline measured (typically) using the first sample in the event. The top row shows data from one event in side~C and the bottom row shows another event in side~A. The tail from a pulse generated in (most likely) the preceding filled crossing can be seen in all the data. 
    }
    \label{fig:FADC_exampleOOT}
\end{figure}

In spite of the substantial improvements made for Run~3 to the transmission of signals from the tunnel to USA15, and in the performance of the electronics, the large dynamic range of the signals seen in the ZDC  means that OOT pileup still can have an impact on ZDC data. In particular, a small signal from a single-neutron (\onen) event could follow a preceding collision that produced more than 50 neutrons in the ZDC. The contribution of OOT pileup is illustrated in Figure~\ref{fig:FADC_exampleOOT} where the tails of pulses from a preceding collision underlie the in-time pulses. The different OOT pileup seen on the two sides indicates that the number of neutrons in the preceding collision were significantly different in the two calorimeters, either due to the physics of the collision or due to the presence of electromagnetic pileup in one of the calorimeters. Pileup such as that shown in Figure~\ref{fig:FADC_exampleOOT} is seen at a rate $\ll 1\%$. With the Run~2 ZDC configuration, the in-time pulses for such events would have been nearly impossible to reconstruct.

\FloatBarrier

%% file: trigger.tex
\label{sec:trigger}
For Run~3, ATLAS uses a two-level trigger system \cite{ATLASTriggerRun3} consisting of a Level-1 (L1) trigger that is synchronous with the LHC clock and implemented using dedicated electronics and a ``high-level trigger'' implemented in commodity PCs running algorithms similar to those used in offline analysis. The L1 trigger has a maximum latency of 2.5~$\mu\text{s}$. 
The ATLAS L1 central trigger processor (CTP) \cite{Spiwoks:889547} is designed to accept "calibration request" triggers that are intended to provide calibration events for different detectors. Also relevant for this paper, the CTP defines and distributes the current number for  "luminosity blocks" which correspond to 1~minute (typically) intervals of data-taking for which ATLAS luminosity data is tabulated and recorded.
\begin{figure}[!tbp]
\centering
\includegraphics[width=0.98\textwidth]{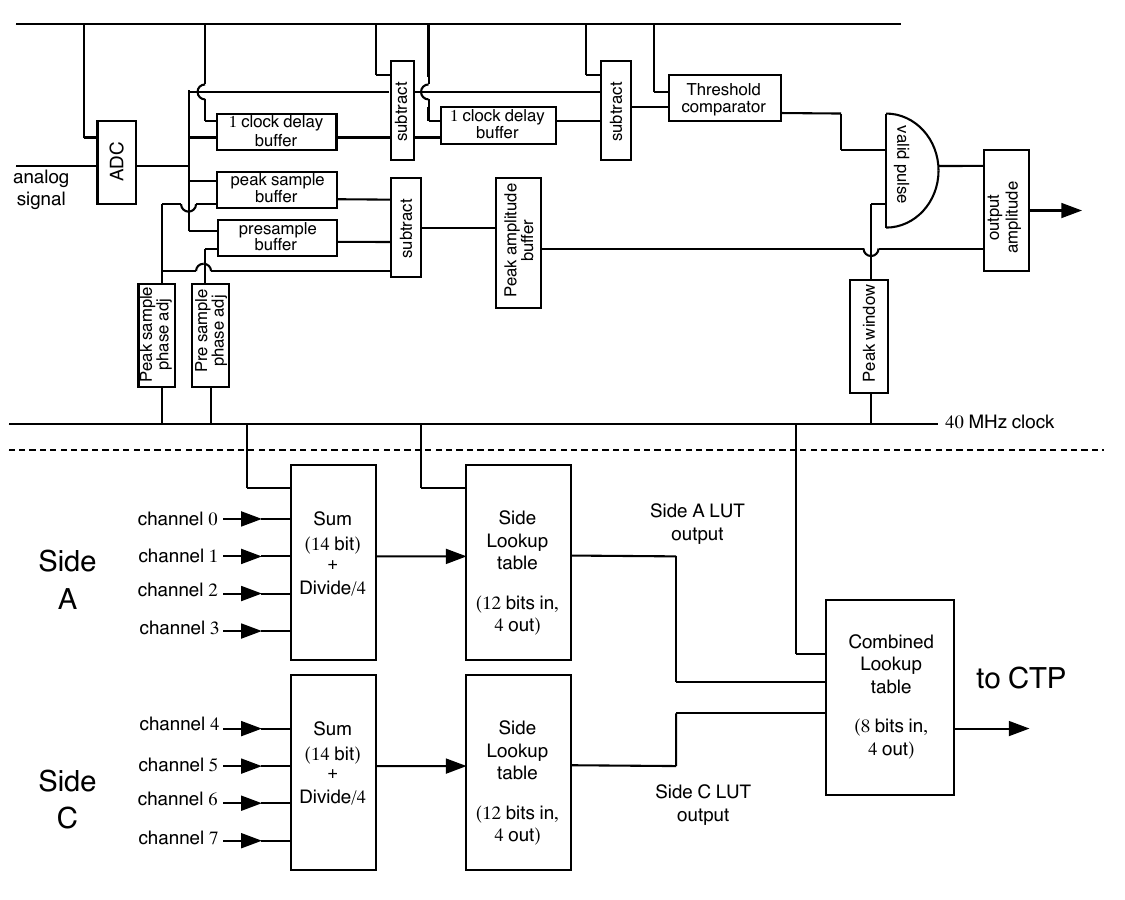}
  \caption{Functional block diagram illustrating the different components of the ZDC trigger firmware. The top portion of the diagram illustrates the single-channel processing with 2nd derivative calculation, baseline subtraction and extraction of the pulse amplitude. The bottom section illustrates the summing of pulses from four channels in each calorimeter and the two-stage lookup tables used to produce the three-bit decision. 
  }
    \label{fig:triggerBlock}
\end{figure}

The ZDC trigger in Run~3 is a fully-digital L1 trigger implemented using lookup tables (LUTs) to provide flexibility in configuration and functionality.
The trigger decision is based on energy (ADC) sums in each of the two calorimeters. 
Three of the available LEMO outputs on the LUCROD are used to transmit signals to the ATLAS central trigger processor; these three lines are used to encode 8 possible trigger decisions.
Although the same firmware is implemented on all
ZDC LUCRODs, the trigger was initially envisioned to be 
implemented only in the high-gain LUCROD, which provides greater separation
between pulses associated with single-neutron events and low-energy/noise signals. However,
as will be discussed below, for the 2024 \PbPb\ run, additional
triggers were implemented using the low-gain LUCROD. 

Noise suppression is provided in the ZDC trigger by a pulse-finding
algorithm that operates in the FPGAch. The algorithm is based on
a pipelined second-derivative calculation. The second derivative
becomes most negative at the maximum of a positive pulse and is more
robust against the presence of OOT pileup than a simple 
threshold. In \BCs where the (negative of) the second derivative
exceeds a user-specified threshold, the maximum FADC value within the
\BC is obtained. The pedestal or ``baseline'' is evaluated from the FADC sample a configurable number of samples prior to that of the maximum sample. This
baseline is subtracted from the maximum ADC to yield the pulse
amplitude. If the second derivative threshold is not crossed in a
given \BC, the amplitude is taken to be zero. If any FADC value within
a \BC is equal to 4095 -- {\it i.e} if the signal saturates -- then no baseline subtraction is performed and
the amplitude is taken to be 4095.

The amplitudes from the eight channels are received in FPGAv where they are summed in groups of four -- channels
0-3 for side A and 4-7 for side C --  to produce 14~bit words. These are
divided by four and the resulting 12 bit values are used as the
address  for ``side'' look up tables (LUTs) that each produce 4-bit words. The words from the two sides are
combined to form a single 8~bit address in the
``combined'' LUT that produces three bits that are transmitted to the CTP. The LUTs are
implemented using sRAM memories connected to FPGAv. The amplitudes per channel and the amplitude sums divided by four for each side are stored in the
event data produced by the LUCROD for offline validation of the trigger.

For \PbPb\ data-taking in 2023 and 2024, the side
LUTs in the high gain LUCROD were configured to define three energy regions corresponding to $< \onen$, $1-4n$ and $>4n$. Because the three-bit output from the LUCROD only allows 8 possible values, the nine possible combinations of the side thresholds cannot all be assigned to unique combined LUT values. To accommodate this limitation, the two possible combinations of $1-4n$ and $\geq~4n$ use the same output value. In the CTP, a variety of minimum-bias and UPC triggers with different neutron topologies were formed using the ZDC inputs to the CTP. 
Also, combinations of the ZDC triggers were formed in the CTP to allow selection of events having $\geq \onen$ on either side ("ZDC\_OR") or  $\geq \onen$ on both sides ("ZDC\_AND"). These triggers were used for both luminosity measurement and to generate a dedicated stream of events from the ATLAS HLT containing
only ZDC data, used for monitoring of the detector and the energy scale. The luminosity data obtained from the ZDC are based on per-bunch monitoring of the ZDC\_OR and ZDC\_AND triggers in the ATLAS CTP that is reported to the ATLAS online luminosity calculator every ATLAS luminosity block \cite{Wierda:2910201}.

In 2024, an additional set of \PbPb\ triggers were implemented in the ZDC low gain
LUCROD to select ``ultra-central'' \PbPb\ collisions. In these
collisions, the number of neutrons observed in each ZDC is small as
there are very few spectator nucleons. However, the produced particle
multiplicities in these collisions are quite large, with the result that the
contribution of energy from produced hadrons, especially neutral pions
and neutral hadrons in the ZDCs is no longer negligible. To suppress
the largest background from $\pi^0$ decays, the second derivative
thresholds on the EM modules were set to 4095 such that these modules
did not contribute to the energy sums. The side and combined LUTs
were used together to form triggers based on the total energy observed
in the two calorimeters with thresholds at 15, 20, 25, 35, and 50~TeV.  
The results were used in the CTP to form L1 triggers that required
more than 4.5~TeV of transverse energy in the ATLAS central
calorimeters and ZDC energy smaller than one of the above
thresholds. The 16 possible values from the side LUTs limit the
precision of the effective energy sum thresholds, so modifications to the
firmware to implement 5 or 6 bit values from the side LUTs are being
considered. 

\begin{figure}[!tbp]
\centering 
\includegraphics[width=0.98\textwidth]{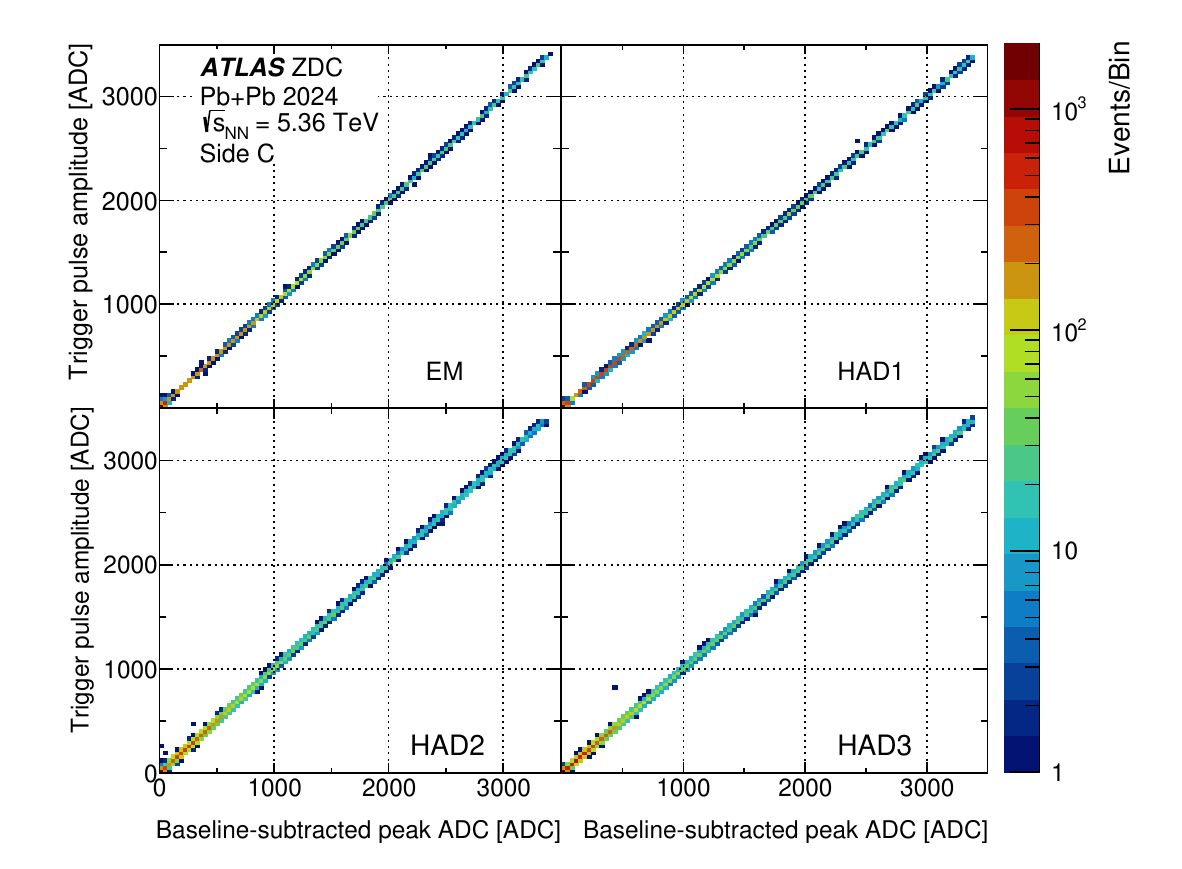}
    \caption{
    Distributions of the amplitude of pulses found by the trigger firmware operating in the FPGAch versus the baseline subtracted peak ADC from the regular data stream. Results are shown for the four modules in the C side.
    }
    \label{fig:TriggerModuleAmpADCPeak}
\end{figure}

\begin{figure}[!tbp]
\centering 
    \includegraphics[width=0.49\textwidth]{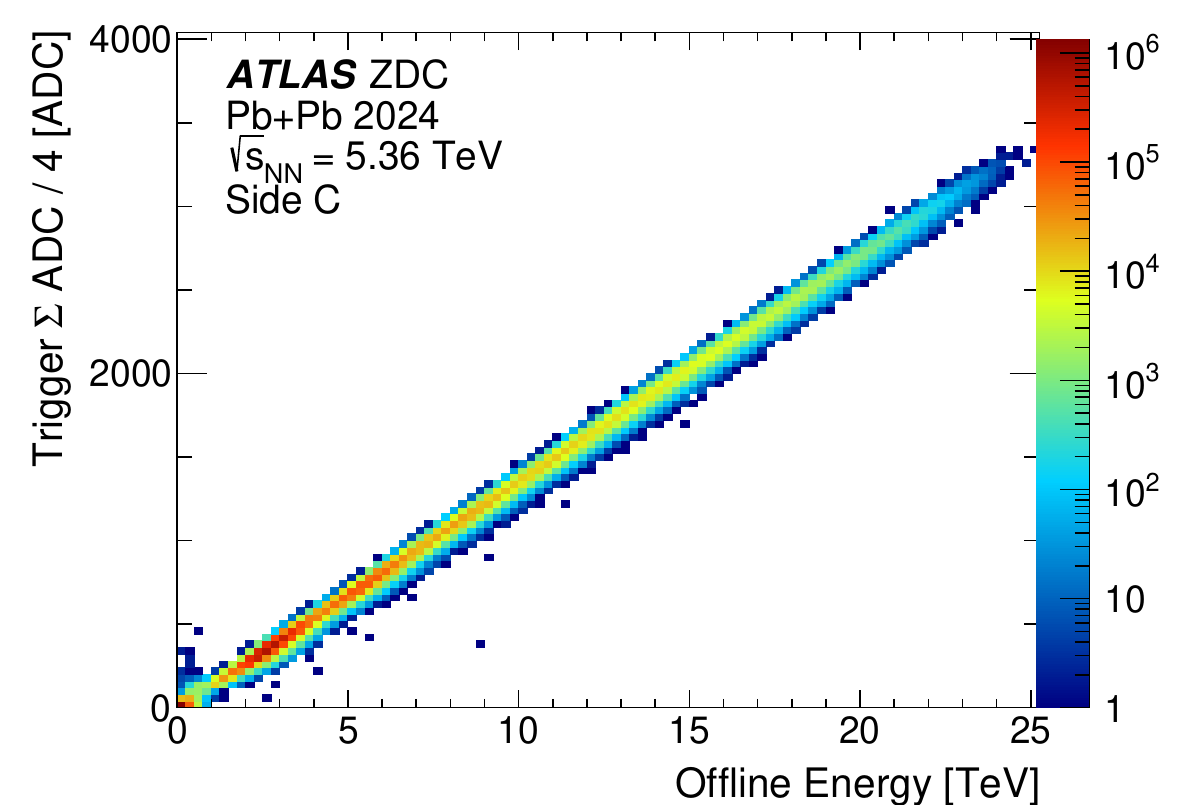}
    \includegraphics[width=0.49\textwidth]{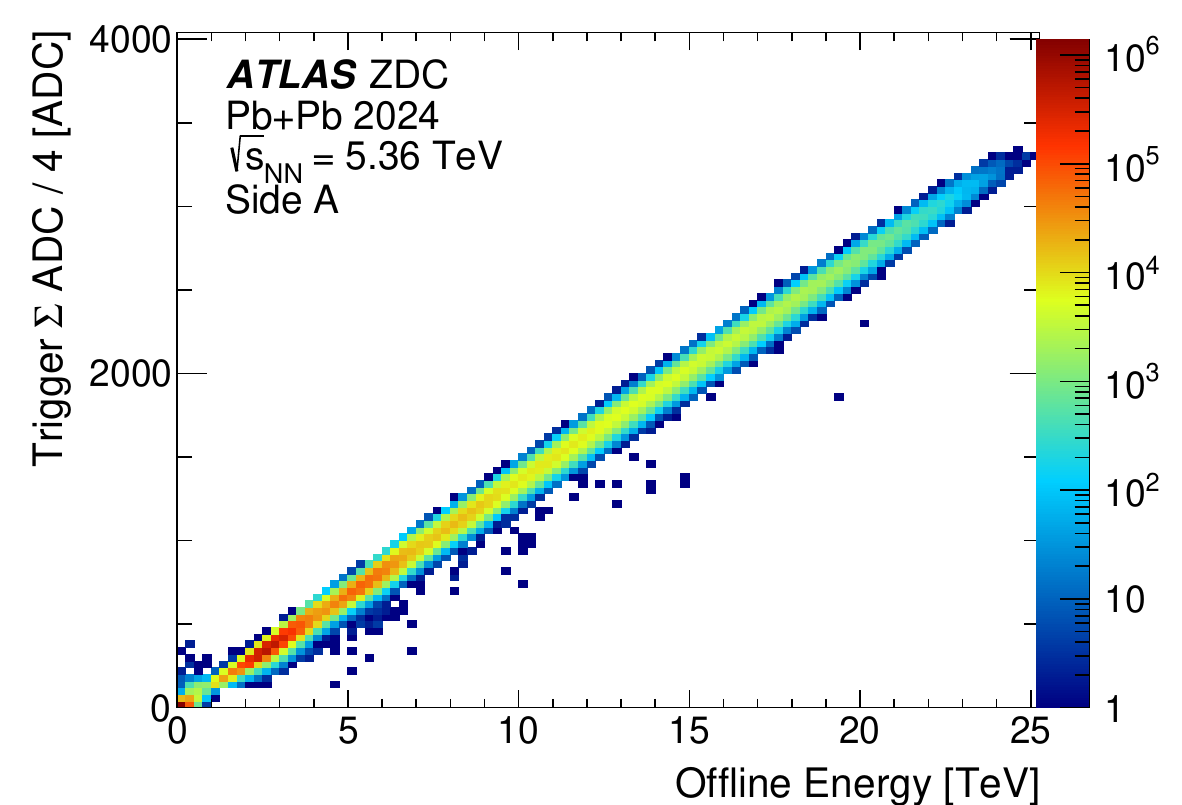}
    \caption{
    Distributions of High Gain \sumadcf versus the offline calibrated energy sum for Side C (left) and side A (right). The shown data are selected by ZDC triggers requiring at least one beam energy neutron in one of the calorimeters and only include events in which none of the modules on the given side overflow.
    }
    \label{fig:TriggerADCSumVsE}
\end{figure}

\begin{figure}[!tb]
\centering 
    \includegraphics[width=0.49\textwidth]{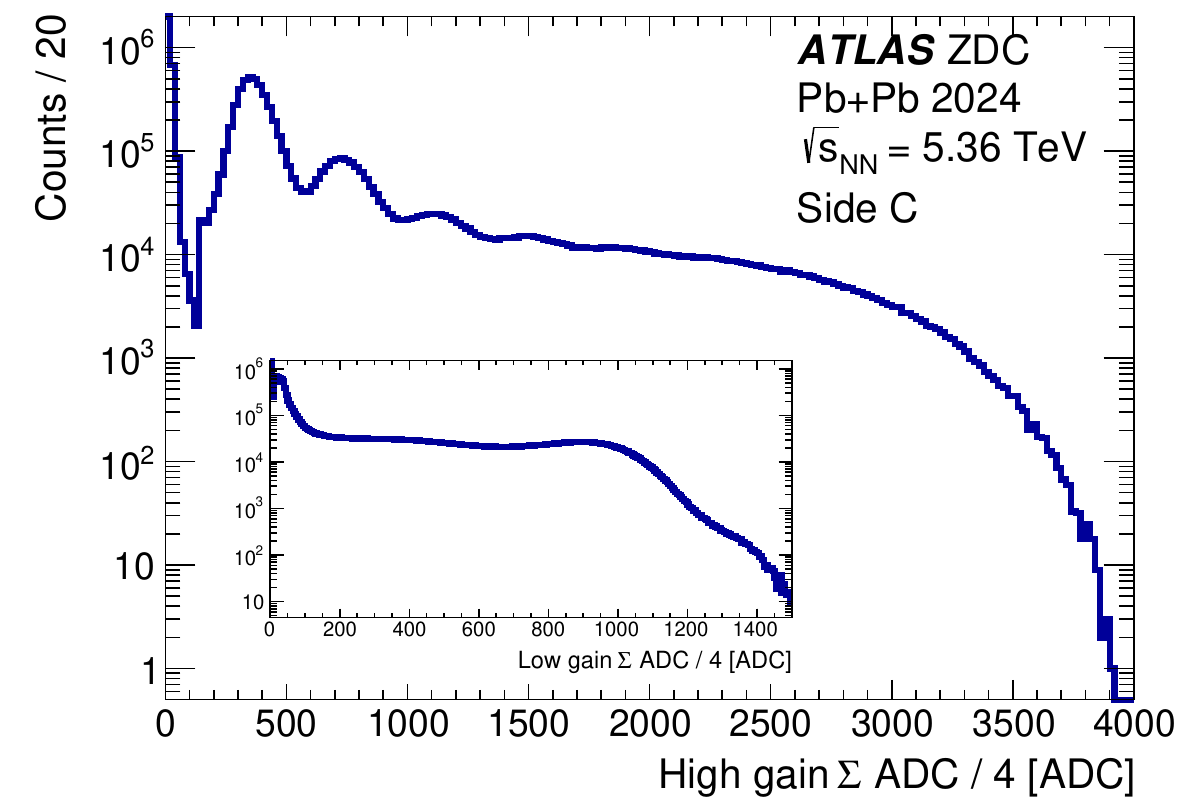}
%
    \includegraphics[width=0.49\textwidth]{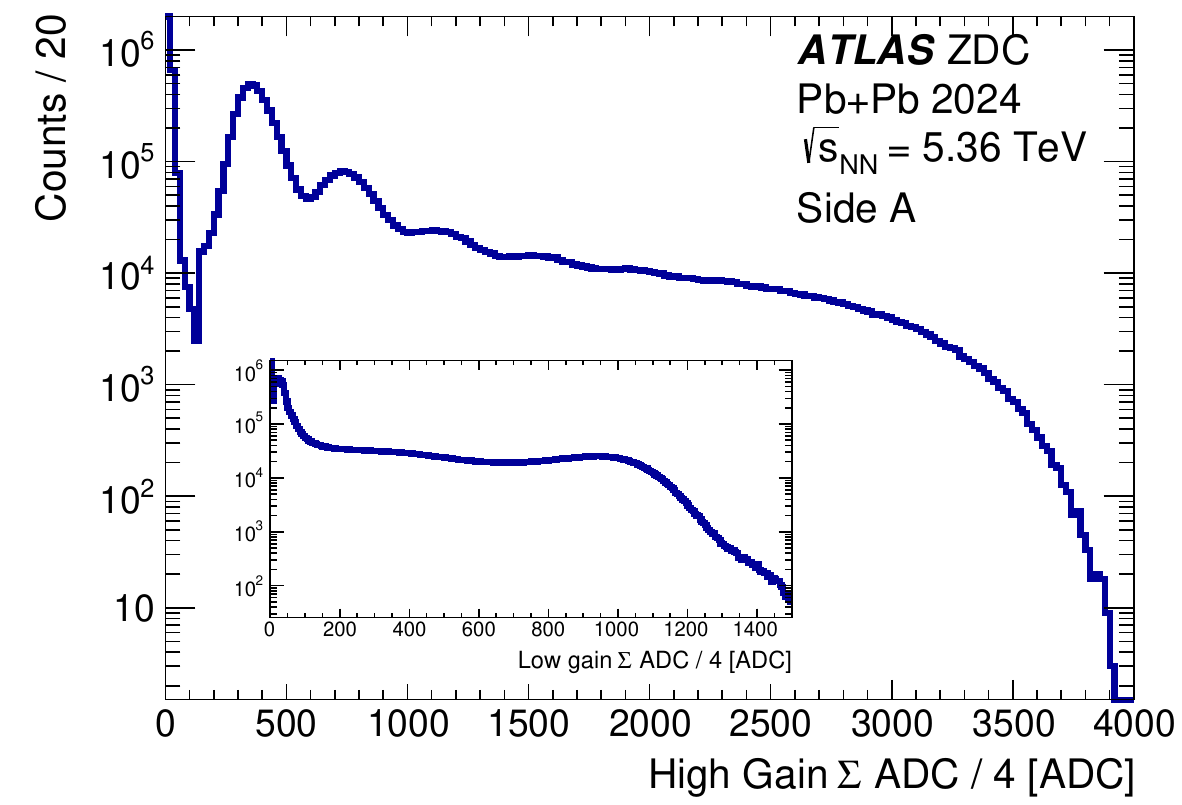}
    \caption{Distributions of trigger \sumadcf for data collected in 2024 \PbPb\ operation on the C (top) and A (bottom) sides. The main distributions show the results from the High Gain LUCROD; the insets show the \sumadcf distributions in the Low Gain LUCROD which intentionally excludes the EM modules from the sums (see text for explanation). The 1-4 neutron peaks can be easily seen in the high gain sums.
    }
    \label{fig:TriggerADCSums}
\end{figure}

The performance of the ZDC trigger is illustrated using data from the 2024 \PbPb\ run in Figures~\ref{fig:TriggerModuleAmpADCPeak}, \ref{fig:TriggerADCSumVsE}, and \ref{fig:TriggerADCSums}. Figure~\ref{fig:TriggerModuleAmpADCPeak} shows a comparison of the trigger amplitudes produced by the firmware to offline amplitudes extracted from the data. Results are shown for the four modules on the C side for events that do not overflow the high gain readout. The offline amplitude is the baseline-subtracted maximum FADC value from samples within $\pm 1$ of the nominal peak position, \adcmax. To best match the trigger processing, the baseline is obtained from the FADC sample three prior to that of the nominal peak position.  The trigger and offline amplitudes typically agree to better than 0.5\% for pulse amplitudes above 100 ADC counts. The finite spread primarily results from events in which the trigger and offline baselines are obtained from different samples, mainly due to the use of the fixed baseline position when evaluating the offline amplitude. The performance on the A side is found to be slightly better than that shown for side~C.

Figure~\ref{fig:TriggerADCSumVsE} shows a comparison of \sumadcf calculated by the trigger and the offline calibrated energy in the same event on the C (left) and A (right) sides. Only events in which none of the modules overflow the high gain ADC range are shown. The correlation between trigger \sumadcf and offline energy sum is excellent. The typical fractional spread in the trigger sum for a narrow range of offline energies varies from 6\% at lower energies -- in the vicinity of the \onen peak -- to 1\% at the highest energies. This spread is dominated by the effect of the offline energy calibration (see Section~\ref{sec:OneNeutcalib}).

Figure~\ref{fig:TriggerADCSums} shows distributions of \sumadcf for $\sqn=5.36$~TeV \PbPb\ collisions for events selected by ZDC triggers. The main panels show results from the high gain LUCROD for events where none of the modules saturate the FADC. The insets show distributions from the low gain LUCROD where the EM contribution is intentionally excluded from the sum for the ultra-central collision trigger. In the high gain data on both sides the \onen, \Nn{2}, \Nn{3}, and \Nn{4} peaks can be clearly seen. The sudden changes in the high gain distributions around  $\sumadcf = 200$ result from the application of \Nn{0.4} thresholds used in the trigger to select events having at least one neutron on one side. 
\FloatBarrier

%% file: monitoring.tex
\subsection{Prompt Monitoring of ZDC and RPD}
\label{sec:promptmonitoring}
Prompt monitoring of the readout of the ZDC and RPD detectors and the operation of the trigger is performed using two separate systems. The first operates in semi-real-time on the single-board computer (SBC) in the VME crate hosting the LUCRODs and is asynchronous with respect to the readout. The second system is the GNAM monitoring system \cite{Zema:886320} which is used to evaluate the integrity of and the content of ZDC and RPD data after it is assembled in the ROS.

The SBC-based monitoring is implemented in a dedicated thread running as part of the ATLAS TDAQ run control application. It continuously loops querying registers on the LUCROD that provide information on the status of the board, snapshots of data being continuously read from the FADCs, and snapshots of results from the trigger processing at all levels. The board-level diagnostics include  
information on the status of external connections such as the TTCrq and \Slink, information regarding the state and possible errors in  the different FPGAs, the depth of various buffers and FIFOs, and a variety of counters. A subset of these data are accessed by the run control program to detect errors and perform automatic resets of the LUCROD while asserting a busy to the ATLAS trigger. 

The snapshots of data and trigger cover 8 BCs at a starting BC number that is set by a register on the LUCROD. The FADC data for the 8 BCs (on FPGAch) and the results of the amplitude sums and LUTs outputs (on FPGAv) are independently ``latched'' until they are read and cleared.
Separately, registers on the FPGAch record the output of the pulse-finding algorithm: the amplitude, before subtraction, of the pulse satisfying the second difference threshold, the baseline as measured in the firmware, and the subtracted amplitude. These data are also latched until they are read and cleared. Monitoring of the pedestal or baseline is performed by the monitoring thread intermittently changing the starting \BC number of the data snapshot to cover \BCs in the abort gap \footnote{The abort gap corresponds to a set of \BCs spanning a 3~$\mu\text{s}$ interval which are never intentionally filled with beam particles. This time interval covers the ramping of the LHC abort kicker magnets.} well separated from those used for calibration pulses (see Section~\ref{sec:calib}). 

As the ZDC GNAM application works on assembled data, it must first decode data and handle errors that occasionally result from missing fragments detected in the ROS. Such errors nearly exclusively occur during heavy ion operation while the LHC is ramping and the frequency of the LHC clock is changing. These  errors are frequently coincident with mis-alignments in   communication internal to the LUCROD. For successfully decoded data, the GNAM application performs an estimate of the baseline using a pre-specified sample number and then estimates the amplitude of a pulse as the maximum FADC value from the 24 recorded samples minus the baseline. Separate estimates of the time of the pulse are obtained from the number of the sample in which the maximum is found and an ADC-weighted mean time computation. The data inserted into the data stream from the trigger -- the individual pulse amplitudes and the sums divided by four -- are extracted and compared to the amplitudes obtained directly from the FADCs to provide validation of the trigger operation. A variety of one and two-dimensional histograms, implemented using the ATLAS Online Histogram \cite{OH} service are created and filled with the results of the above analysis by the GNAM application. The ZDC GNAM application does not make use of any offline software, however simple combinations of data from different channels are made to provide monitoring of the uncalibrated energies in each calorimeter and the centroid position of showers as measured in the RPD.

 One important component of the monitoring system is the monitoring of the calorimetric energy scale in the ZDC trigger. It is done by looking at the ZDC sum amplitude over the different data streams, in order to gain insights into possible effects and biases. The ZDC Calibration Stream is fed by events satisfying ZDC triggers, and GNAM monitors the spectrum of both same-side and opposite-side triggers. These histograms show the effect of the ZDC trigger thresholds on the energy spectrum and allow correction for drifts in the calorimetric energy scale due to radiation damage and other environmental factors. To ensure that the ZDC triggers do not bias the monitoring of the energy spectrum, the two-sided spectrum from a minimum bias stream is also monitored in the same way. The trigger monitoring tracks the average position of the 1n peak as well as its variation over the course of the run. The tracking of the \onen position is performed by a separate application that: receives the ZDC sum amplitudes per side;  performs a single-Gaussian fit over a narrow predetermined range to obtain an initial estimate of the uncalibrated 1 neutron peak position; performs a 4-Gaussian fit over the range covered by the $1-3$ neutron peaks; uses the result of the four-Gaussian fit to determine the position of the single-neutron peak and its relations to the \Nn[2] peak. The resulting information is then propagated into new histograms to be tracked for drifts and sudden changes in the detector response to the 1 neutron peak.

%% file: OperatingPoints.tex
\subsection{ZDC input from test beams}
\label{sec:zdc_testbeam}
\begin{figure}[!tb]    
\centering 
\includegraphics[width=0.8\textwidth]{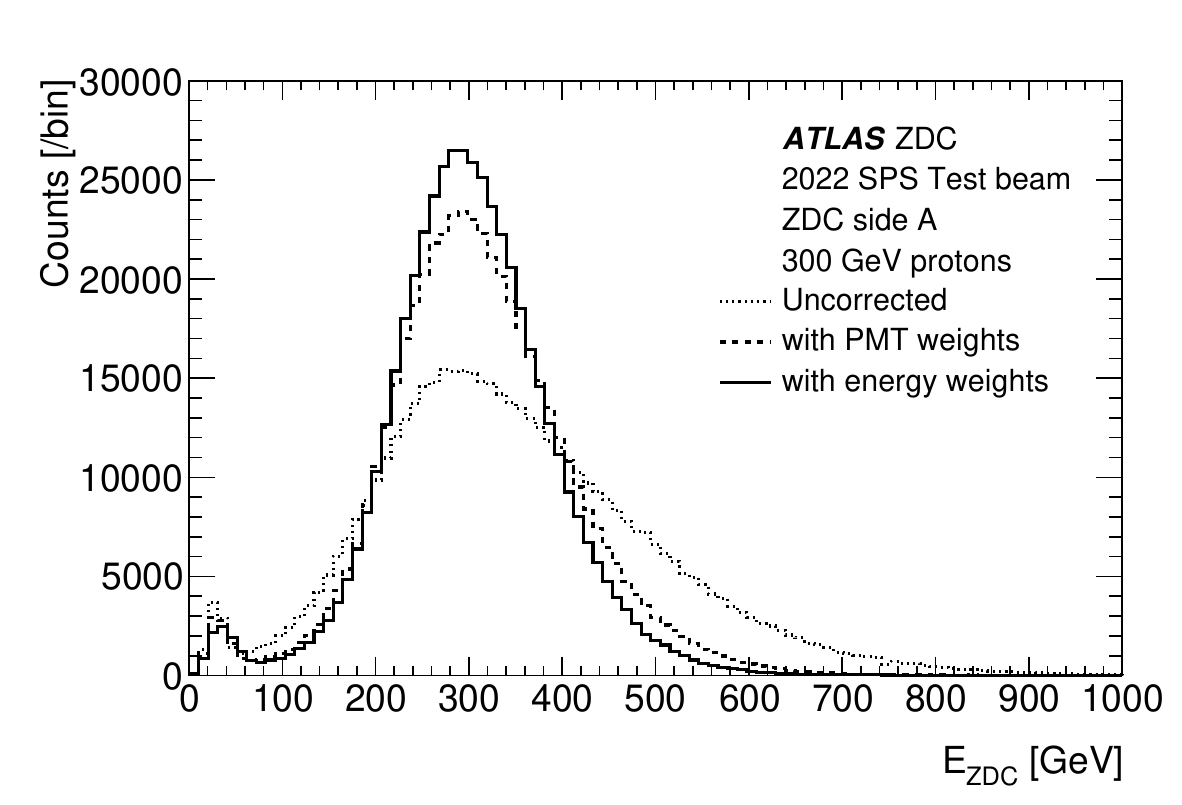}
\caption{Calorimeter energy sums for the  ATLAS side A ZDC obtained from test beam measurements performed in 2022 in the H4 test beam at the SPS using 300~GeV protons. Results are shown for three different stages (see text) of energy calibration applied to each of the four modules prior to the sum.}
\label{fig:TB_300GeV_ZDCA}
\end{figure}

In 2022, the ZDCs were exposed to protons, hadrons and electrons at a variety of energies in 
a test beam at the CERN SPS North Area\cite{Banerjee:2774716}.
Data from this run, digitized using the LUCROD readout system,
were used to calibrate the variation of the ZDC PMT response as a function of the anode high voltage. 
Separately, data collected at fixed HV were analyzed to study the detector energy response. 
Results from 300 GeV protons in ZDC side A are shown in Figure~\ref{fig:TB_300GeV_ZDCA}.
Without correction for the tube-to-tube response discussed in Section~\ref{sec:calor}, the ZDC energy sums provided poor response: the distribution was both excessively broad and asymmetric. 
Correcting each tube for its particular anode sensitivity yields a much narrower energy-sum response.
Finally, additional per-module corrections that account for the shower development in the calorimeter were obtained using a optimization procedure similar to that applied to data (see Section~\ref{sec:OneNeutcalib}) that further improved the ZDC response.  After normalizing the weight to the first module, the relative weights were found to be 1.14, 1.38, and 1.76, and the same weights were applied to both sets of four modules.

\subsection{ZDC working points}
\label{sec:ZDCwp}
For ZDC \PbPb\ operation in 2023, the PMT response and test beam-based shower correction factors were combined to produce desired values for the relative response of the different modules. The HV response curves, measured in test beam for each module using electron beams from 150-250 GeV, were then used to obtain HV settings that would provide approximately equalized energy response for the different modules. While the detailed shower development for multi-TeV neutrons is different than for 300~GeV protons, applications of the offline calibration showed only small (O(10\%)) 
variations in the optimal per-module weights. The equalized gain scale was chosen in 2023 with the intention that the single-neutron ADC sum in the trigger would be centered at 1000. During operation, however, the actual position of the single-neutron peak was at $\sim$~1250. In 2024, the equalized HV working point was increased to produce a 10\% increase in gain to correct for the observed radiation damage to the PMTs during the 2023 run. However, the result of this change was that the \onen ADC sum increased to ~1450. This result is attributed to presumed annealing of the radiation damage to the PMT windows between operation periods; similar, though weaker, annealing has been observed during periods of LHC downtime. However, a rigorous quantitative assessment of the impact of the annealing process on the ZDC PMT response and its time-dependence has not yet been performed.

\subsection{RPD working points}
\label{sec:RPDWP}
\begin{figure}[!tb]    
\centering 
\includegraphics[width=0.47\textwidth]{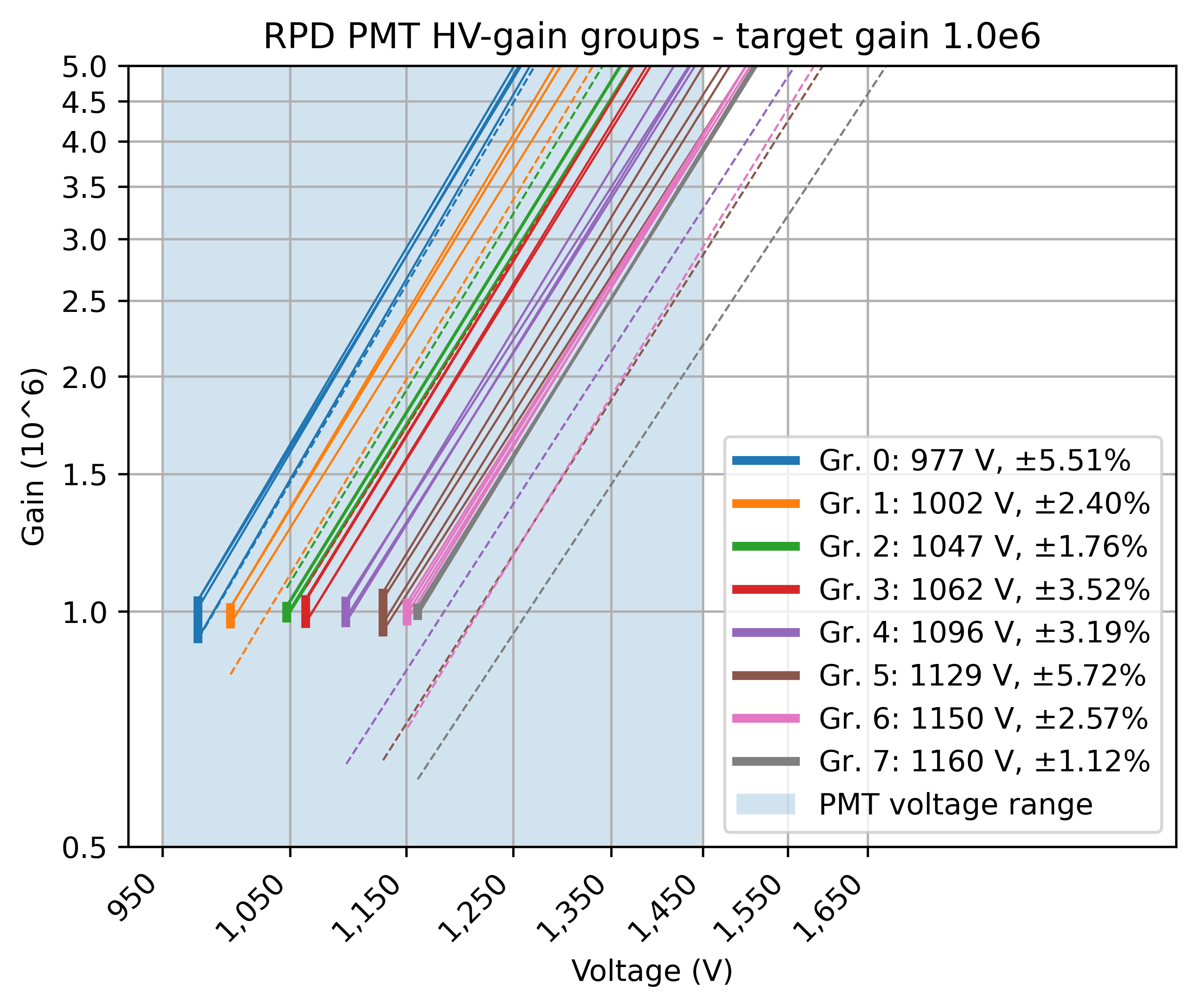} 
\includegraphics[width=0.52\textwidth]{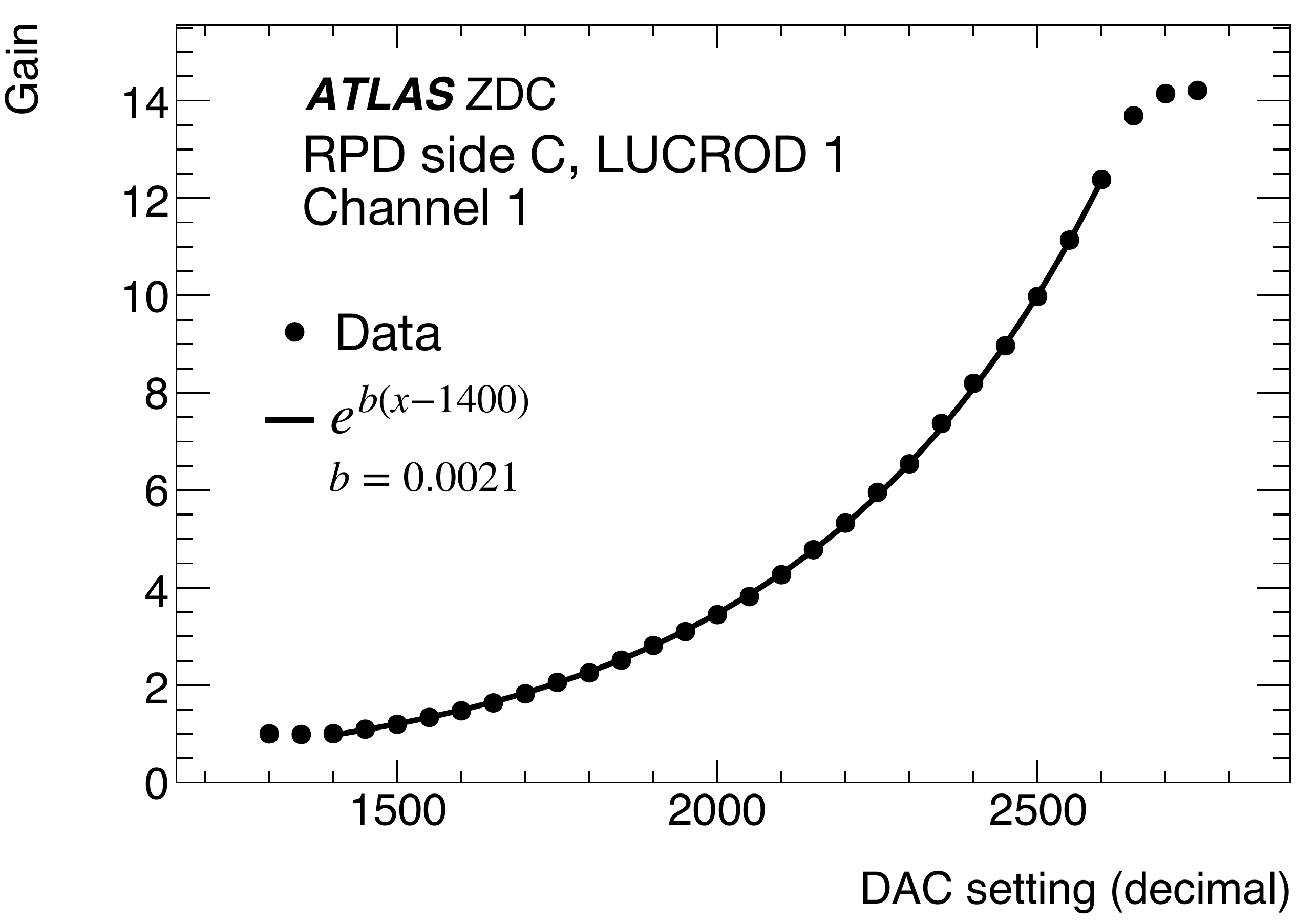} 
\caption{Left: HV-gain curves for all the 40 RPD PMTs. Different colors represent different PMT groups. The dashed lines indicate spare tubes for a given group, which were not installed in the detector. The solid boxes indicate the gain spread of a single group at the optimal voltage to minimize deviations from a target voltage ($1\times 10^6$ in this specific case). The spread is quantified by the percentage values written in the legend. The shaded area denotes the operating range of the PMTs. Right: example of gain calibration of one of the RPD LUCROD channels, providing the correspondence between DAC value and gain for the channel. The gain is unity for all DAC values below 1400.}
\label{fig:rpd_operating_points}
\end{figure}

A set of forty R1635 PMTs (sixteen per side, plus four spares per side) was used to instrument the RPD readout. 
As described in Section~\ref{sec:HV}, the tubes are powered in groups of four, where each group corresponds to a specific column in the detector.
To minimize inherent variations in detector response, the PMTs are grouped based on their HV-gain characteristics, provided directly by Hamamatsu, as shown in Figure~\ref{fig:rpd_operating_points} (right panel). Each group consists of four PMTs installed in the detector and one spare. However, this grouping method has finite precision, as HV-gain curves still exhibit slight variations even within the same group. The gain spread ranges from 1\% to 8.5\%, depending on the target operational gain, which is set between $[1-2.5]\times 10^6$. Residual channel-to-channel discrepancies are corrected at the readout gain level, using the LUCROD per-channel capabilities described in Section~\ref{sec:readout}.

The gain of each channel amplifier in the LUCROD boards was characterized through dedicated measurements. As an example, the results of the gain calibration for one RPD channel are shown in Figure~\ref{fig:rpd_operating_points} (left panel). The data are fitted with an exponential function which well-describes the dependence of the gain on the DAC setting. The PMT HV gain characterization and the LUCROD channel gain characterization are combined to determine the optimal HV (per group of four) and LUCROD (per channel) settings required to achieve a given target operational gain. This procedure ensures that the response of all sixteen RPD channels within one detector is calibrated to the same gain. 

%% file: LED.tex
\label{sec:led}
Both the ZDC and RPD were equipped with an LED flasher system, which is triggered in the LHC abort gap during physics running to monitor the response of the photomultiplier tubes. That response can vary with time due to radiation damage to the PMT windows, high current draw in the tubes, and/or ambient conditions  in the tunnel that can affect the gain of the PMTs. The LED calibration system consists of two LED boards each for the 
A and C detectors, one for the ZDC and another for the RPD. 
The LED boards are integrated into the design of the RPD but  are external to and separated from the ZDC modules. The boards are based on a system developed for the PHENIX MPC detectors \cite{Koster:2011bfs}. 
Each board is equipped with three LEDs (two blue and one green).  The blue LEDs were chosen to provide redundancy for wavelengths close to the UV peak of the Cherenkov spectrum, while the green LED was intended to provide sensitivity to  potential wavelength dependent changes in the response of the PMTs or the rods due to radiation damage. 

For each  ZDC, the LED board is mounted on short aluminum light-guide, designed to convey the LED light onto the face of an optical fiber. The optical fiber is then passed through a 1-to-4 splitter with each output routed to a ZDC module. The fibers are coupled through the module housing to an internal fiber that directs the light onto the edge of the PMT window. For the RPD, the LED board is mounted on a hollow Teflon homogenizer box, designed to uniformly distribute the LED light onto the face of a fiber bundle with 16 fused silica fibers, one per PMT, whose ends are mated to the individual RPD PMTs via an interface plate (see panels C, D and E on Figure~\ref{fig:RPD_Details}). 

\begin{figure}[b]    
\centering 
\includegraphics[width=0.98\textwidth]{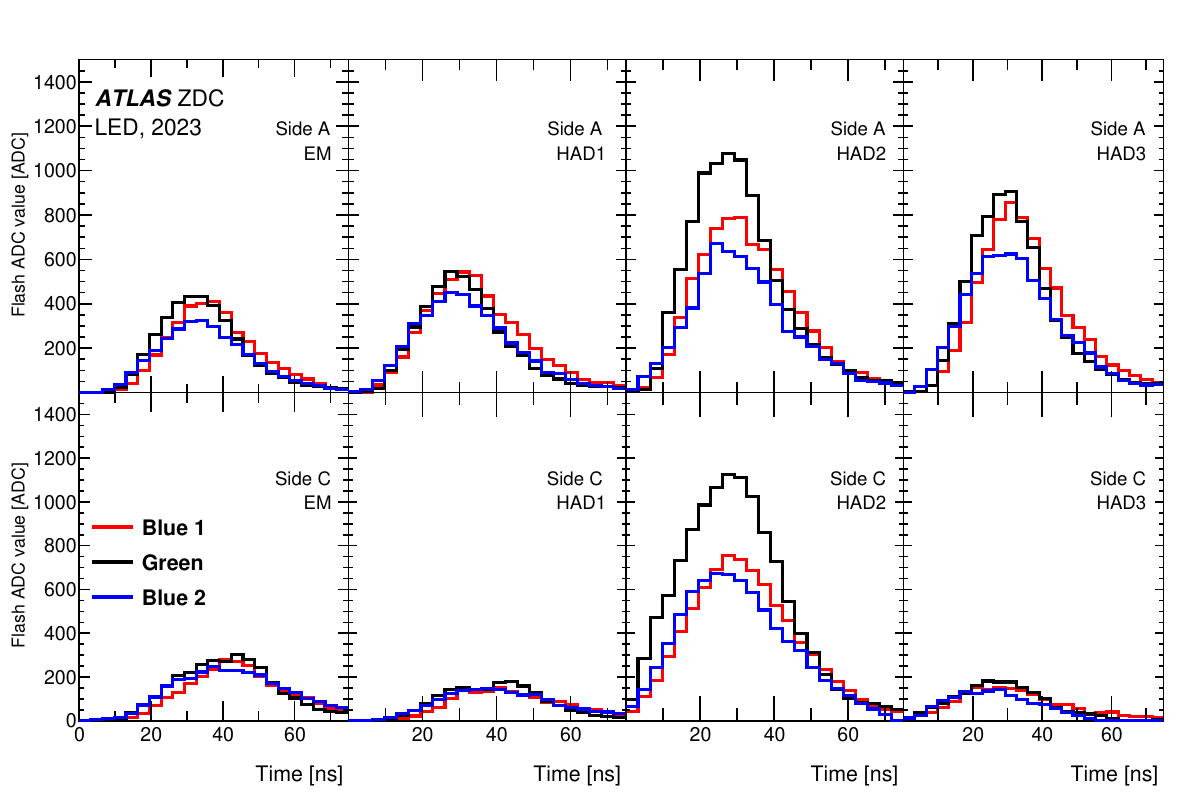} 
\caption{Example LED events measured in the ZDC in 2023. Baseline-subtracted FADC data are plotted versus time (3.125 ns/sample) for each ZDC module. The three colors show different events, one for each LED color.}
\label{fig:led_Pulses}
\end{figure}
The LED system is implemented on both detector sides for a total of four LED boards and twelve independent LEDs. The LEDs are driven by signals generated using three BNC Model 588 pulser boards \cite{bnc:588series} mounted in NIM modules. Each board has four channels for which the pulse amplitude, width, and delay can be separately configured; the maximum amplitude is 10~V. Each board can be independently triggered, and a single board
provides the driving signals for one LED color for both ZDC and RPD and both sides of the detector. The signals are sent from USA15 to the detector over preexisting 140~m cables: CC50 for the RPD and CB50 for the ZDC. The actual pulse amplitudes are substantially attenuated during transmission from USA15 to the detectors. For the ZDCs in particular, the combination of driver amplitude loss over the CB50 cables and the light reduction in the 1-to-4 optical splitting was such that maximum amplitude driving pulses were needed to produce the desired LED pulse amplitudes. 

\begin{figure}[b]    
\centering 
\includegraphics[width=0.85\textwidth]{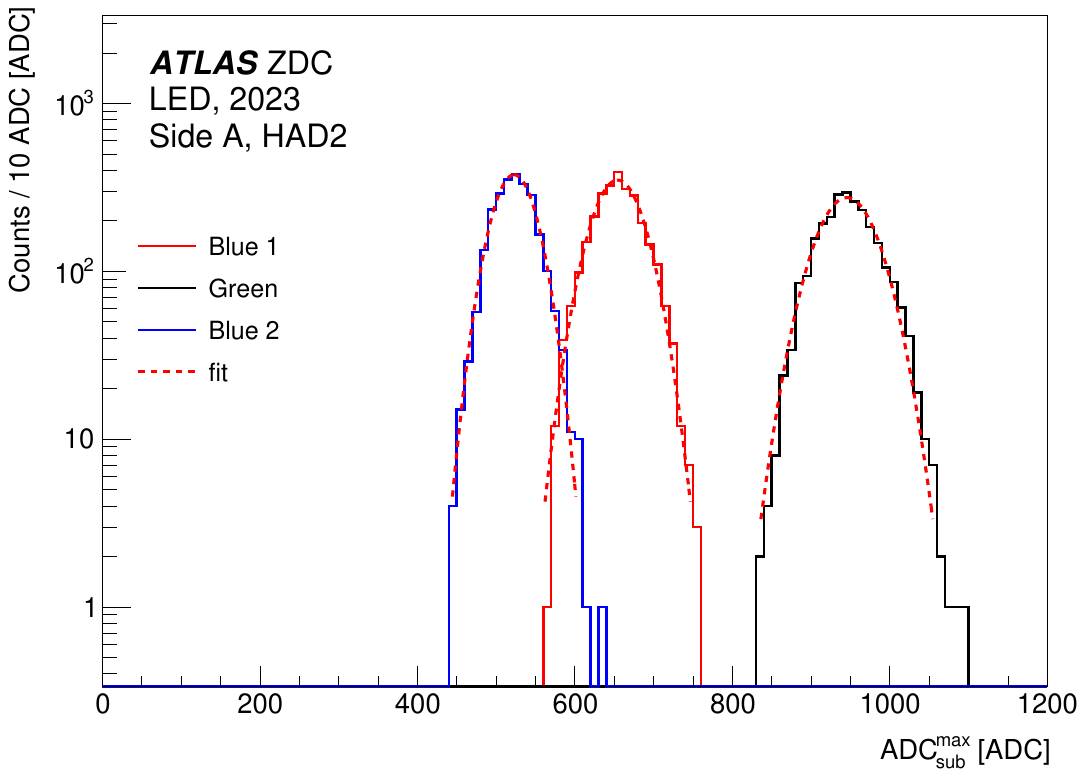} 
\caption{Distributions of LED pulse amplitudes, characterized by the baseline-subtracted maximum ADC values, \adcmax,  recorded in the A-side HAD2 module during a single luminosity block during 2023 \PbPb\ operation. The dotted lines show the results of Gaussian fits to the distributions. These describe the LED amplitude distributions well.}
\label{fig:led_sum_distribution}
\end{figure}
The LED boards are triggered using signals produced by a sequencer or "pattern generator" that is synchronized to the LHC clock running in an ATLAS ALTI \cite{ALTI} module. The triggers are timed such that the LED pulses arrive at the PMTs in \BCs corresponding to the LHC abort gap. In addition to the signals needed to fire the LED pulsers, the ALTI sends ``calibration requests'' (CALREQs) to the ATLAS CTP. The ATLAS calibration request mechanism is implemented in such a way that a CALREQ from a detector can only be accepted once every 16 LHC orbits. The three different LEDs are fired in successive cycles of 16 orbits. For the orbit in which a CALREQ is accepted, an associated L1 trigger is generated and the digitized pulses are read from the LUCRODs in the same manner as physics events.

The LED System was successfully commissioned in the Fall 2023 Heavy Ion Run and was used extensively in the 2023 and 2024 LHC heavy ion runs. 
Figure~\ref{fig:led_Pulses} shows FADC data for three separate events taken in a similar time interval (one luminosity block in an experimental run from 2023).
It is observed that the signals have similar amplitude and time structure between the three colors, but the amplitude varies strongly module-to-module, for the reasons described above. 

In analyses of the LED data, the FADC sample with the largest magnitude after baseline subtraction is taken as the amplitude of the signal, and is referred-to below as \adcmax. The baseline is obtained from the first FADC sample for the given module in the event. Distributions of \adcmax obtained for a single module (side A HAD2) for the three LED colors, recorded during a single luminosity block during 2023 \PbPb\ data-taking are shown in Figure~\ref{fig:led_sum_distribution}. The distributions are well described by Gaussians -- indicated by dashed lines in the figure -- with relative widths of 4-5\%.

\begin{figure}[!tb]
\centering
\includegraphics[width=0.94\textwidth]{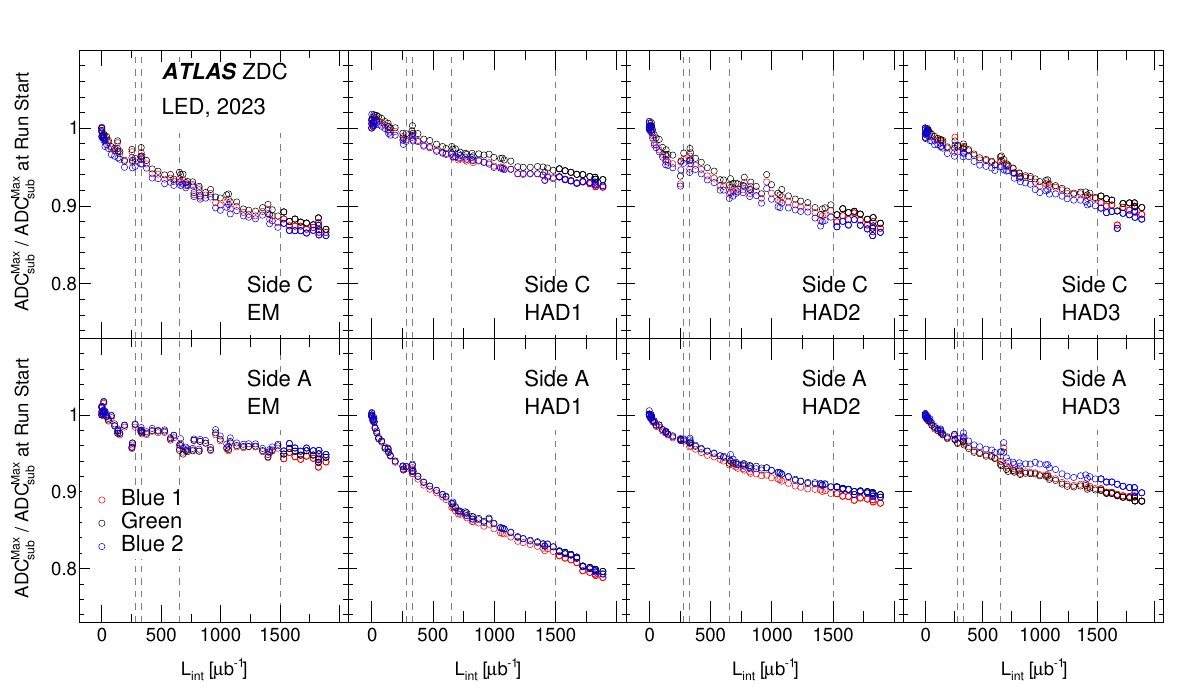} 
\caption{
The mean LED \adcmax measured during standby data-taking (see text) in different ATLAS runs plotted as a function of the integrated luminosity preceding the run, which should scale linearly with the received dose.  The different color markers correspond to the different color LEDs.  The vertical dashed lines indicate when the PMT voltages were increased to adjust for observed reductions in the ZDC energy scale.}
\label{fig:LED_corrected}
\end{figure}

Figure~\ref{fig:LED_corrected} shows the long-time 
variations of ZDC PMT response to the LEDs during 2023 \PbPb operations. The figure shows the average \adcmax  measured during ``standby'' data-taking -- the period in each LHC fill when ATLAS is fully operational but before stable collisions are achieved. Results are shown after correction for discontinuities caused by intentional changes in HV settings. Specifically,
each distribution is fit to a piecewise exponential, and the \adcmax values in each interval are scaled so that the fits yield a continuous function.  
After corrections, an overall exponential decrease of the LED signal amplitude is observed as a function of increasing luminosity. 

Barring radiation damage to the LED system (see Section~\ref{sec:calib}), the LED signal is expected to be constant and the LED light is coupled directly to the PMT window. 
Thus, the observed monotonic reduction in the LED signal amplitude is plausibly a consequence of a changing response from the PMTs themselves.  
The dependence of the signal degradation  on luminosity further suggests that this is caused by radiation damage to the borosilicate PMT windows. 
It should be noted that the decrease in each module is never perfectly exponential, partly reflecting extended delays between some runs during which the response of the PMT is observed to increase. This behavior is understood to result from self-annealing of the radiation damage to the PMT window.
\begin{figure}[!tb]    
\centering 
\includegraphics[width=0.99\textwidth]{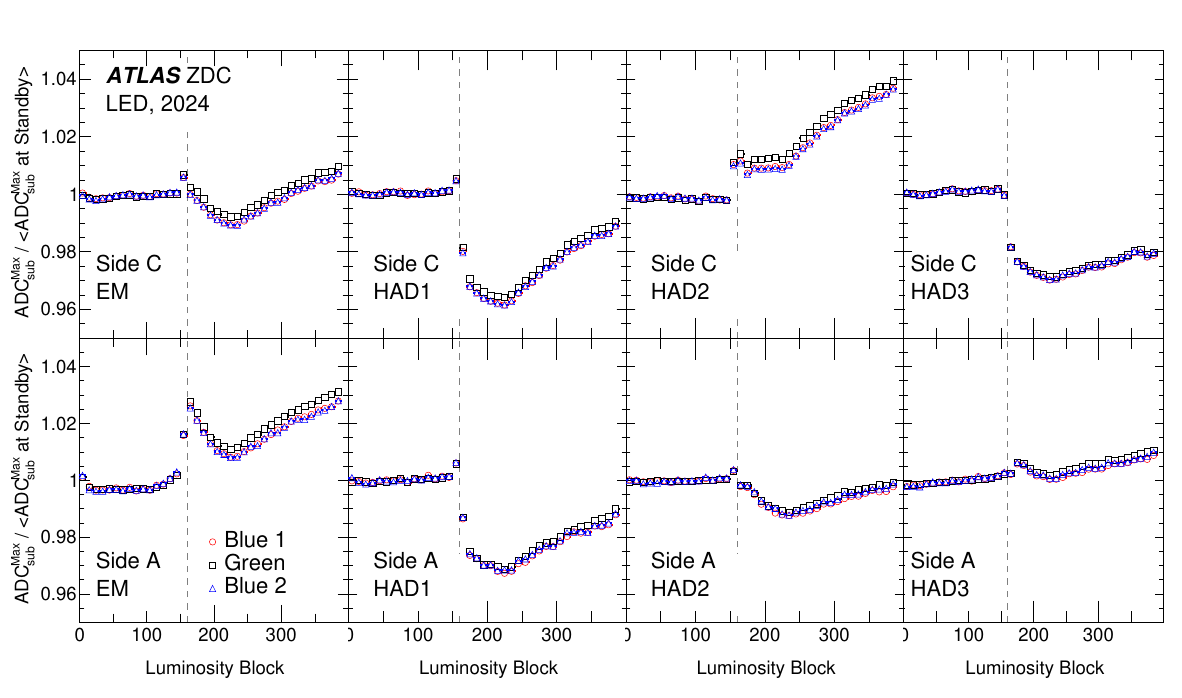} 
\caption{The mean LED \adcmax  divided by the mean \adcmax observed during standby data-taking preceding a fill, plotted as a function of the ATLAS luminosity block number. The different color markers correspond to the different LED colors. The variations in the ratio reflect the relative change in PMT response observed during an LHC fill. A sudden change is observed in most of the modules just after beams are brought into collision; that time is indicated by the vertical dashed lines. Only small differences are observed between the different LEDs, primarily between the green and the two blue LEDs.   }
\label{fig:ZDC_MaxADC_means_norm}
\end{figure}

In addition to the observed damage accruing over longer timescales (weeks), and so visible on the time scale of the \PbPb running period, variations  of the LED response are observed over much shorter time-scales, e.g. over minutes or hours. These variations depend on the instantaneous luminosity, and may reflect the residual impact of large dynode currents, not fully  compensated by the PMT boosters.

Both the long-term and short-term behaviors in the LED data are accounted for using a time-dependent calibration.
A per-luminosity block correction is obtained by comparing the mean measured LED amplitude over a  luminosity block to a reference value  measured during standby operations in early data-taking in the given year of  \PbPb operation. In other words, the correction factors  
\[C_{\text{LED}}(LB) = \frac{\langle \adcmax\rangle_{LB} }{\langle \adcmax\rangle_{\text{ref}} }, \]
are computed and applied to physics data by dividing measured ZDC pulse amplitudes with the module-specific  $C_{\text{LED}}$. An example of the obtained correction factors are shown in Figure~\ref{fig:ZDC_MaxADC_means_norm}. 
Results are shown for all luminosity blocks  with a single ATLAS run that started with three hours of no-beam LED data-taking  followed by five hours of \PbPb collisions. The correction factors vary significantly between ZDC modules and can be both greater and less than unity. The largest correction differs from unity by about 4\%.


%% file: pulser.tex
\begin{figure}
    \centering
    \includegraphics[width=0.98\linewidth]{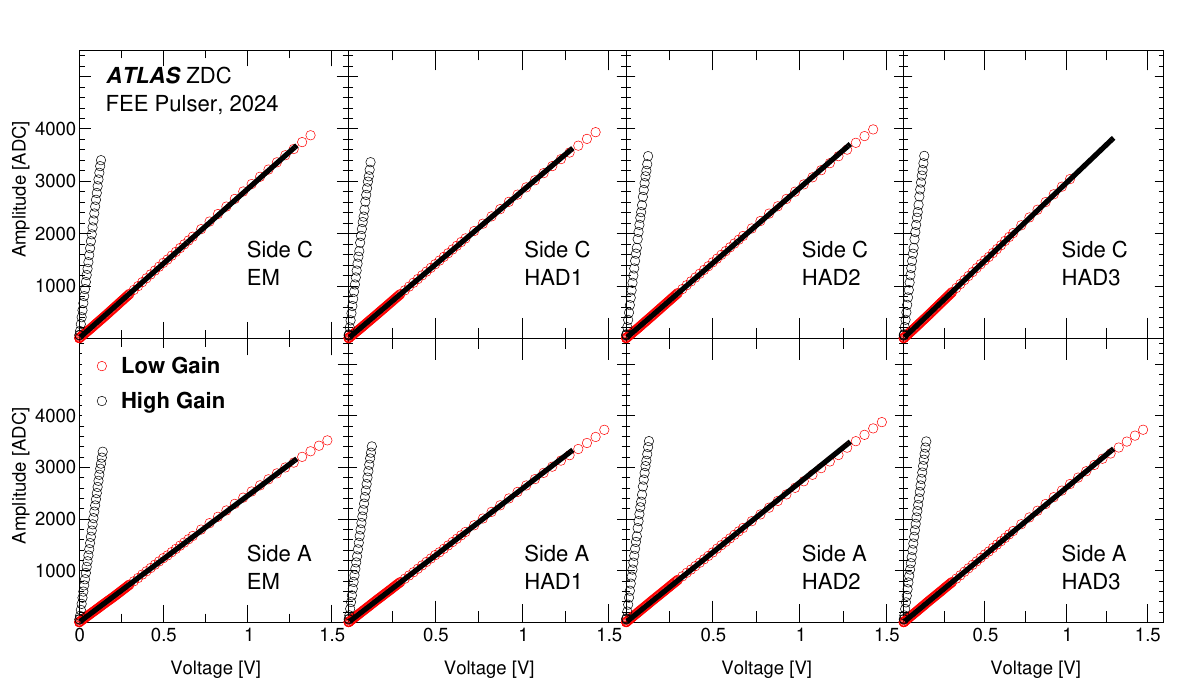}
    \caption{
    Results from one full scan of the LUCROD response to injected pulses as a function of the amplitude of the pulse in volts. The amplitude is obtained from offline analysis. Separate results are shown for high gain and low gain readout. Data are not shown for pulser settings where the FADCs overflow. The solid lines show the results of linear fits to the data.
    }
    \label{fig:pulser_curves}
\end{figure}

Comparisons of the high and low gain readout in data recorded in 2023 suggested deviation from linearity in the response of the analog cards and/or the FADCs in the LUCROD for amplitudes larger than $\sim 70\%$ of the saturation voltage. 
To provide direct, absolute calibration of the response of the FEE, an ``injected pulse'' system was implemented for the ZDC that used pulses generated by a Tektronix AFG3252 pulse generator operated in custom pulse mode. 
Both channels of the pulse generator were used, one attenuated by a factor of 100, though only one channel was enabled at a time. 
The generated pulses were combined with (added to) pulses from the detector using two Lecroy model 428F linear fan-in/fan-out modules. Separate outputs from the fan-in/fan-out channels were used to provide signals to both the high gain and low gain LUCRODs. 

Calibration events were produced 
with the pulse generator in a manner similar to that described above for the LEDs. Namely, the 
ALTI sequencer was programmed to generate logic signals needed to fire the pulser and to send CALREQ requests to the CTP. The generated pulses were timed to arrive at the LUCRODs at \BCs in the abort gap. During the 2024 \PbPb\ data-taking typically 100-200 pulser events were recorded per second. The amplitude of the pulses was adjusted every minute, timed to match the ATLAS luminosity blocks. Sequences of amplitude steps were chosen to map out the LUCROD response to pulses with amplitude varying from 0.5~mV to 1.5~V. The lower portion of the amplitude range was primarily used to measure the response of the high gain LUCROD while the upper portion of the range was useful only for the low gain LUCROD. 

Results from a complete scan of pulser amplitudes are shown in Figure~\ref{fig:pulser_curves}. Solid lines show the results of linear fits to the data. For the both low- and high-gain data, a deviation from strict linearity is observed for higher voltage values though the high-gain deviation is not illustrated in the figure. A complete determination of non-linear corrections from the pulser data and the application of these in offline analysis of is under development.

%% file: calib.tex
\begin{figure}[p]    
\centering 
 \includegraphics[height=0.35\textheight]{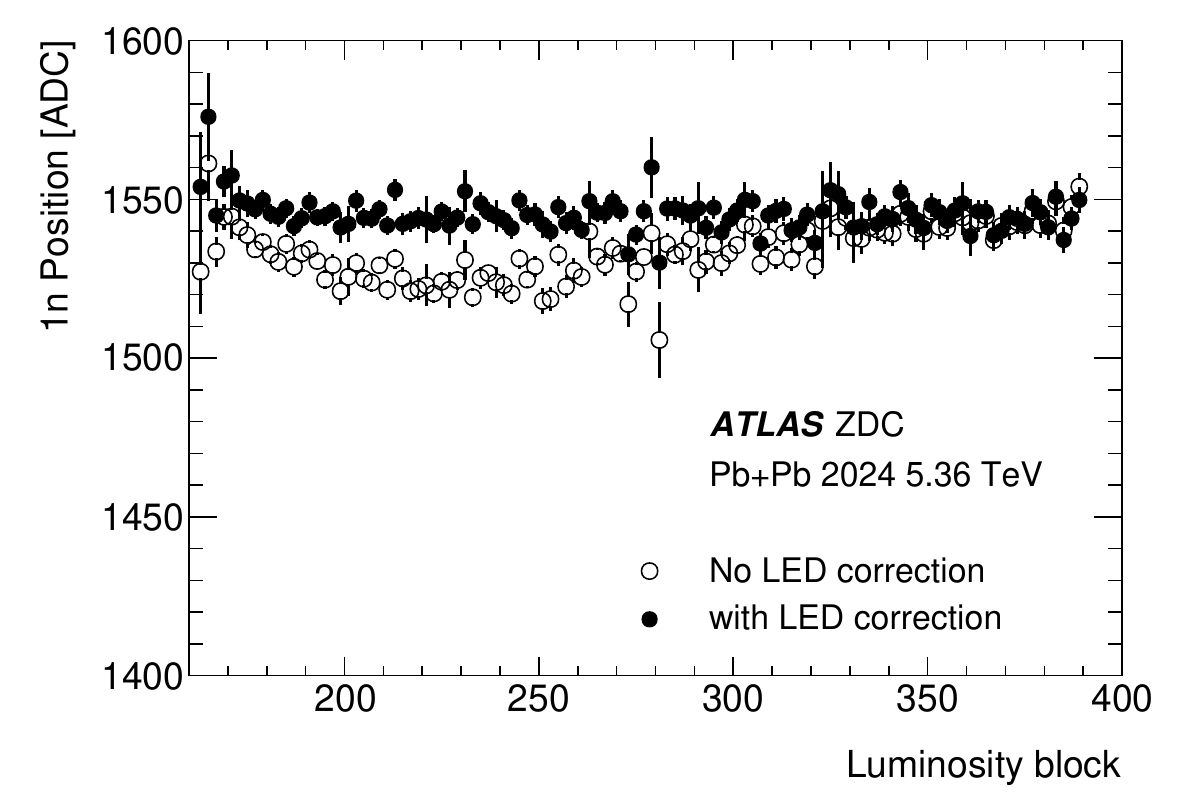}
\caption{
The position of the single neutron peak, determined by a Gaussian fit, before final energy calibration,
as a function of ATLAS luminosity block number (each with a duration of about one minute). The data were taken over a five hour period of $\snn=5.36$~TeV \PbPb operation in 2024. The open points show the peak position without application of LED-based corrections; the closed points show the results after correction, demonstrating that the correction removes most of the time dependence.}
\label{fig:LED_corr_short}
\end{figure}

As discussed in Section~\ref{sec:ZDCwp} the high voltage working points for the ZDC are established accounting for variations in the PMT anode sensitivity and the hadronic shower response as measured in test beam. The resulting pulse amplitudes, when summed without additional calibration factors, yield uncalibrated shower energies that provide clear identification of the \onen, \Nn[2], \Nn[3], and higher peaks. This fact is essential to operation of the ZDC trigger in Run~3. However,
results obtained with the LED system demonstrate clear impacts on the calorimeter energy response from both radiation damage -- observed by the dependence on integrated luminosity-- and large anode currents -- demonstrated by the rapid variation of the LED response at the start of and within an LHC fill. Both of these effects are addressed by corrections derived from the ZDC LED system, described in subsection~\ref{sec:LED}.
The LED-based corrections were derived and applied as a function of ATLAS luminosity block.  The effect of these corrections on the time-dependent response of the ZDC is illustrated in Figure~\ref{fig:LED_corr_short}, 
which shows the luminosity block dependence of the uncalibrated one-neutron peak before and after LED corrections for a single LHC fill. Prior to correction, the ZDC response as indicated by the \onen peak drops as the LHC establishes collisions  and then recovers as the luminosity decreases. This pattern was repeated for essentially all \PbPb\ fills. The response of the ZDC after LED-based corrections is nearly constant as a function of luminosity block, or equivalently time within an LHC fill.

\begin{figure}[p]    
\centering 
\includegraphics[height=0.35\textheight]{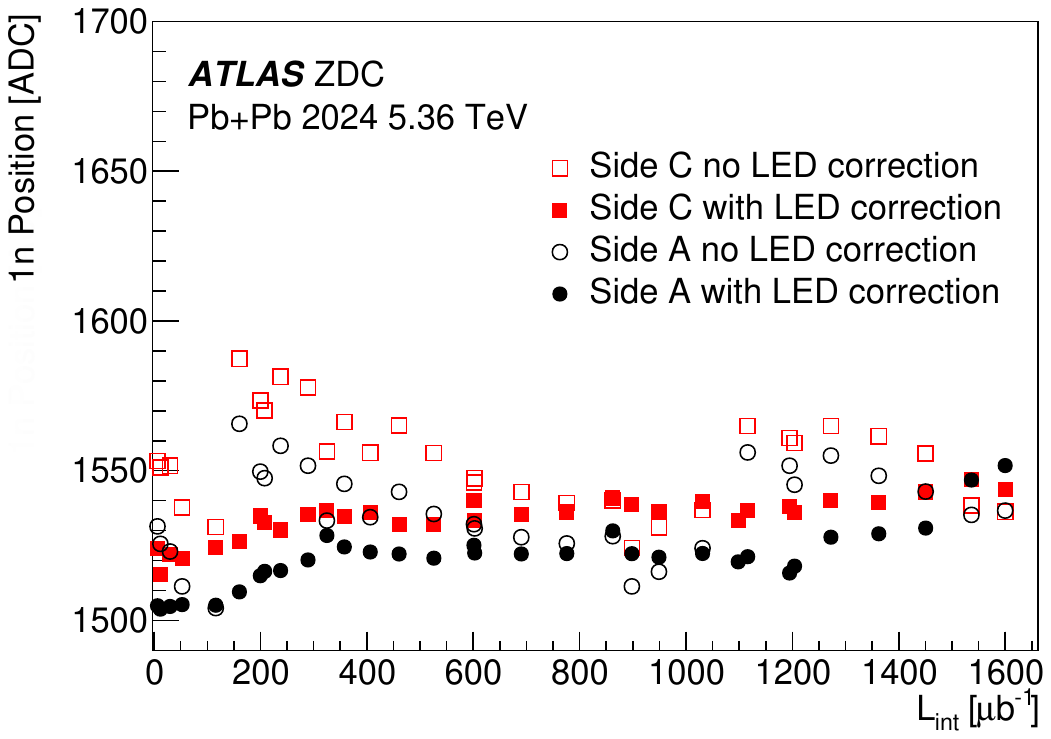} 
\caption{
The position of the single neutron peak, determined by a Gaussian fit before final energy calibration, for each experimental run, both before and after LED calibration, as a function of the integrated luminosity.  Data from both ZDC sides is shown, with Side A with black marks and Side C with red.
On each side, the LED-calibrated mean \onen energies are found to be approximately independent of luminosity within $\pm 1\%$, while the uncorrected values show the clear impact of radiation damage, periodically compensated by increasing the ZDC PMT HV. The results from the two sides also agree within 1\%.}
\label{fig:LED_corr_long}
\end{figure}

After performing the above-described LED-based  corrections, which account for both the long-timescale
and short-timescale effects, the ZDC energy spectrum in each ATLAS data-taking run is fit to a Gaussian in the vicinity of the observed single-neutron peak to extract the mean \onen energy. The resulting \onen mean energies for data collected in 2024 are shown in the right panel of Figure~\ref{fig:LED_corr_long} as a function of the integrated luminosity from the start of 2024 \PbPb operation. The plot covers a time-range of approximately 3 weeks. Results are shown prior to and following the application of corrections derived from the LED data.
Prior to correction, the \onen peak position decreases systematically with increasing luminosity, though the uncorrected data also show sudden increases in \onen peak position due to adjustments that were made to the PMT HV values  during data-taking to ensure stable operation of the trigger. Subsequent to the LED-based corrections the \onen mean energy is observed to be constant within $\pm 1\%$.  

To summarize, corrections derived from the LEDs correct both the long-term and short term deviations in the PMT response in a manner that can be evaluated using the \onen peak position.  
The short-term behaviors are clearly associated with the instantaneous luminosity, changing rapidly when beams come into collision, varying by several percent over a run, and not always in the same direction for different ZDC modules.
In contrast, the long-term variations (shown in Fig.~\ref{fig:LED_corrected}) are primarily downward by 5-20\% and scale nearly exponentially with the integrated luminosity. The integrated luminosity should be approximately proportional to the dose received by the ZDCs. Based on the previously measured sensitivity of the fused silica radiator rods and the known sensitivity of the borosilicate windows on the PMTs, the radiation damage effects are interpreted as being primarily associated with the PMTs. Based on anecdotal observations of increased response in the ZDC after periods of extended (multi-hour) LHC down-time, it appears that the radiation damage to the PMTs may be self-annealing on the time-scale of hours. However, proper, controlled studies of this annealing are not currently available.

The results presented in this section demonstrate that the LED-based corrections simultaneously account for both short-term and long-term effects and maintain the position of the single-neutron peak position to 1-2\% accuracy.  The source of residual variation is not yet understood, but it is possible that it results from radiation damage to components of the LED system, for example the optical fibers carrying the LED light to the PMTs or the LEDs themselves. 

\begin{figure}[t]
    \centering
    \includegraphics[width=0.8\textwidth]{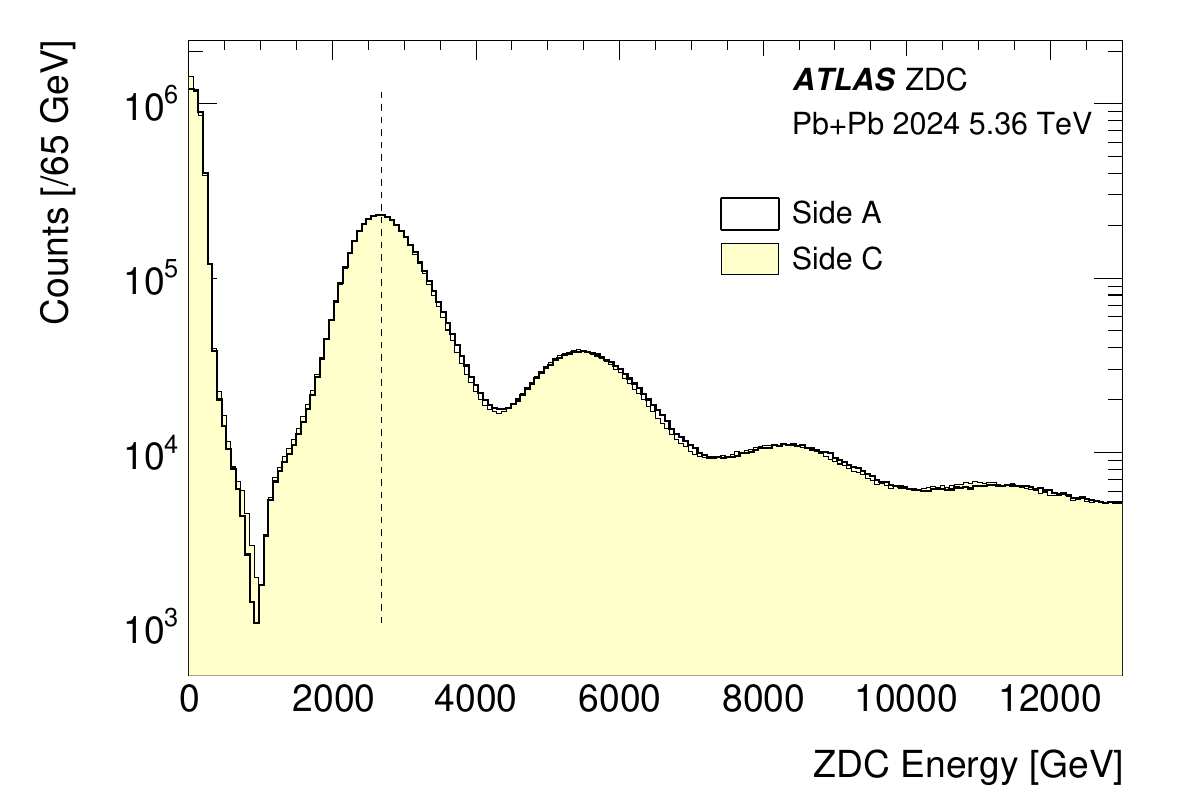}
    \label{fig:ZDC_energies_A_C_490223}
    \caption{Calibrated energies in GeV measured by the calorimeters on the two sides in data recorded by ATLAS on November 23, 2024 over a period of about 5 hours.  The data are fully calibrated using the methods discussed in the text. The dashed line indicates the per-nucleon beam energy of 2680~GeV.}
\end{figure}

The final step in the ZDC energy calibration procedure is a determination of a single multiplicative factor for each calorimeter in each run, that adjusts the \onen mean energy to the match the nominal per-nucleon beam energy of 2680~GeV.  
Distributions of fully-calibrated ZDC energy in the range corresponding to $0-4$ neutrons 
are presented in Figure~\ref{fig:ZDC_energies_A_C_490223}.
The shown data were measured during a five hour run period near the end of the 2024 \PbPb run. The fully-corrected energy distributions necessarily match in the vicinity of the \onen peak but are in good agreement over the entire shown energy range.

The above-described procedures, combined with the initial choices for the ZDC HV working points provide the calorimetric energy scale calibration needed by ATLAS to accomplish its heavy-ion physics goals. A more involved, data-driven energy calibration procedure, described in the appendix, was used in Run~2 to determine per-module gain factors, or "weights" based on a global optimization of the \onen peak position and resolution. In Run~3, that procedure is not needed, though it is still occasionally used to check that the chosen HV working points yield uniform energy response in each of the ZDC modules.

%% file: rpdcalib.tex
\begin{figure}[b]
    \centering
\includegraphics[width=0.49\linewidth]{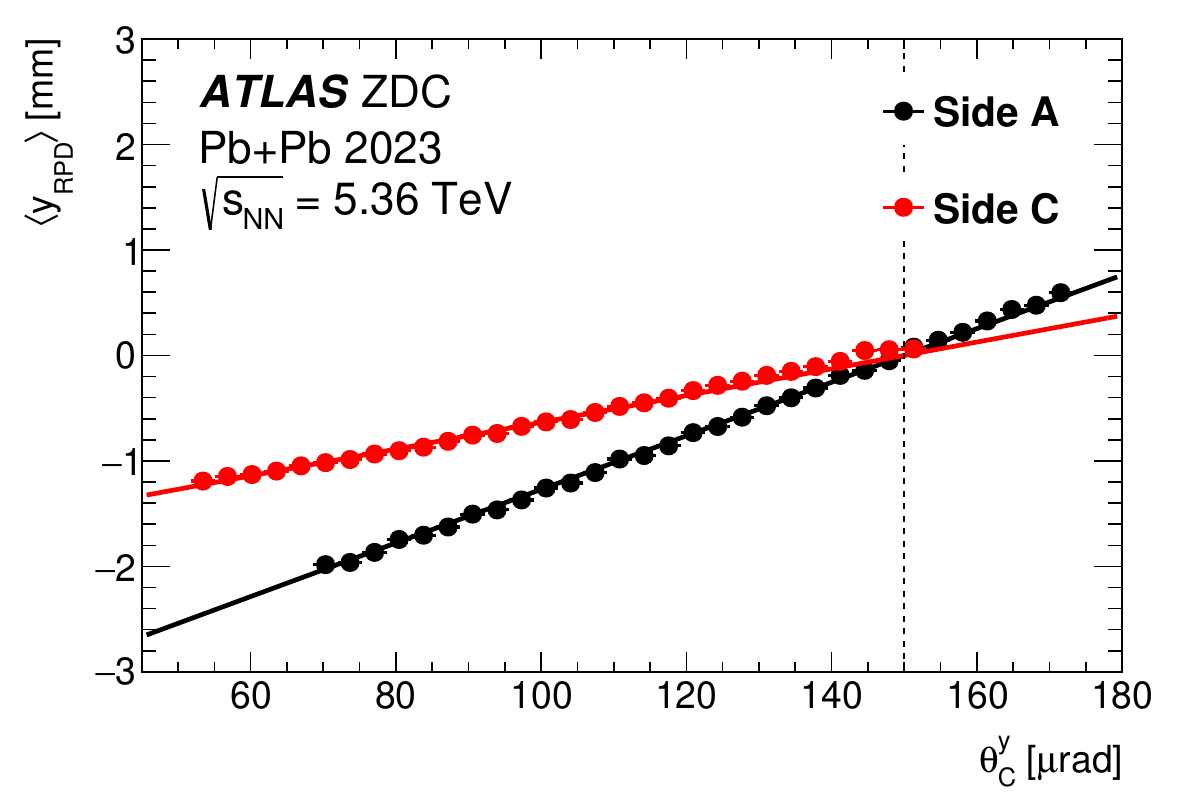}
\includegraphics[width=0.49\linewidth]{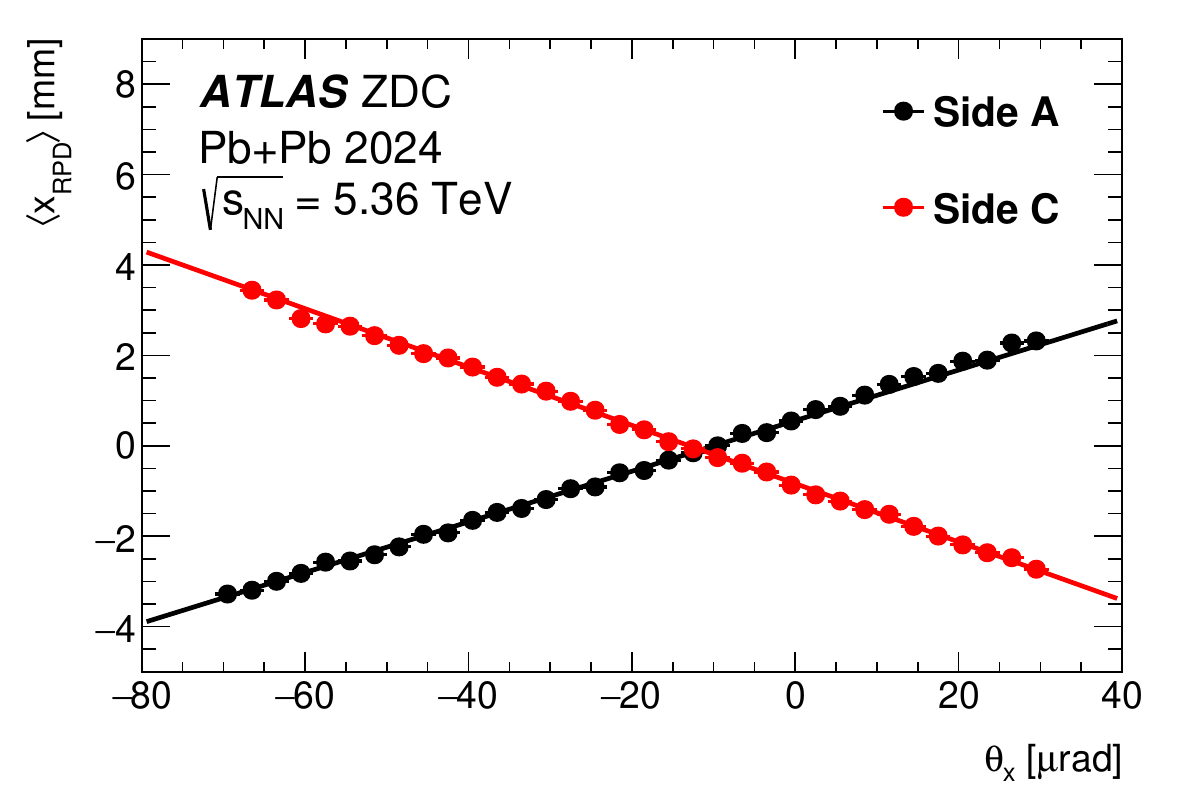}
   \caption{Left: Correlation between shower centroid position measured by RPD, $\mathrm{y}_\mathrm{RPD}$, and the beam vertical crossing angle ($\theta_{C}^\mathrm{y}$). Data are from LHC fill 9310 during which the crossing angle was varied for machine development purposes. Data from the A (C) side of the detector are displayed in black (red). A linear fit to the correlation between shower centroid and the beam vertical crossing angle is also reported.\\ Right: Correlation between shower centroid position measured by the RPD, $\mathrm{x}_\mathrm{RPD}$, and the variation of horizontal tilt angle ($\theta_\mathrm{x}$). Data are from LHC fill 10294, during which the beam tilt in ATLAS was varied by the LHC to investigate the residual horizontal tilt observed during the 2023 run. Data from the A (C) side of the detector are displayed in black (red). A linear fit to the correlation between shower centroid and the beam tilt is also reported. A correction of +15~$\mu$rad to the horizontal beam tilt was determined from this analysis and implemented for the 2024 Heavy Ion run.}
    \label{fig:RPD_xing}
\end{figure}

\label{sec:rpd_calib}
During the 2023 heavy ion run, the beam position was varied along the vertical direction, over a range of approximately 100 $\mu$rad for machine development (MD) purposes. During this hour of MD, the RPD collected data to be used for position calibration purposes. The time evolution of the beam vertical crossing angle was retrieved from the data posted by the DOROS beam position monitor, and correlated with the RPD data on a timestamp basis. The analysis of the sample collected was used to assess the correlation between the beam vertical crossing angle and the $y$-coordinate measured by the RPD. As one can see in Figure~\ref{fig:RPD_xing} (left panel), the detector response to changes in the vertical crossing angle of the beam exhibits high linearity.
However, the slope of the linear fit to the data is below unity, as expected from the detector design. This result is also consistent with the fact that the current analysis does not yet account for the known vertical dependence of the photon capture efficiency in the detector. 

During the 2023 run, both the DOROS and the RPD provided evidence of a residual horizontal crossing angle at IP1 during the heavy ion data taking. 
In 2024, during the Pb beam commissioning, a dedicated horizontal beam tilt scan at IP1 was carried out to evaluate this angle. The scan included five different variations ($\pm$50 $\mu$rad, $\pm$25 $\mu$rad, +15 $\mu$rad) of the horizontal beam tilt relative to the nominal configuration, corresponding $\theta_x$=0 in the right panel of Figure~\ref{fig:RPD_xing}. By analyzing online data collected at $\pm$50 and $\pm$25 $\mu$rad, the optimal beam position to cancel the residual horizontal crossing ($+15~\mu$rad) was identified and requested as the final step of the scan, demonstrating that this configuration results in an average horizontal position centered on both the RPDs. The results of an offline analysis of these data, shown in the right panel of Figure~\ref{fig:RPD_xing}, demonstrate the linear detector response with respect to variation of the horizontal beam position.

%% file: MC.tex
\subsection{Implementation of ATLAS ZDC simulations}
The ZDC description in the ATLAS \Geant  Monte Carlo \cite{SOFT-2010-01,SOFT-2022-02} has been completely overhauled for Run~3. The improvements consist of three main parts, the conversion of the geometry construction to a modular format with the addition of the RPD and BRAN modules, the move to using \Geant processes for the simulation of Cherenkov photons, and the addition of calorimeter calibration Sensitive Detector (SD) classes.
The addition of the RPD and BRAN modules required a modular design allowing for simulation of the multiple arrangements of the detectors described in Section~\ref{sec:zdc_overview}. 

The ZDC simulation was improved for Run~3 to include \Geant Cherenkov
and optical processes, to accurately simulate the production and
propagation of Cherenkov photons within the ZDC rods and RPD
fibers. This required the addition of optical properties to some
materials within the detector, which were taken from manufacturer data
sheets where available. Although the generation of Cherenkov photons
in \Geant has very little effect on the simulation time, the
propagation of the photons can lead to a dramatic increase. For this
reason the SD performs calculations after each photon's first step in
the optical volume to determine if the photon will be detected by the
PMT. If not, the photon is removed immediately, eliminating the cost
of propagation through the optical medium. For each ZDC channel, the
total number of photons determined to reach the PMT is converted to a
pulse amplitude using calibration factors chosen to reproduce the
pulse amplitudes seen in the data. The time structure of the digitized
pulses is produced using the same waveform used to fit pulses in the
ZDC offline analysis.

To evaluate ``truth'' energy deposits in components of the ZDC, a new SD class has been developed based on one used by the ATLAS LAr calorimeter. This class is used to record ``calibration hits'' that keep track of several categories of energy deposits in each ZDC, RPD and BRAN volume:
\begin{itemize}
    \item \textit{EM energy}    : visible energy loss by EM processes
    \item \textit{Non-EM energy}: visible energy loss by non-EM processes, e.g. pion $\text{d}E/\text{d}x$.
    \item \textit{Escaped energy}: energy which escapes from a given volume cell, primarily via production of neutrinos or escaping muons.
    \item \textit{Invisible energy}: invisible energy loss, typically due to nuclear binding energy.
    \item \textit{Visible energy}: the sum of the EM and  non-EM visible energies.
    \item \textit{Total energy}: the sum of Visible, Invisible, and Escaped energies
\end{itemize}
These deposits are recorded for each material in each ZDC module, 
  allowing for detailed studies of their effect on the energy resolution 
  of the ZDC.

\subsection{Event generation for ZDC Simulation\label{sec:mc_generation}}
As most Monte Carlo event generators do not properly simulate neutron emission in nuclear collisions, the simulations used here 
rely on a stand-alone event generator that produces a fixed number of neutrons on each side with a Fermi momentum distribution \cite{Fermi} in the nuclear rest frame. 
For processes involving $^{208}$Pb ions, neutrons are sampled uniformly in $p^2\text{d}pd\Omega$ up to the Fermi surface ($p_F = 265$~MeV). To account for nuclear binding, the energy of each neutron in the nuclear rest frame is reduced by 31~MeV.  This value was determined by requiring that the (average) total energy of 208 nucleons with the same momentum and energy distribution would have an equivalent mass equal to that of a Pb nucleus. 

The (off-shell) neutron three-momenta and energies are boosted from the rest frame to the lab frame according to the velocity of the Pb nucleus and the resulting 4-momenta are put on shell. Two different methods of applying the on-shell condition -- adjusting the energy or adjusting the momentum magnitude  -- yield results that are indistinguishable. With this procedure, the generated neutrons in the laboratory have a mean energy consistent with the nominal energy and a standard deviation corresponding to 13\% of the mean. The broadening of the neutron energy distribution due to the intrinsic momenta of the neutrons in the Pb nucleus represents a substantial irreducible contribution to the resolution on the measurement of the number of neutrons in the ZDC.

\subsection{Simulations of ZDC calorimeter response }
\begin{figure}
    \centering
    \includegraphics[width=0.99\linewidth]{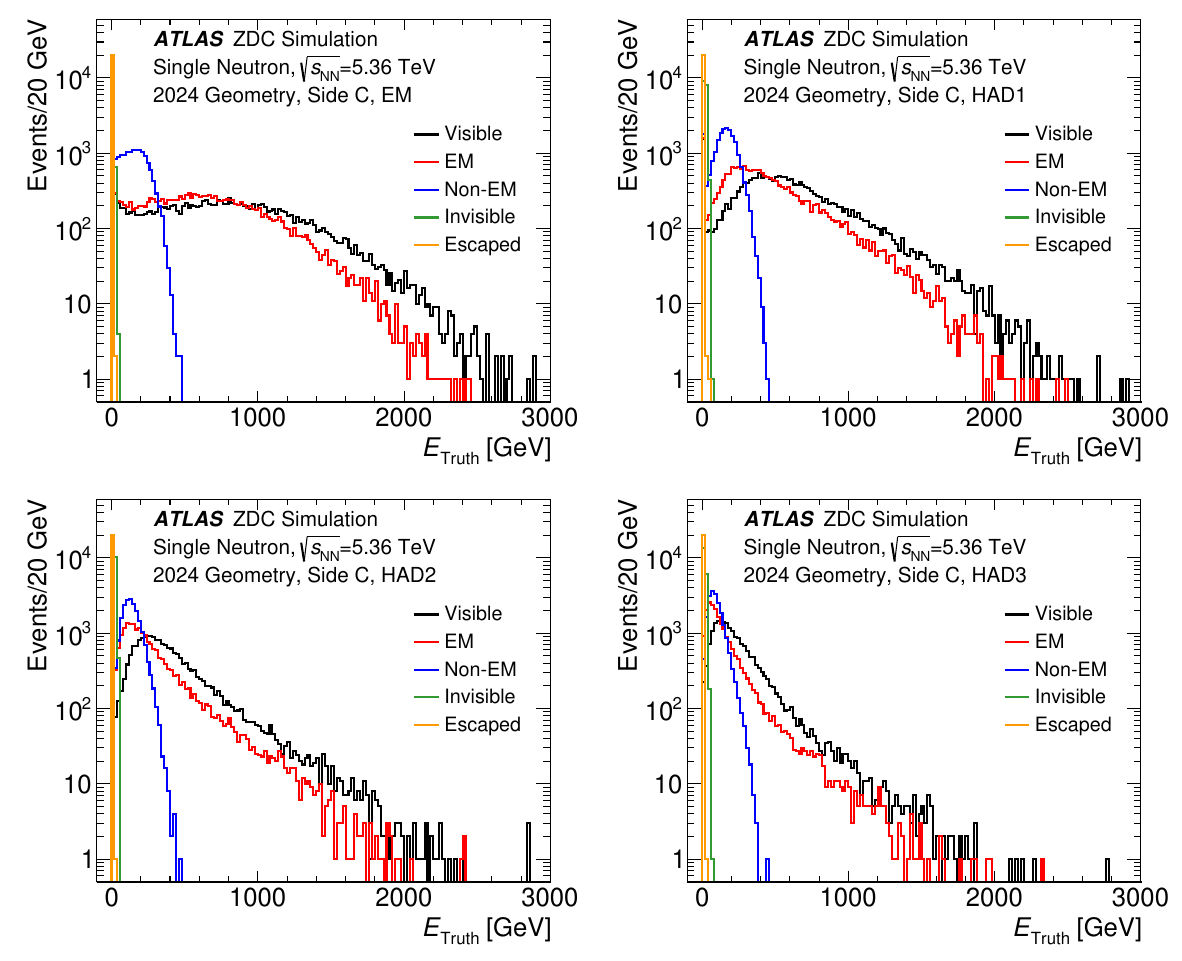}
    \caption{
      Breakdown of the truth-energy deposited 
      by 2.68~TeV single neutrons in each of 
        the four ZDC-Modules on the C Side, for 
        the 2024 ZDC configuration. Distributions are shown of the different categories of truth energy deposited within each module. The horizontal axis has been shifted slightly to make visible the distributions of invisible and escaped energy, which are highly compressed near 
        zero.
    }
    \label{fig:ModuleEnergyBreakDownSideC}
\end{figure}

\begin{figure}
    \centering
    \includegraphics[width=0.99\linewidth]{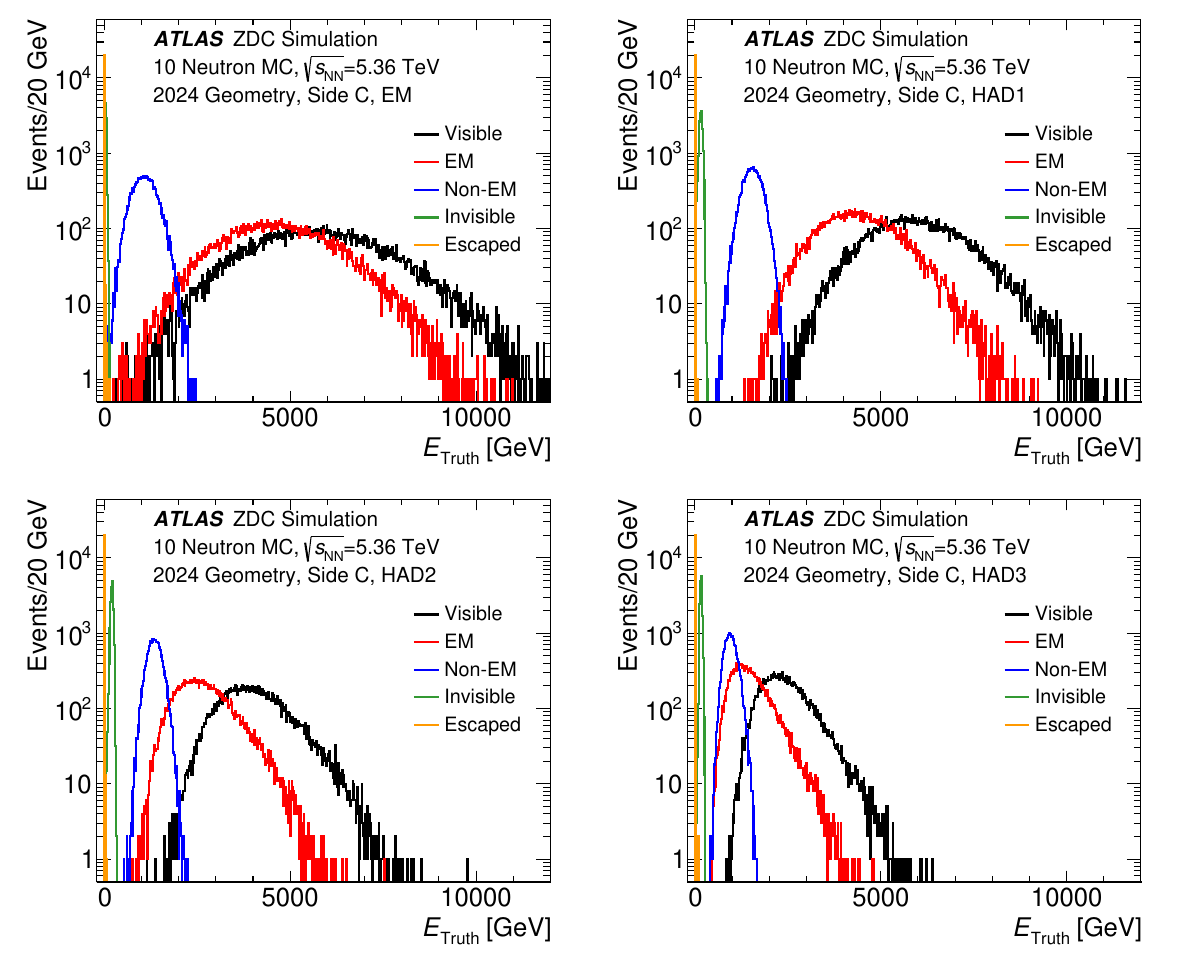}
    \caption{
      Breakdown of the truth-energy deposited 
      by ten-neutron events in each of 
        the four ZDC-Modules on Side-C, for 
        the 2024 ZDC configuration.  Distributions are shown of the different categories of truth energy deposited within each module. The horizontal axis has been shifted slightly to make visible the distributions of invisible and escaped energy, which are highly compressed near 
        zero.
    }
    \label{fig:ModuleEnergyBreakDownSideC_5n}
\end{figure}

Monte Carlo simulations for the ZDC were performed using the event
generation framework described above in
Section~\ref{sec:mc_generation}.  Samples of 20000 events were
generated separately for single-neutron and ten-neutron events, at
\snn=5.36~\TeV or, equivalently, with single-neutron energies of
2.68~TeV, for the different ZDC configurations in 2023 and in
2024. The simulations were performed using the full ATLAS detector geometry including,
beam-pipes, and other beam-line components upstream of the ZDC so that contributions from showers initiated upstream of the ZDC are included in the simulation.
We note that there is a small, but persistent asymmetry in the ZDC response on the A and C sides when performing simulations of the ZDC in the ATLAS GEANT4 ATHENA environment. Extensive checks of the GEANT4 description of the ZDCs and the generation methods described above rule them out as potential causes of the asymmetry. Comparisons between data and MC, an example of which is shown below, suggest that the simulation of the C side response is to be better trusted. Thus, in the following, results will be shown primarily for the C side, though results for the A side will be used when comparing 2023 and 2024 geometries. 

\begin{figure}
    \centering
    \includegraphics[width=0.99\linewidth]{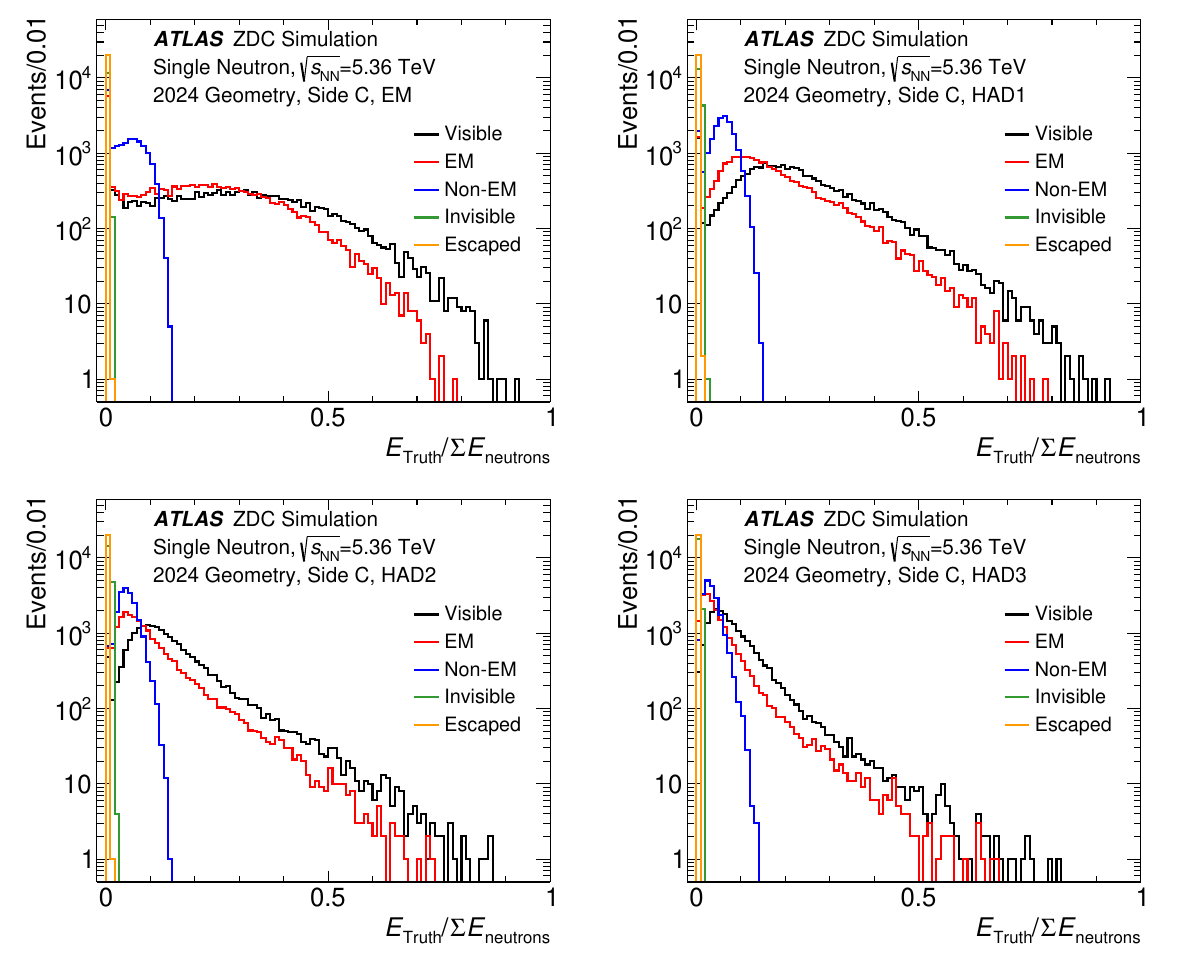}
    \caption{
      Breakdown of the truth-energy fraction deposited by single-neutrons in each of the 
        four ZDC-Modules on Side-C, for the 2024 ZDC configuration. 
     Distributions are shown for the different categories of truth energy 
       deposited within each module.
     The horizontal axis has been shifted slightly to make visible 
       the distributions of invisible and escaped energy, 
       which are highly compressed near zero.
    }
    \label{fig:ModuleScaledEnergyBreakDownSideC_1n}
\end{figure}

\begin{figure}[p]
    \centering
    \includegraphics[width=0.9\linewidth]{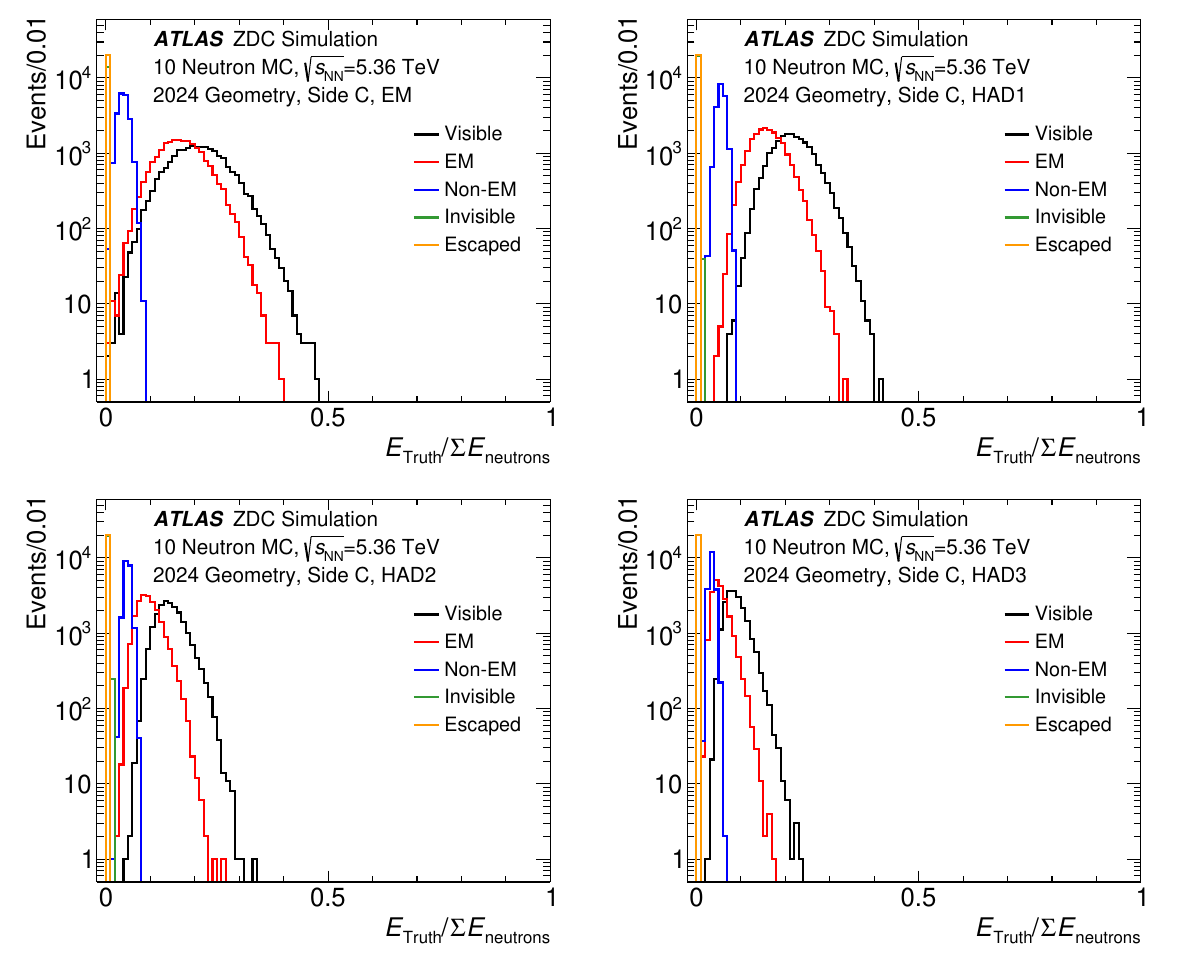}
    \caption{
      Breakdown of the truth-energy fraction deposited by ten-neutron events in each of the 
        four ZDC-Modules on Side-C, for the 2024 ZDC configuration. 
     Distributions are shown for the different categories of truth energy 
       deposited within each module.
     The horizontal axis has been shifted slightly to make visible 
       the distributions of invisible and escaped energy, 
       which are highly compressed near zero.
    }
    \label{fig:ModuleScaledEnergyBreakDownSideC_5n}
\end{figure}

\begin{figure}[p]
    \centering
    \includegraphics[width=0.45\linewidth]{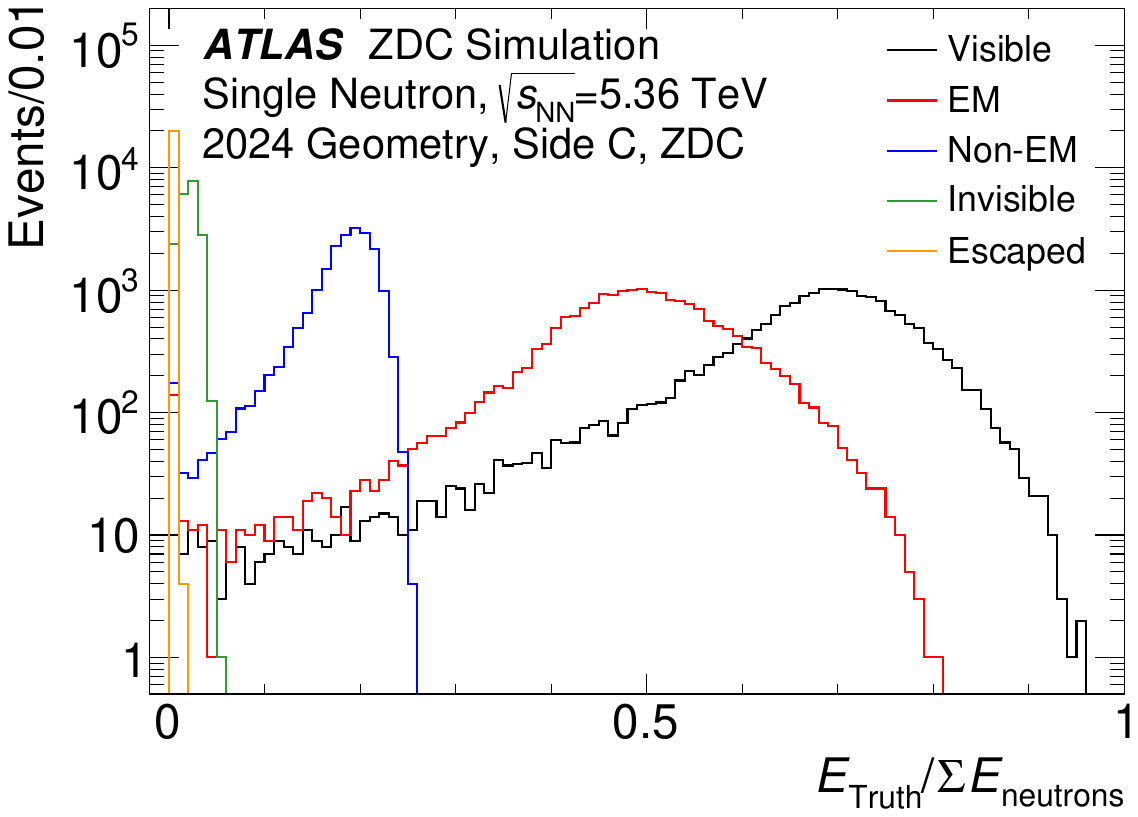}%
    \includegraphics[width=0.45\linewidth]{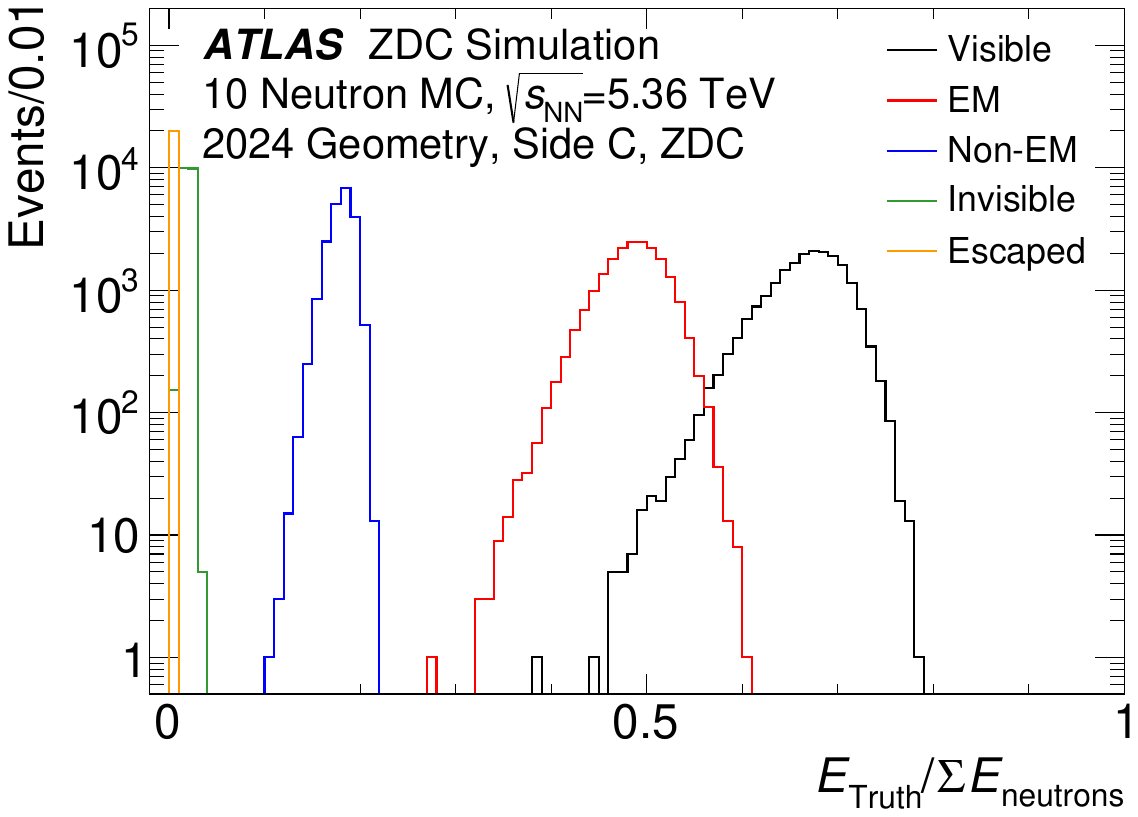}
    \caption{
      Breakdown of the truth-energy fraction deposited by single-neutron (left) and ten-neutron (right) events in the ZDC on Side-C, for the 2024 ZDC configuration.  
      Distributions are shown for the different categories of truth energy. 
      The horizontal axis has been shifted slightly to make visible the distributions of invisible and escaped energy, which are highly compressed near zero.
    }
    \label{fig:ModuleEnergyBreakDownSumSideC}
\end{figure}

Figure~\ref{fig:ModuleEnergyBreakDownSideC} shows the distribution of the
  Visible, EM, non-EM, Invisible and Escaped energy in the four ZDC modules
  on Side~C, for the 2024 ZDC configuration and for the single-neutron 
  MC sample.
Figure~\ref{fig:ModuleEnergyBreakDownSideC_5n} shows similar distributions for 
  the ten-neutron sample.
As expected, the mean values of the EM and non-EM visible energies decrease 
  systematically with depth -- from the EM to the HAD3 module. 
However, the decrease in the EM energy is much faster than that for the 
  non-EM energy.
The fractions of Invisible and Escaped energy are negligible compared to 
  the Visible energy.
Furthermore, the mean values of the different components are found to scale
  linearly with the number of neutrons used in the simulation.
Figures~\ref{fig:ModuleScaledEnergyBreakDownSideC_1n}
  and~\ref{fig:ModuleScaledEnergyBreakDownSideC_5n} 
  show similar distributions but with the energy scaled by the sum of the energy 
  of the incident neutrons.
This scaling takes out the effects of the smearing of the neutron energy
  due to the Fermi-momentum which was discussed in Section~\ref{sec:mc_generation}.
Figure~\ref{fig:ModuleEnergyBreakDownSumSideC} shows similar distributions for
  the sum of the four ZDC modules on Side-C, for the single-neutron and ten-neutron samples.

\begin{figure}[!tb]
    \centering
    \includegraphics[width=0.85\linewidth]{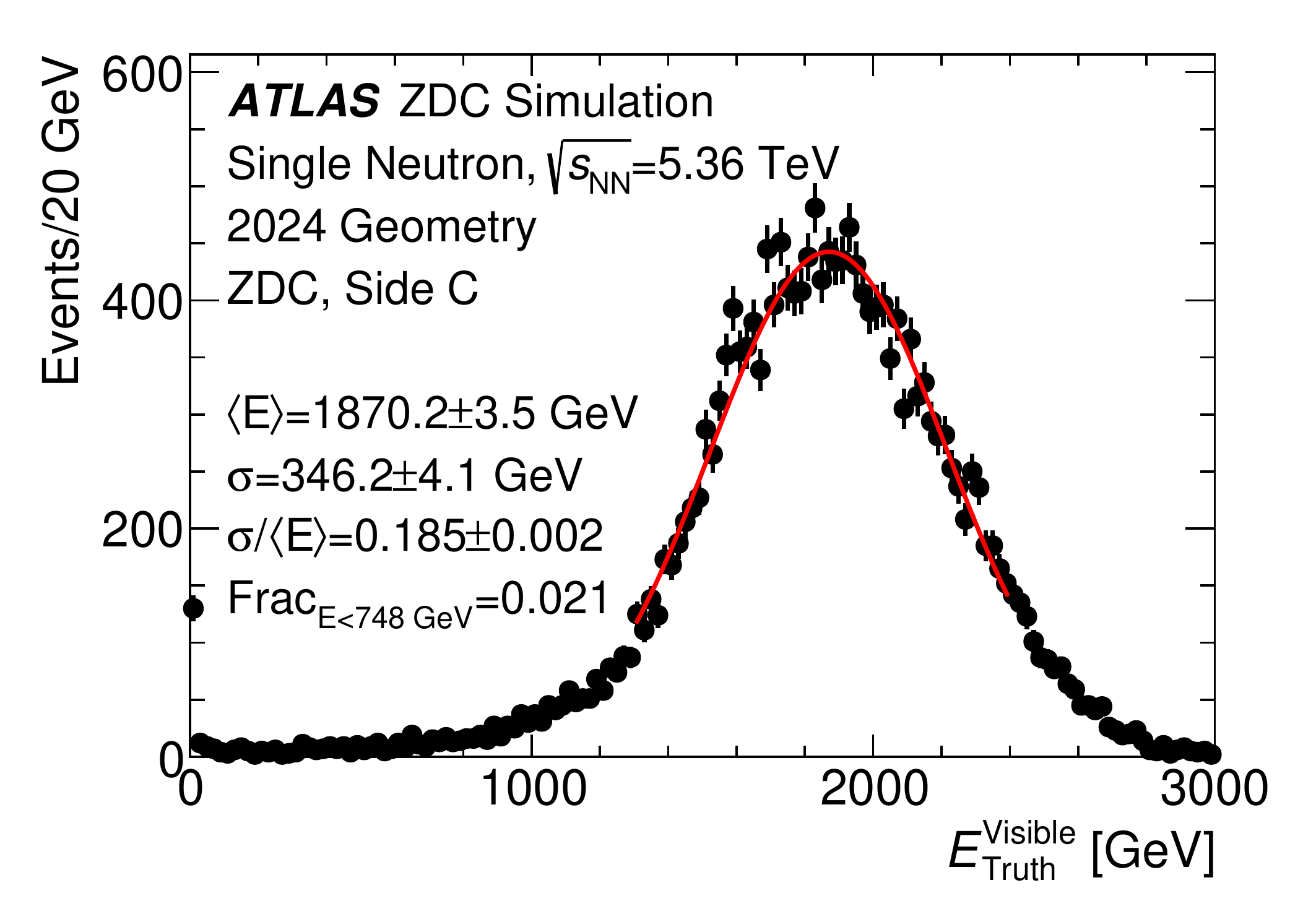}
    \caption{
      The sum of the Visible energy deposited in the four modules of 
        the ZDC, in single 2.68~TeV neutron MC simulations.
      The plot corresponds to the 2024 ZDC configuration.
      The distribution is fit with a Gaussian function (red line) 
        to obtain the energy resolution of the ZDC and an estimate of the
        inefficiency introduced by threshold requirements imposed 
        in the offline analysis (see text).
    }
    \label{fig:ZdcTruthSumReso}
\end{figure}

Figure~\ref{fig:ZdcTruthSumReso} shows the sum of the Visible energy
  deposited in the four modules of the ZDC, for single-neutron events.
The core of the distribution is fit with a Gaussian function
  to obtain the mean and the standard deviation of the distribution. The mean deposited energy of 1.87~TeV corresponds to 70\% containment of the hadronic showers. Previous Monte Carlo studies indicated that the 30\% lost energy is, on average, approximately-evenly split between transverse and longitudinal leakage. However, the fluctuations in the former have negligible impact on the ZDC response while the longitudinal fluctuations, primarily due to late showers, have a much larger impact. 
The ratio of the standard-deviation to the mean energy, 
obtained from the fit, 
  is $0.185\pm0.002$.  This is comparable to, although slightly larger than, the single-neutron resolution of 17\% reported for ATLAS ZDC measurements during Run~2 \cite{ATLAS:2024mvt}.
In \PbPb\ measurements, the separation between \Nn{0} and \onen events is typically obtained by applying a threshold at 40\% of the single-neutron energy. Based purely on the containment of the showers shown in Figure~\ref{fig:ZdcTruthSumReso},
such a selection would have an irreducible inefficiency of $\sim2\%$.
Since the width of the single neutron peak measured in the ZDC is dominated by the combination of intrinsic momentum spread and longitudinal shower fluctuations demonstrated in Figure~\ref{fig:ZdcTruthSumReso}, this inefficiency is likely to be very close to that incurred when applying the \Nn{0.4} requirement to data.

\subsection{Comparison of 2023 and 2024 ZDC configurations}
\begin{figure}[tbp]
    \centering
    \includegraphics[width=0.44\linewidth]{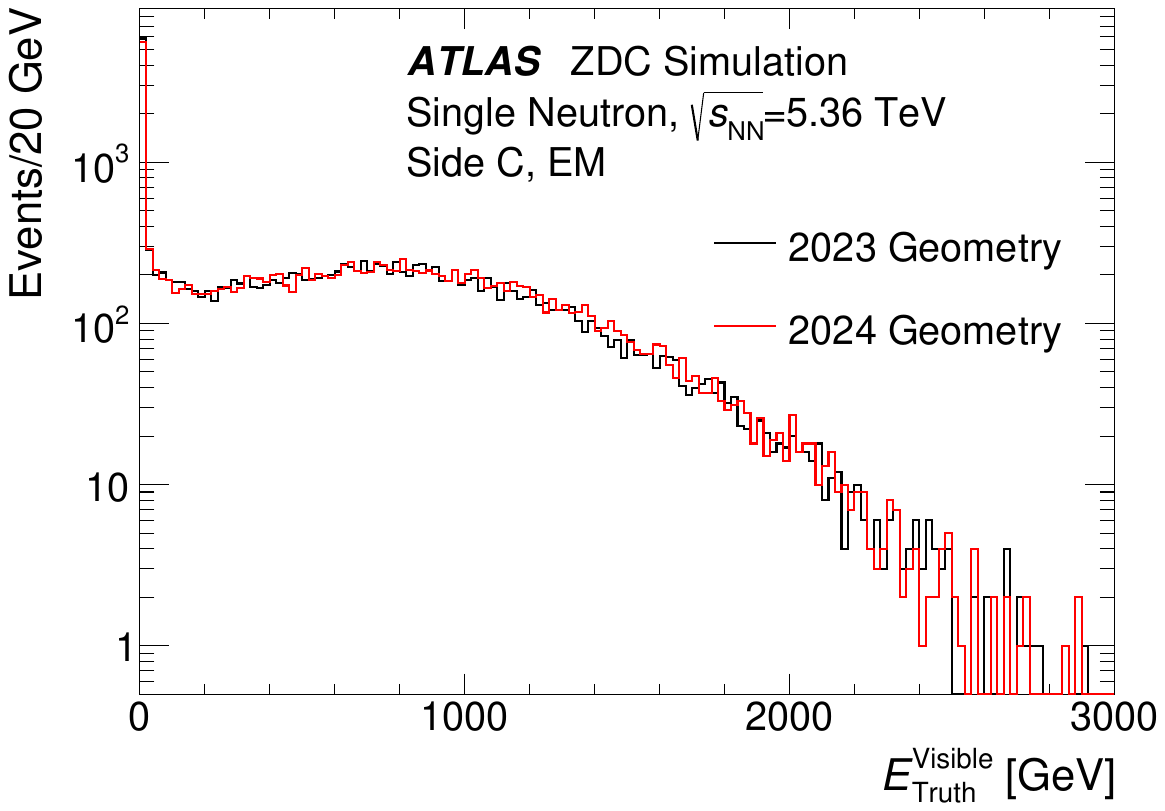}%
    \includegraphics[width=0.44\linewidth]{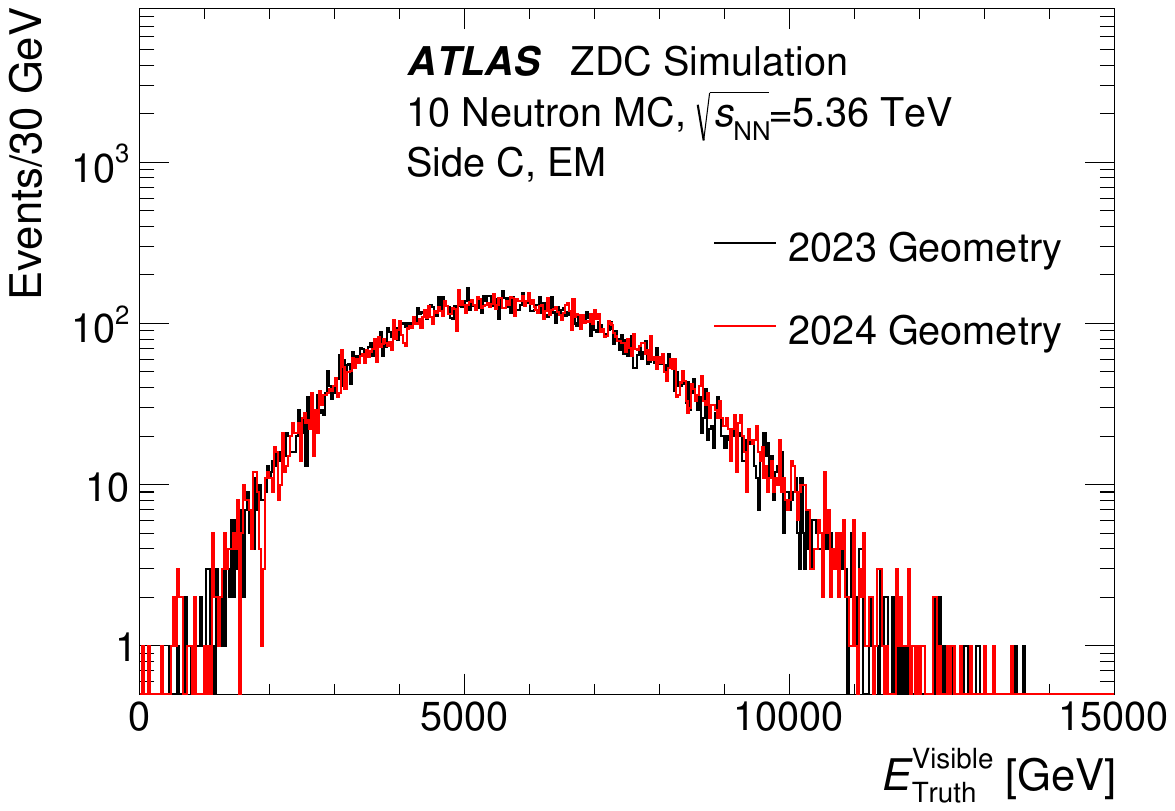}
    \includegraphics[width=0.44\linewidth]{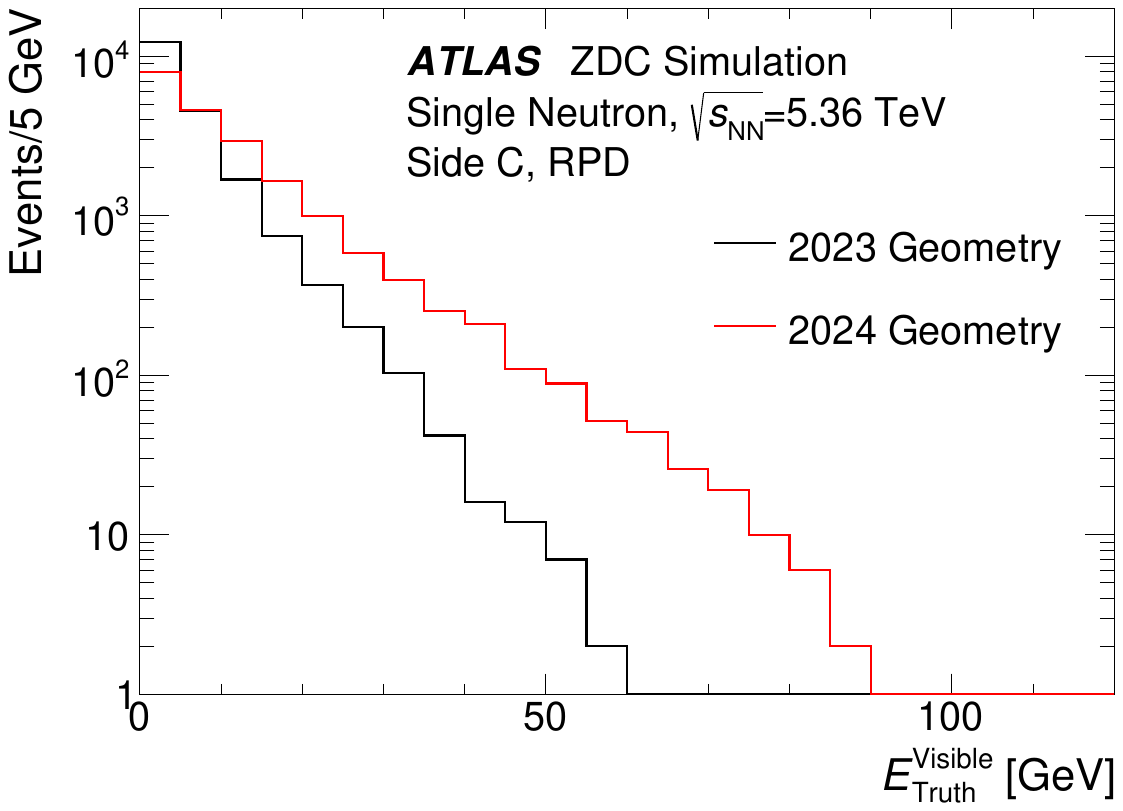}%
    \includegraphics[width=0.44\linewidth]{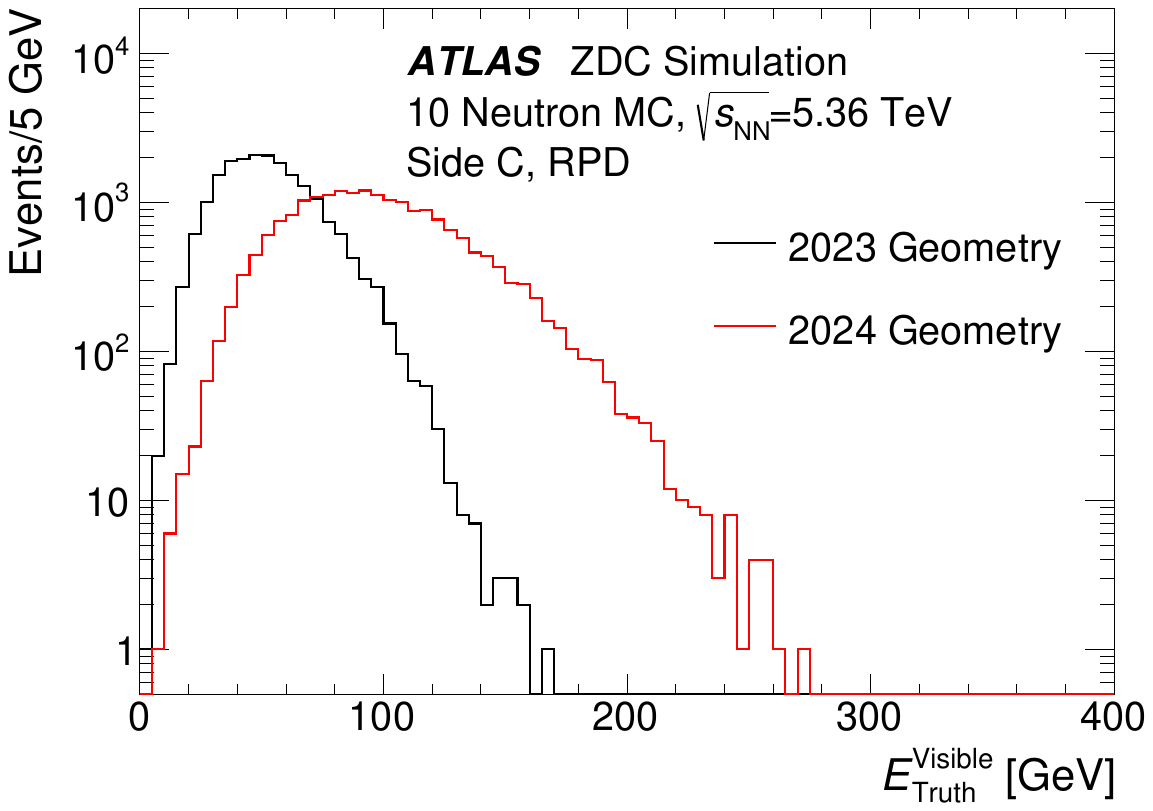}
    \includegraphics[width=0.44\linewidth]{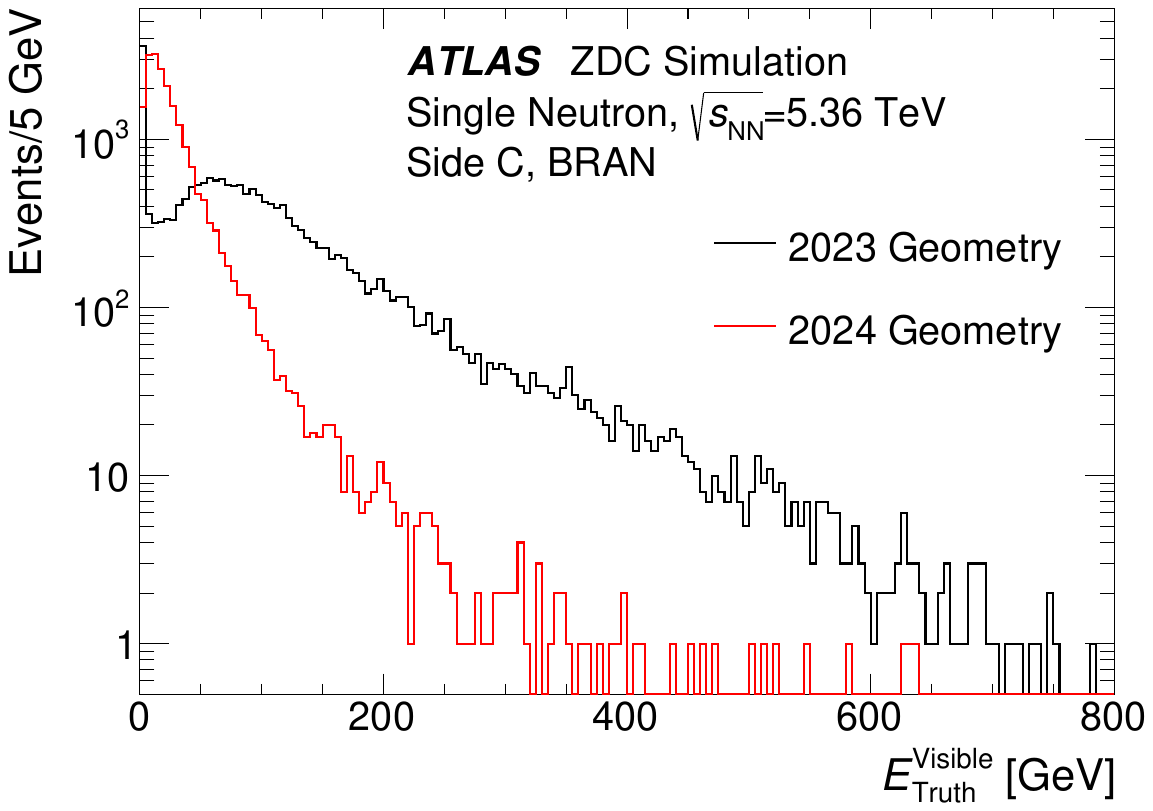}%
    \includegraphics[width=0.44\linewidth]{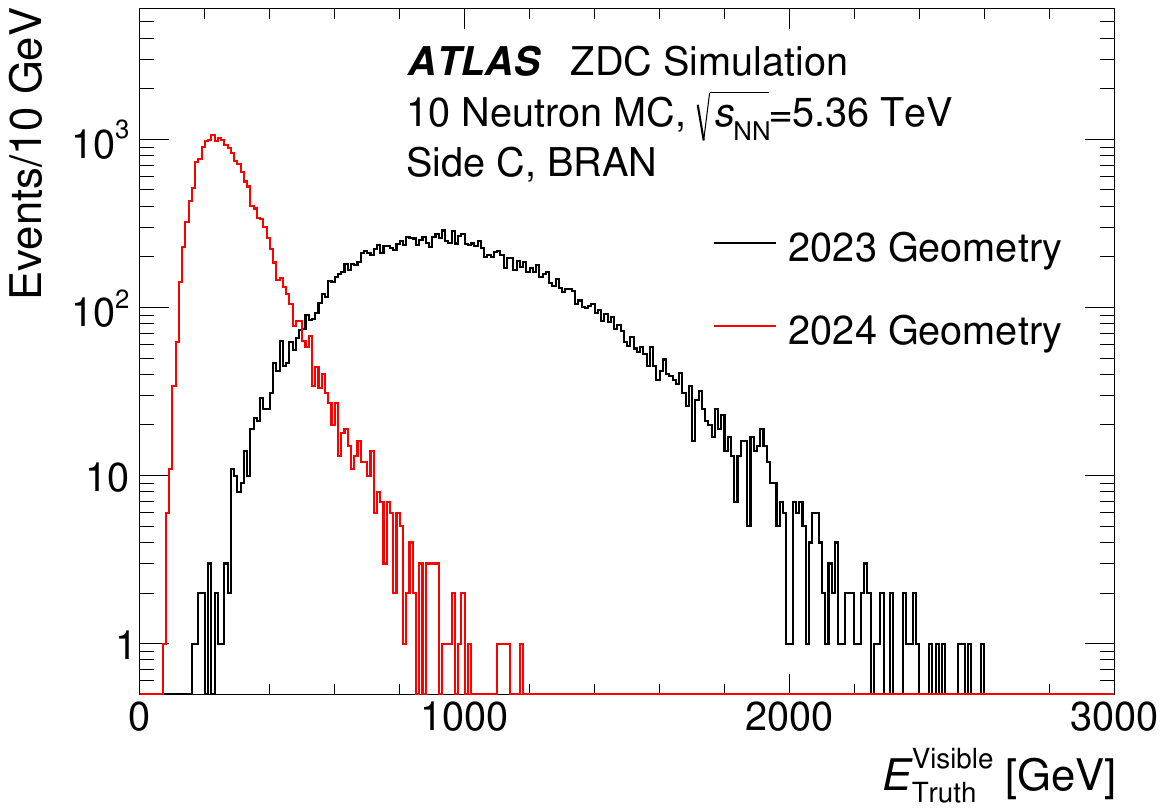}
    \includegraphics[width=0.44\linewidth]{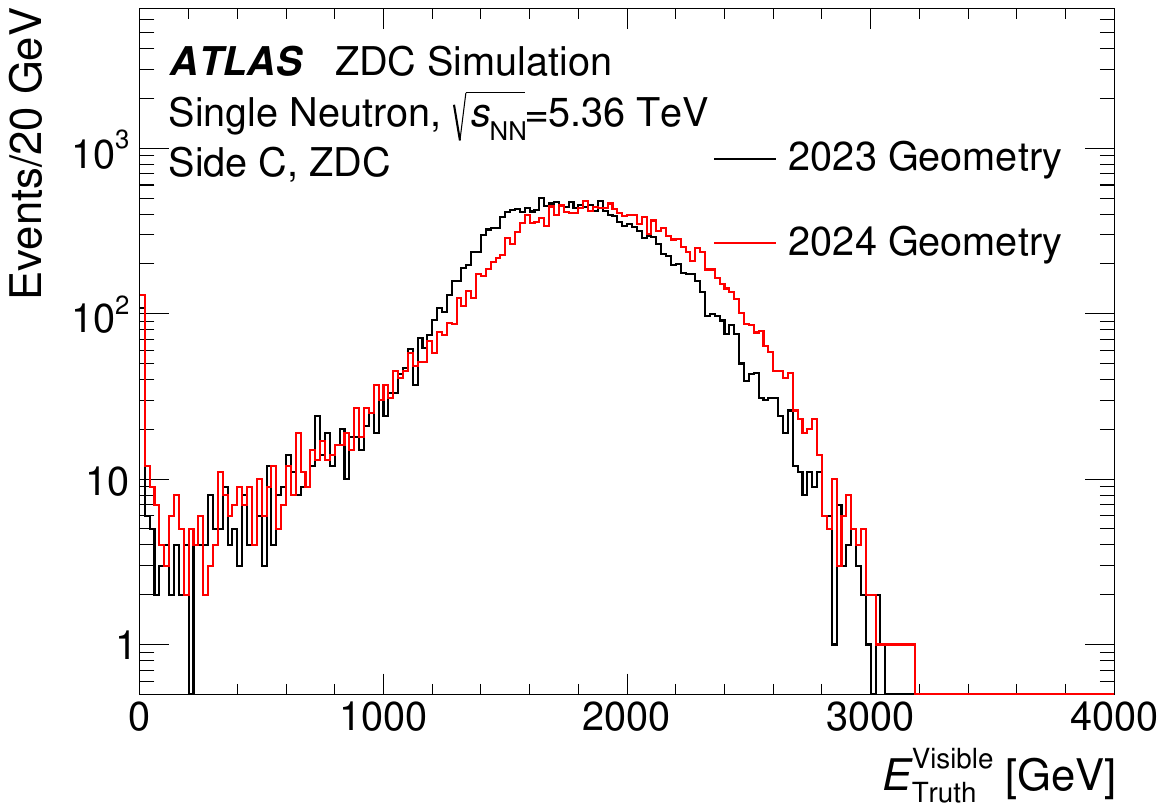}%
    \includegraphics[width=0.44\linewidth]{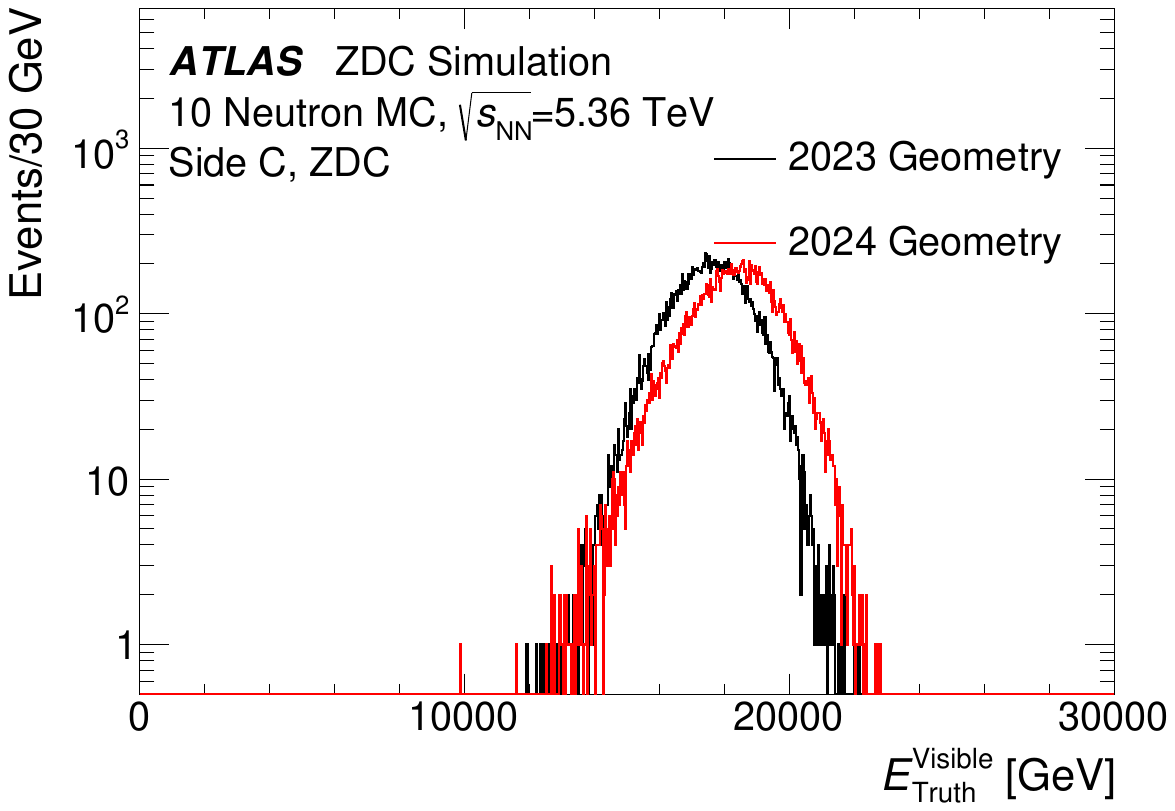}
    \caption{
      Comparison of the total visible energy deposited in the EM Module 
        (Top row), RPD (second row), BRAN (third row) and the 
        combined ZDC (bottom row) in the 2023 and 2024 ZDC configurations.
      Plots are for the Side C ZDC.
      The left and right panels correspond to the single 2.68~TeV neutron and 
        ten-neutron MC samples, respectively. The difference in the BRAN energy deposits results from the relocation of the BRAN behind the ZDC prior to the 2024 run (Section~\ref{sec:zdc_overview}). The change in the RPD response between 2023 and 2024 results from the replacement of the original C EM module with the original HAD3 module, which made the Side C calorimeter symmetric with the A side.  
    }
    \label{fig:CompareTruthEnergyTwoGeoms}
\end{figure}

\begin{figure}[tbp]
    \centering
    \includegraphics[width=0.48\linewidth]{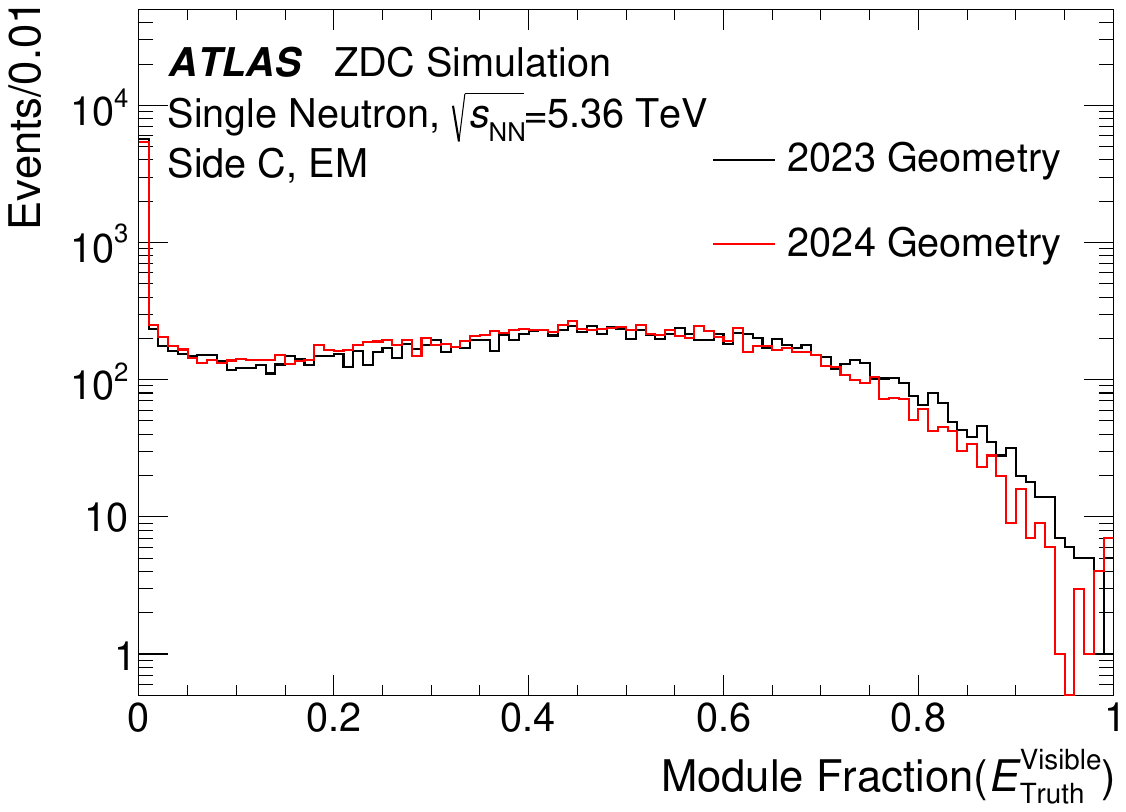}%
    \includegraphics[width=0.48\linewidth]{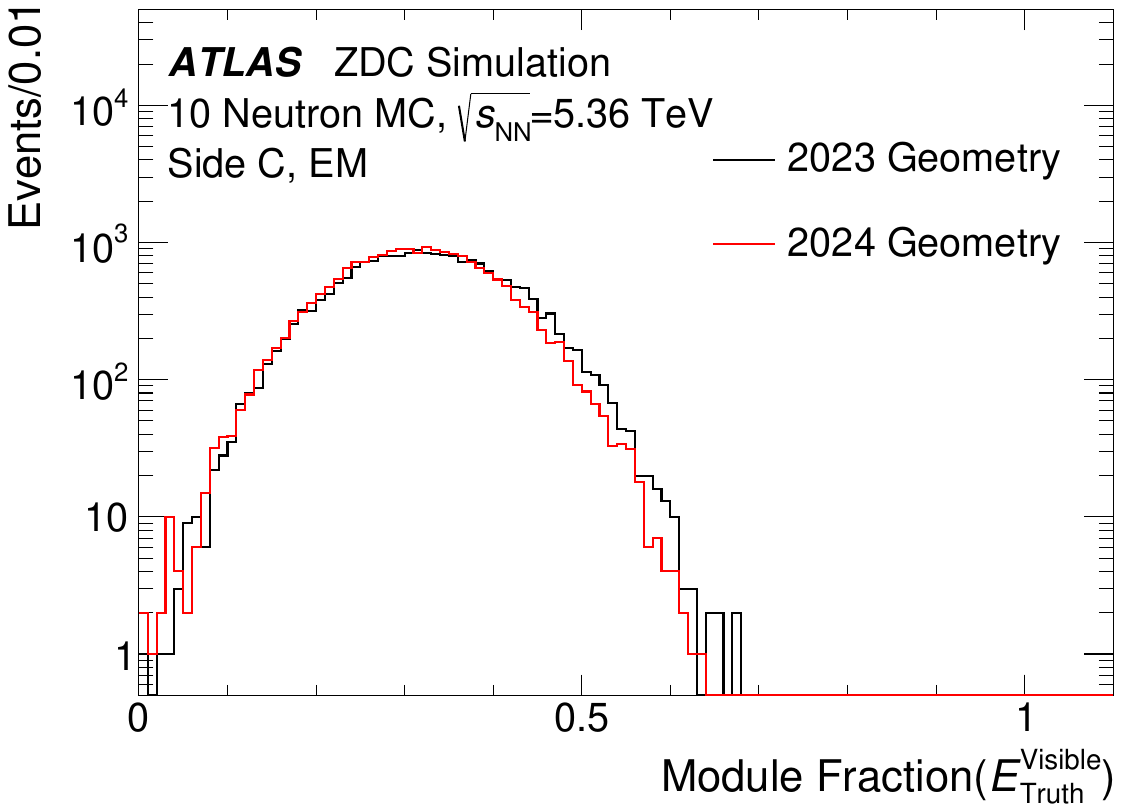}
    \includegraphics[width=0.48\linewidth]{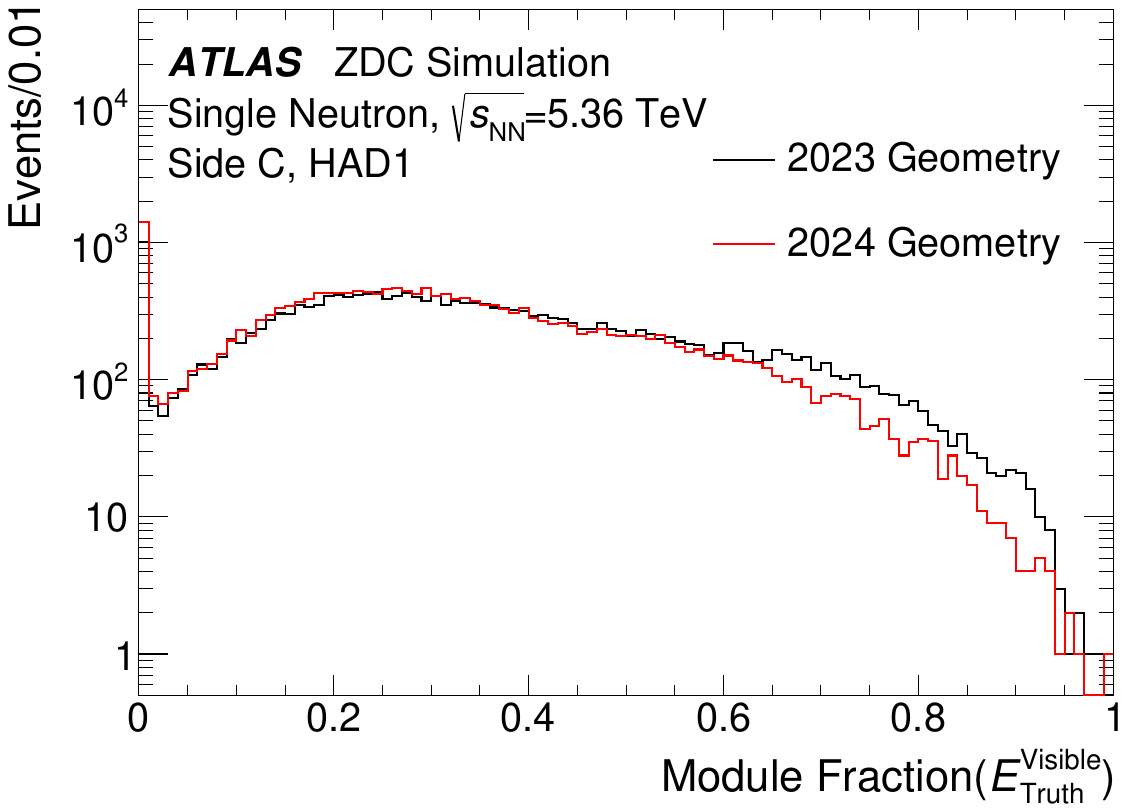}%
    \includegraphics[width=0.48\linewidth]{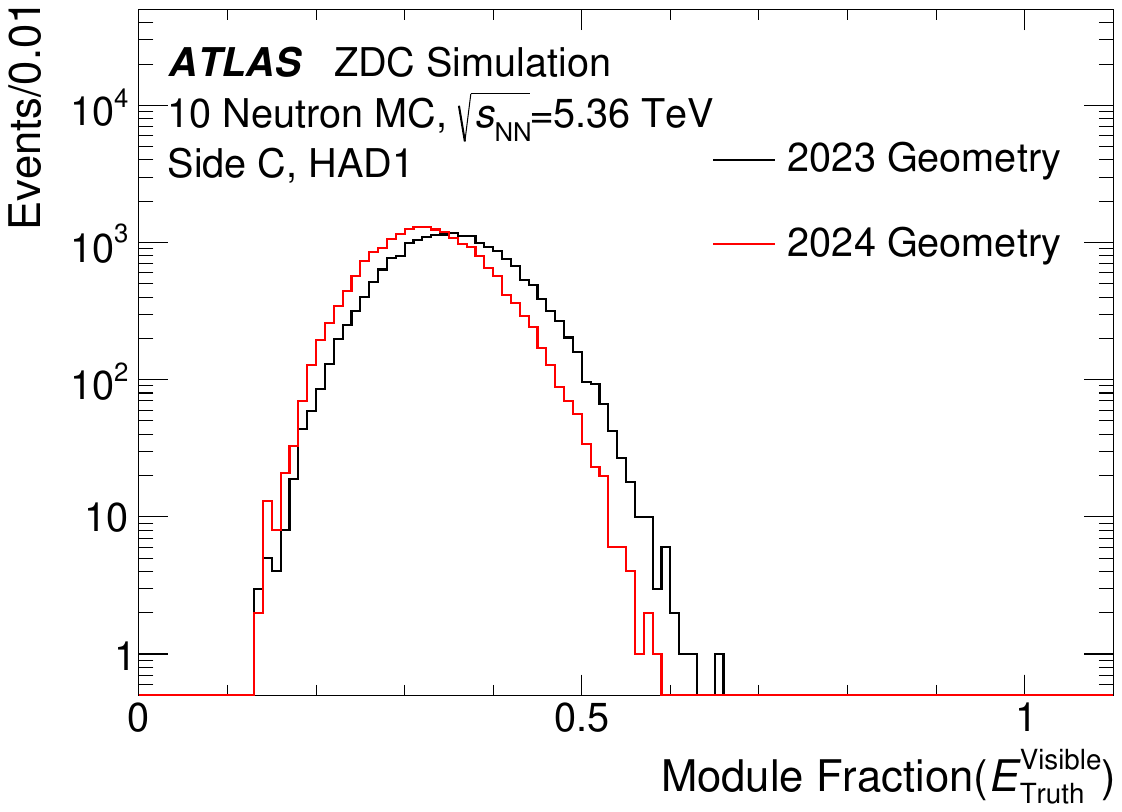}
    \includegraphics[width=0.48\linewidth]{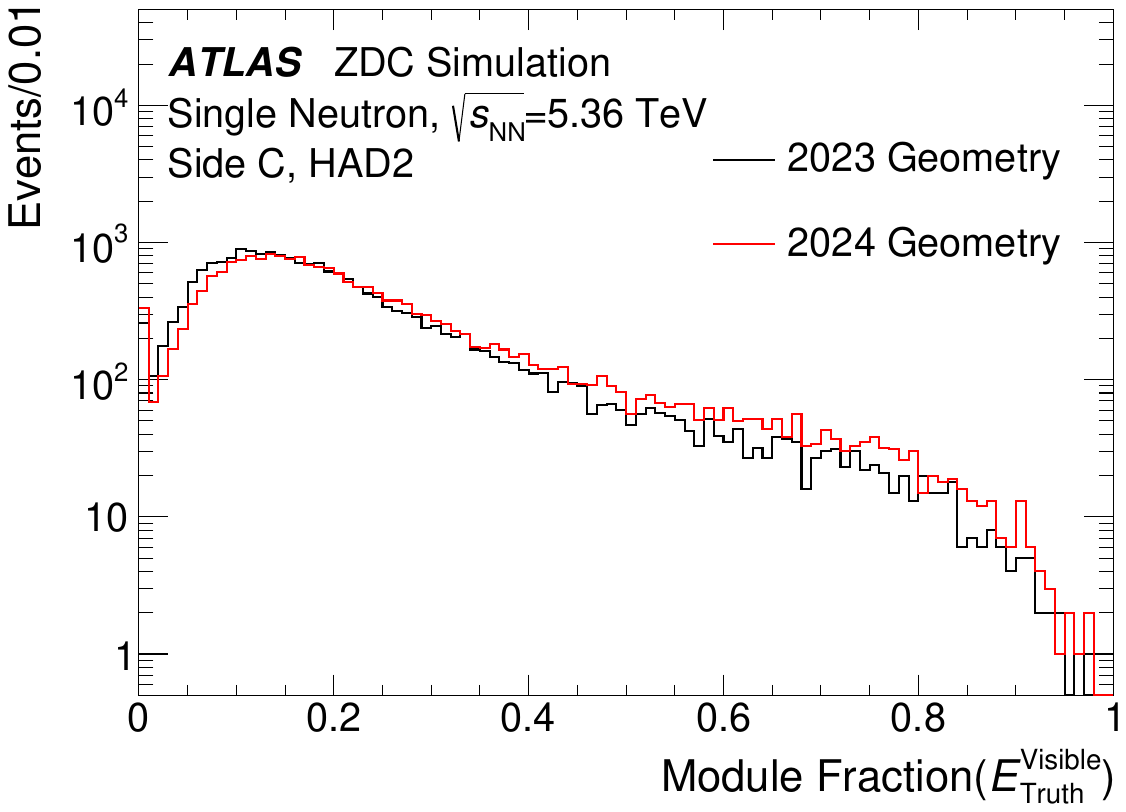}%
    \includegraphics[width=0.48\linewidth]{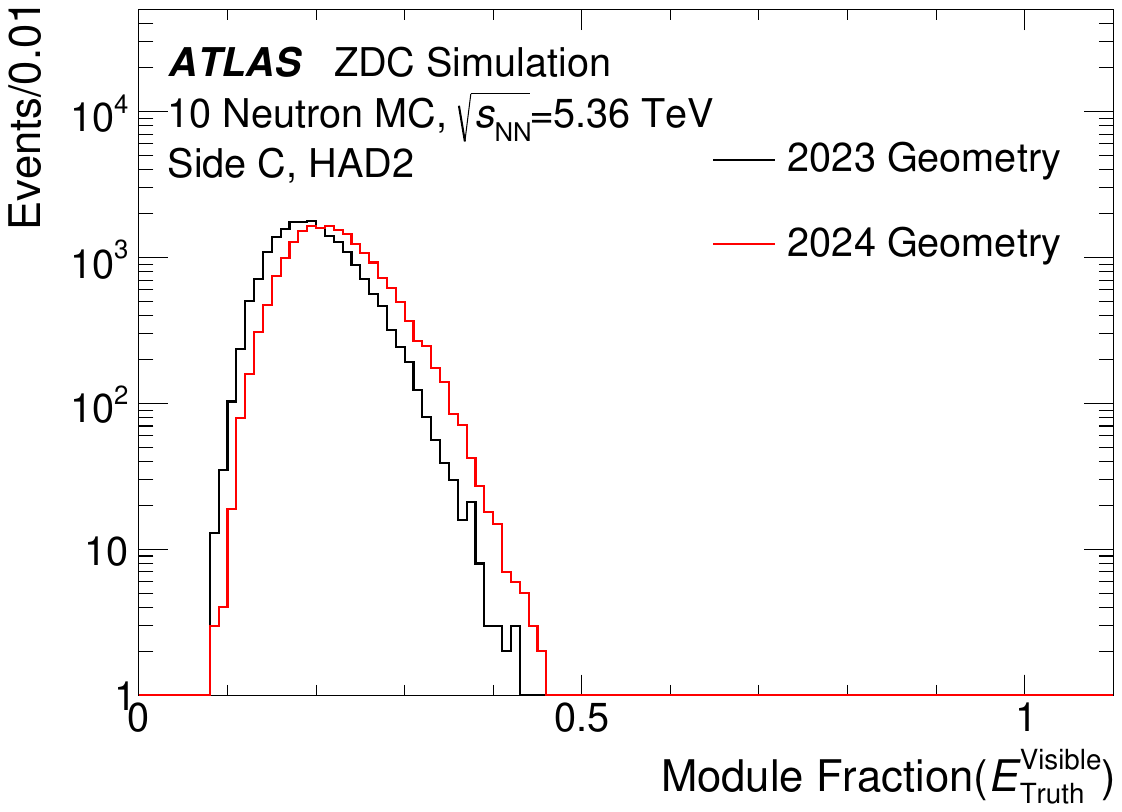}
    \includegraphics[width=0.48\linewidth]{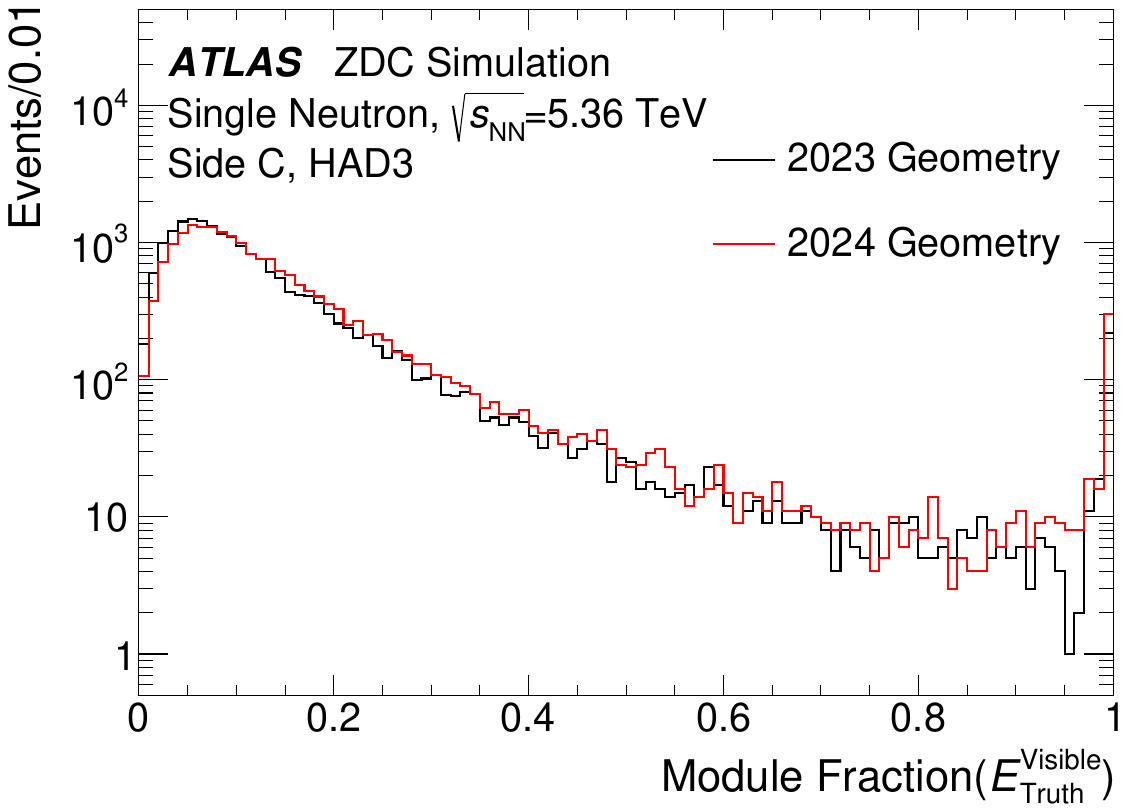}%
    \includegraphics[width=0.48\linewidth]{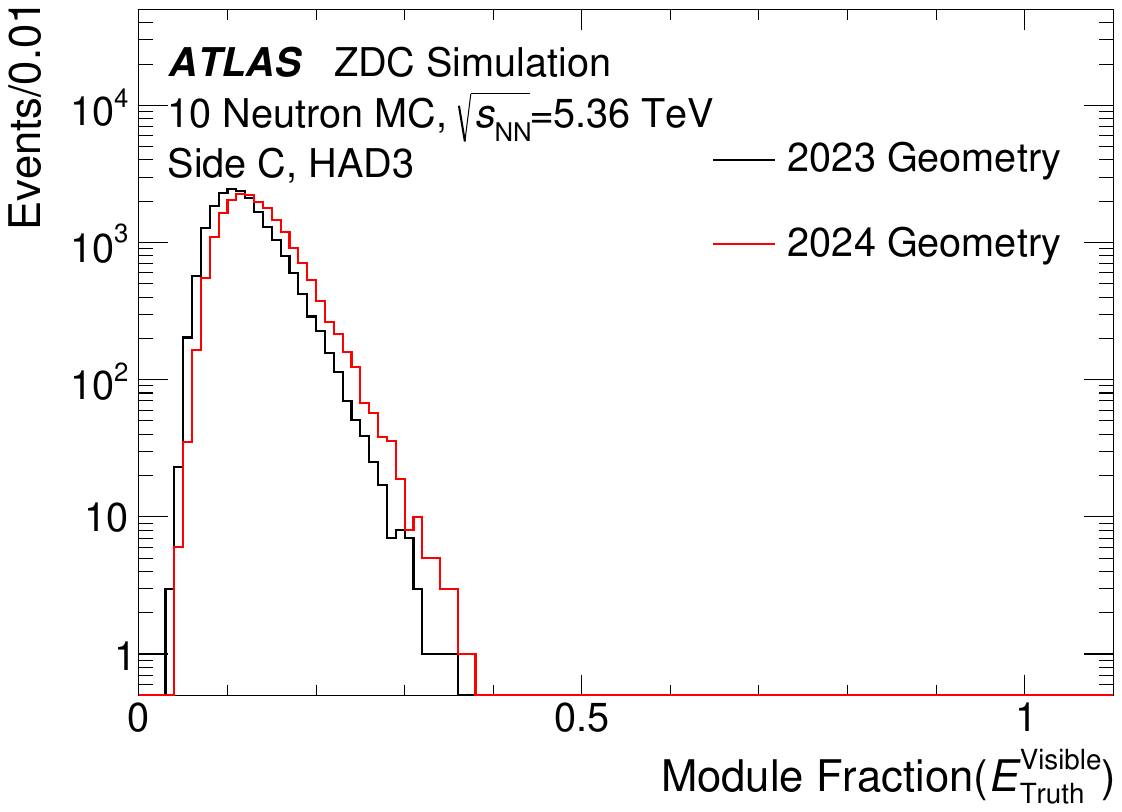}
    \caption{
      Distributions of the fraction of the total calorimeter visible energy deposited in the Side C EM  (Top row), HAD1 (second row), HAD2 
        (third row) and the  HAD3 (bottom row) modules for the 2023 and 2024 
        ZDC configurations.
      The left and right panels correspond to the single-neutron and 
        ten-neutron MC samples, respectively.
    }
    \label{fig:RPD_CompareTruthEnergyTwoSides}
\end{figure}

Figure~\ref{fig:CompareTruthEnergyTwoGeoms} compares the Visible energy
  deposited in the EM module, RPD, BRAN and the sum of all ZDC modules on 
  Side~C, for the 2023 and 2024 ZDC configurations.
The comparisons are shown for the single-neutron and ten-neutron samples.
As expected, the change in the configuration (2023 vs. 2024) does not affect
  the energy distribution in the EM module.
However, significant differences are observed for the RPD and the BRAN.
In particular for the RPD, the increase in the Visible energy is about
  84\% in the 2024 configuration as compared to the
  2023 configuration.
The change in the energy deposition in the BRAN is even more significant,
  with the mean energy deposited in the BRAN decreasing by $\sim$72\% after it is moved to the back of the TAN in 2024.
The configuration change also improves the total energy deposited 
  (i.e. increased shower containment) in the C~side of the ZDC, 
  as shown in the last row  
  in Figure~\ref{fig:CompareTruthEnergyTwoGeoms}, where the total
  Visible energy in the four modules on side C is shown to increase. A breakdown of the contributions to each module, comparing the 2023 and 2024 geometries, is presented in Figure~\ref{fig:RPD_CompareTruthEnergyTwoSides} which shows the fraction of the total Visible truth energy deposited in each module for single-neutron (left) and ten-neutron events (right). The effect of the calorimeter re-arrangement -- particularly the move of the BRAN to the end of the calorimeter -- can be see in the reduction of the energy deposited in the HAD1 module and in the increased energy deposits in the HAD2 and HAD3 modules. The former results from the showers being less developed by the time they reach HAD1 when the BRAN is no longer in front of the module. The latter results from the fact that energy is no longer deposited in the dead material of the BRAN when it is located at the end of the calorimeter. 
  
\begin{figure}[tbp]
    \centering
    \includegraphics[width=0.5\linewidth]{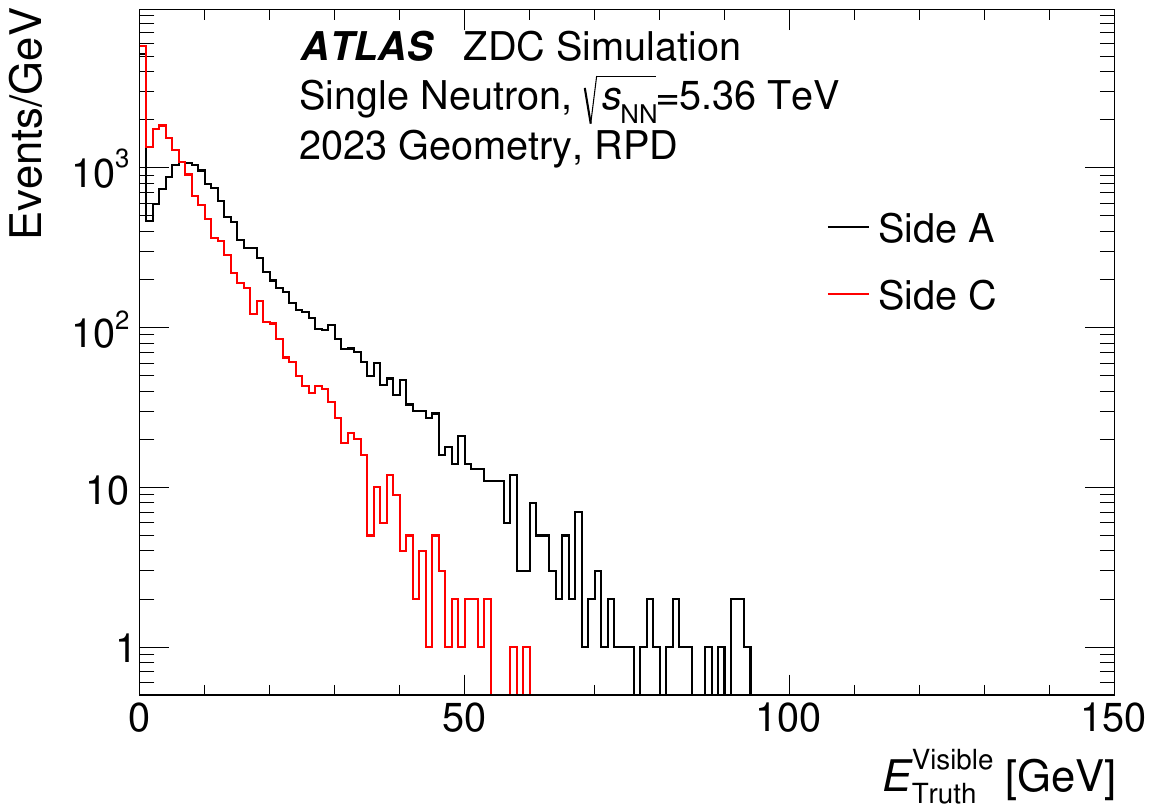}%
    \includegraphics[width=0.5\linewidth]{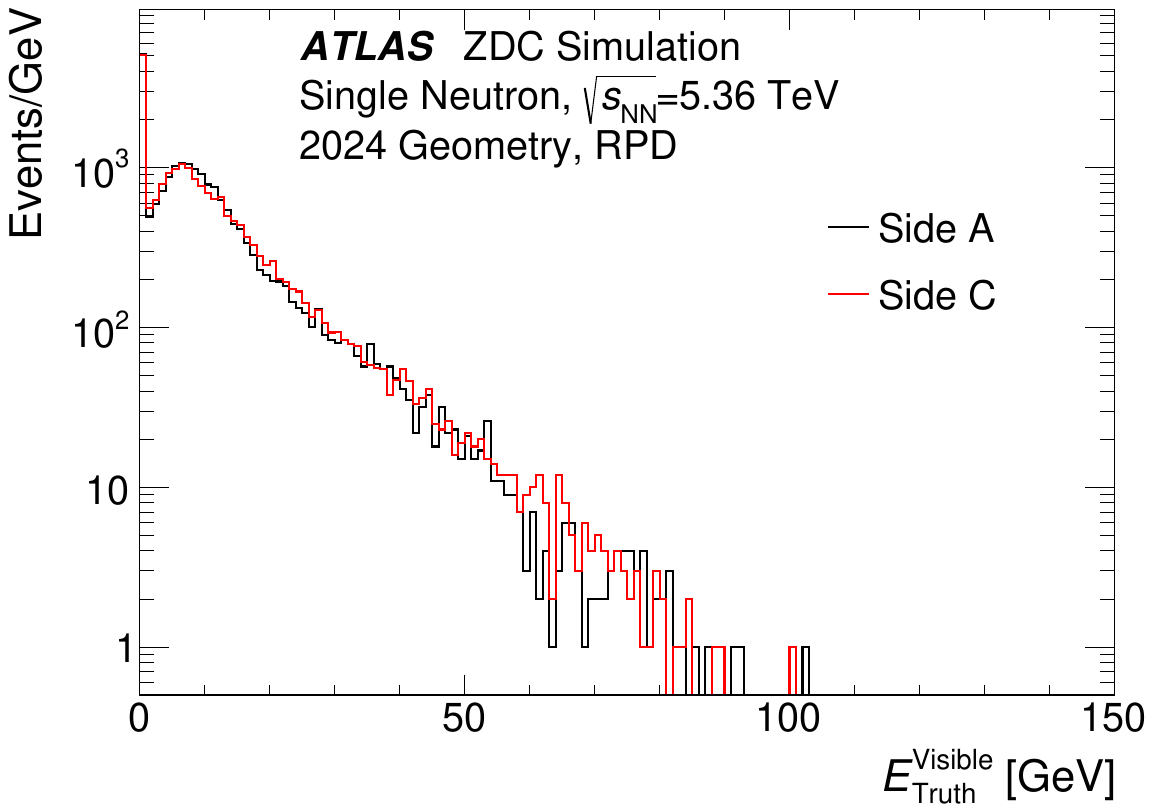}
    \caption{
      Distributions of the Visible energy deposited in the RPD on Side A and C in single 2.68~TeV neutron simulations.
      The left and right panels correspond to the 2023 run and
        the 2024 run, respectively. The difference between A and C sides in 2023 arises from to the extra air gap in the  C-side EM module; re-ordering the calorimeter prior to the  2024 \PbPb\ run (Section~\ref{sec:zdc_overview})  restored the symmetry between the two sides.
    }
    \label{fig:CompareTruthEnergyTwoSides}
\end{figure}

The improved symmetry in the response of the two sides of the RPD in the 2024
  configuration is demonstrated in Figure~\ref{fig:CompareTruthEnergyTwoSides}. The asymmetry due to the different geometries of the EM module in 2023 -- particularly the extra longitudinal space containing mostly air in the original C-side EM module -- is absent in the 2024 configuration. 
In the data analysis, the increased sensitivity to hadronic showers on the C side leads to significant improvement in the RPD performance on that side as well as more consistent reaction plane resolutions for the two RPD detectors.

\FloatBarrier

\subsection{ZDC MC and data comparison}
\begin{figure}[!tb]
   \centering
    \includegraphics[width=0.99\linewidth]{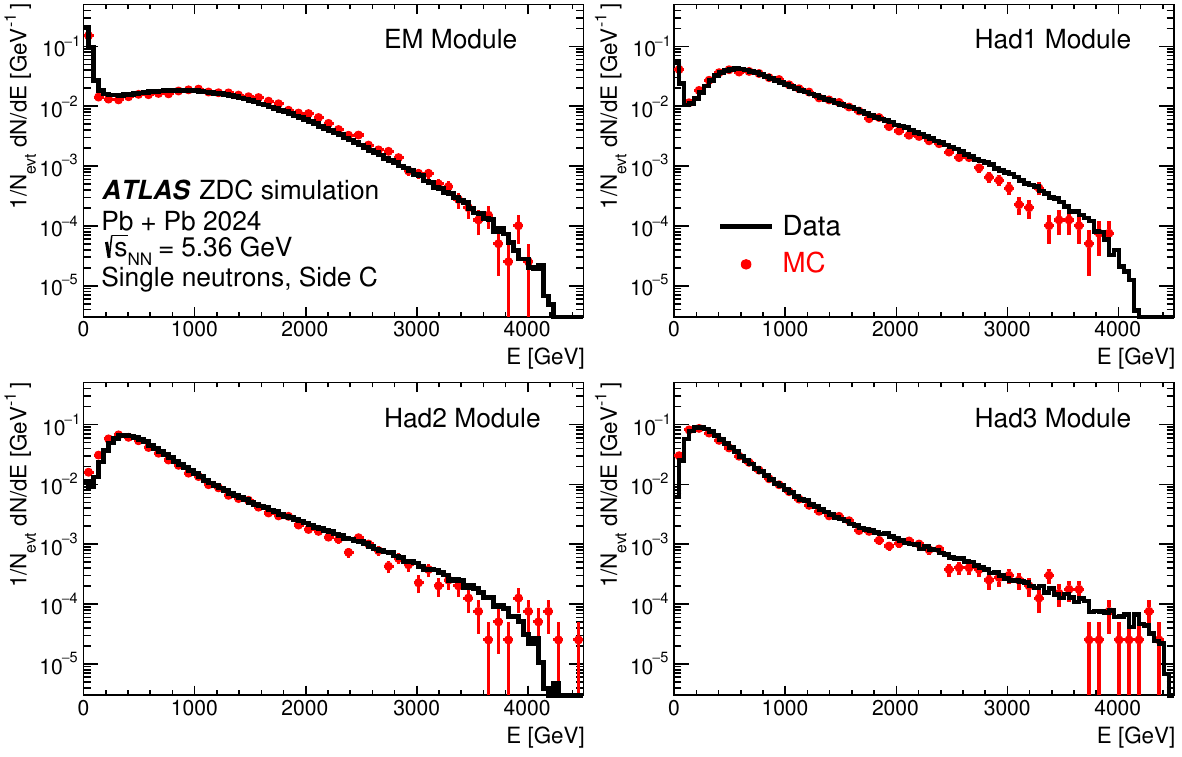}
    \caption{
      Comparison of the energy distributions in data and MC in each of the four calorimeter modules on the C side for single-neutron events. For this figure, a total of 50000 MC events were generated as described in Section~\ref{sec:mc_generation} and fully simulated using GEANT4 within the ATLAS Athena framework. The digitization was configured to provide the same average energy in each module as seen in single-neutron events in the data. The data for this figure were obtained by selecting events with calibrated energies within 2.5 standard deviations of the nominal single neutron position in data from 2024. 
    }
    \label{fig:datamccompareC} 
\end{figure}

Figure~\ref{fig:datamccompareC} shows a comparison between data and MC of the energy distributions in the four C-side calorimeter modules in single-neutron events. The MC sample used for the figure consists of 50000 single-neutron events simulated using the 2024 geometry; the data were obtained from 2024 \PbPb\ events with a calibrated energy within $\pm 2.5\sigma$ of the \onen peak position. The digitized pulse amplitudes were set to produce identically the same average energy in each module as that seen in the data for the single-neutron events. The distributions for both data and MC are normalized to have unit integrals. Because of the limited statistics of the MC sample, the MC distributions have twice the bin width as the data. The MC distributions agree sufficiently well with the data that they are mostly obscured by the data. A small difference between the data and MC is seen in the tail of the energy distribution in HAD1; elsewhere the MC distributions agree with the data within statistical uncertainties.

\subsection{RPD Light acceptance simulations}
\begin{figure}
    \centering
    \includegraphics[width=0.9\linewidth]{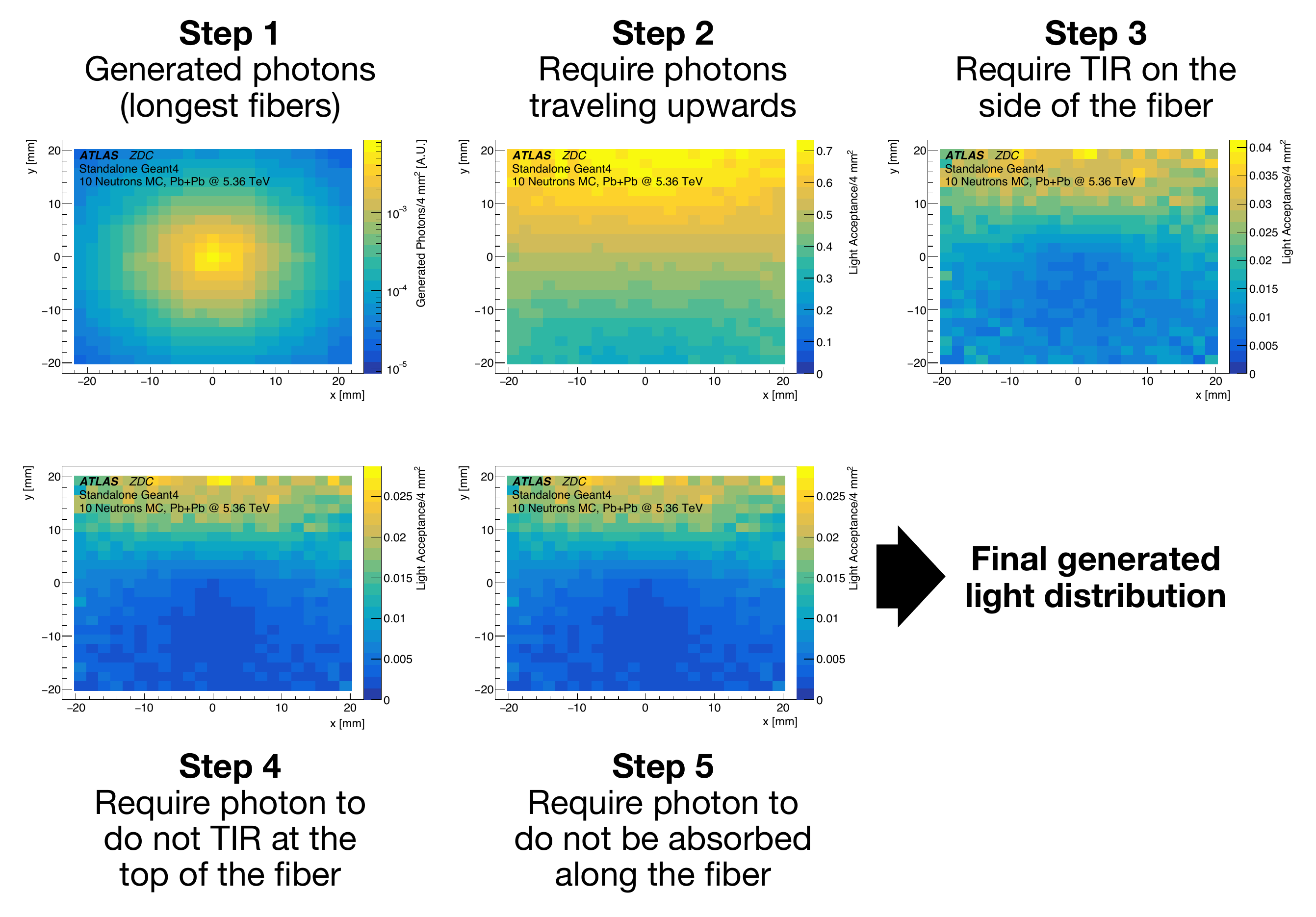}%
    \caption{Demonstration of the evaluation of the fractional photon acceptance in the RPD, obtained with a stand-alone \Geant simulation. The top left panel presents the distribution of initially-generated Cherenkov photons versus $x-y$ position. For each step, moving from left to right and top to bottom, the fractional acceptance, updated to account for the cumulative effect of each requirement, is shown. The acceptance varies strongly with vertical position ($y$) due to the
     numerical acceptance of the fibers and the boosted development of 2.68~TeV neutron showers.           }
    \label{fig:rpd_lacc}
\end{figure}

As described in Section~\ref{sec:RPD}, the RPD is comprised of 256 fused silica core fibers with a numerical acceptance of $\sim$0.22. Such a small numerical acceptance implies that the efficiency for detecting Cherenkov photons generated in the fiber core can be sensitive to details in the development of showers in the ZDC and the geometry of the detector. To evaluate the photon detection efficiency, a dedicated \Geant simulation was performed that implemented precise tracking of the generated Cherenkov photons and allowed a detailed evaluation of different contributions to the efficiency loss.  These characteristics introduce significant implications on the acceptance of Cherenkov light generated into the fiber cores. All of the effects affecting the light generation and retention within the fiber were studied using a dedicated stand-alone \Geant\ simulation which allowed for precise tracking of all the optical particles and a detailed analysis of the detector expectations. Figure~\ref{fig:rpd_lacc}
summarizes the changes in acceptance due to different conditions applied to determine whether the photons should be counted as signal or not. After (1) generating the distribution of light, (2) the photons are required to travel upwards, e.g., there is no simulation of the light going downwards. Then, (3) total internal reflection (TIR) is required for the upward traveling photons to be retained in the core of the fiber and travel to the PMT and (4) the condition that the photon would be able to escape the fiber from the top surface -- {\em i.e.} that it would not undergo TIR -- is applied. Last, (5), wavelength-dependent absorption of the fibers is simulated  according to the data-sheet of the FBP fibers \cite{FBP_Datasheet}. The resulting light yield is provided to the RPD digitization algorithms in Athena. 

\begin{figure}
    \centering
    \includegraphics[width=0.49\linewidth]{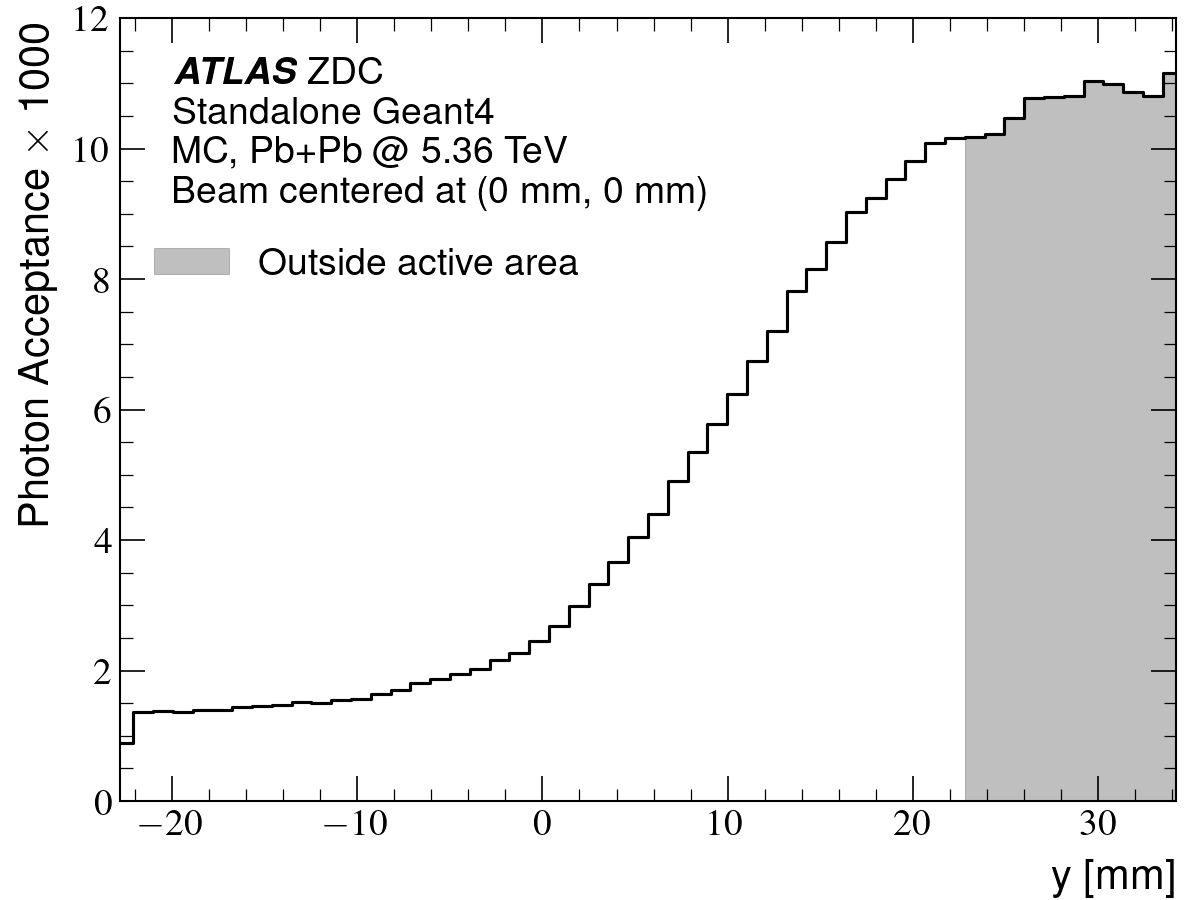}%
    \includegraphics[width=0.49\linewidth]{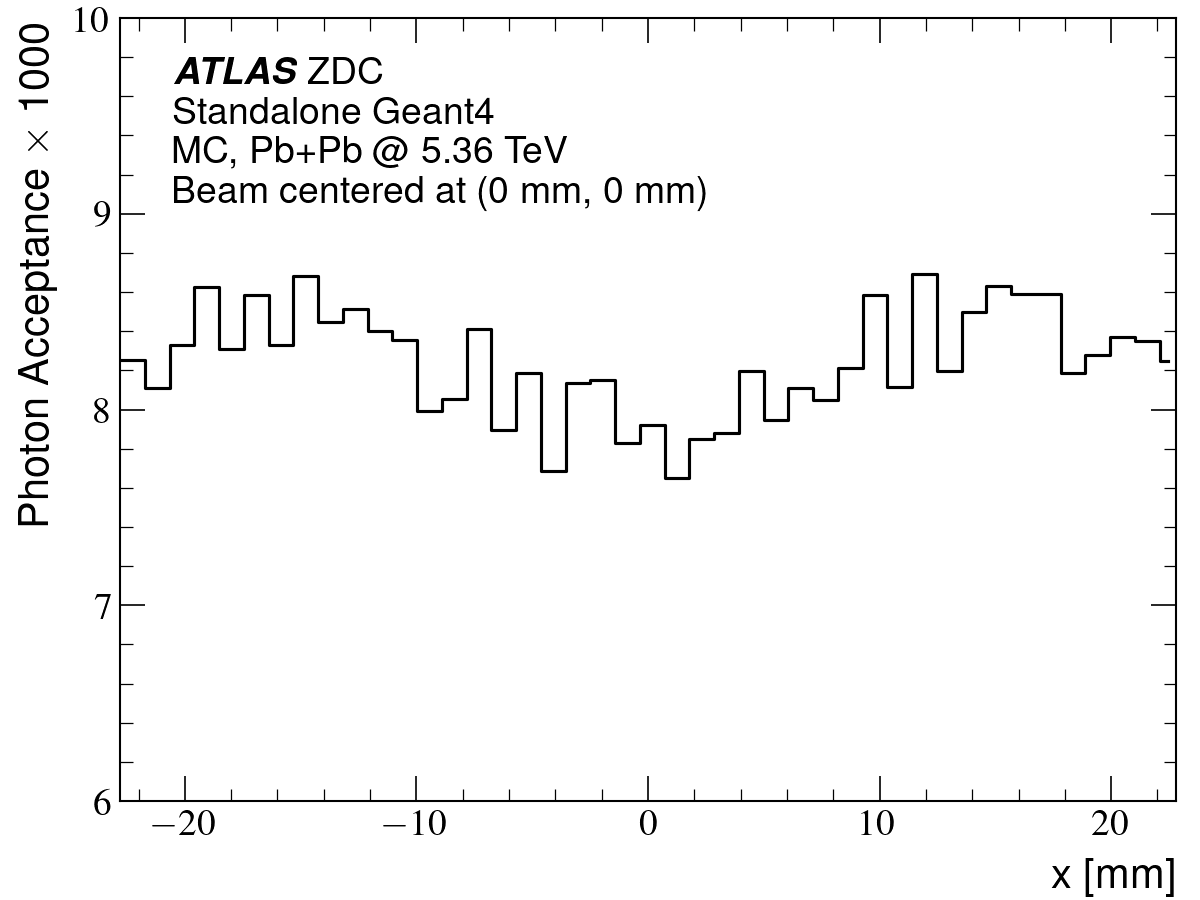}%
    \caption{Demonstration of the $y$ (left panel) and $x$ (right panel) dependence of the light acceptance in the RPD, obtained with a stand-alone \Geant simulation where the beam is centered at (0,0). The shaded area in the $y$-dependence plot distinguish the region above the active area, which extends up to the PMTs. The light acceptance is nearly constant as a function of $y$ for $y$ > 35~mm. }
    \label{fig:rpd_1Dacceptance}
\end{figure}

The results of this study show that the light acceptance is strongly biased in the vertical direction, with a larger acceptance for photons in the top section of the active area than the bottom, as well as for lateral compared to the central regions. For a more intuitive understanding of these effects, Figure~\ref{fig:rpd_1Dacceptance}
shows one dimensional projections of the light acceptance for a beam centered on the active area. 

In both distributions, a distinct pattern emerges around the central position of the neutron beam. In the vertical direction, photons generated below the beam axis exhibit a relatively tamed $y$-dependence of the acceptance, whereas those generated above it show a linearly increasing probability of being retained in the fibers. The top of the active area is characterized by 5 times more light acceptance compared to the bottom region. Notably, the acceptance continues to increase with $y$ even beyond the active area. Despite the progressive reduction in showers impinging on this region with increasing $y$, this effect contributes to the positive $y$ bias observed in the raw detector response. 

\begin{figure}
    \centering
    \includegraphics[width=0.49\linewidth]{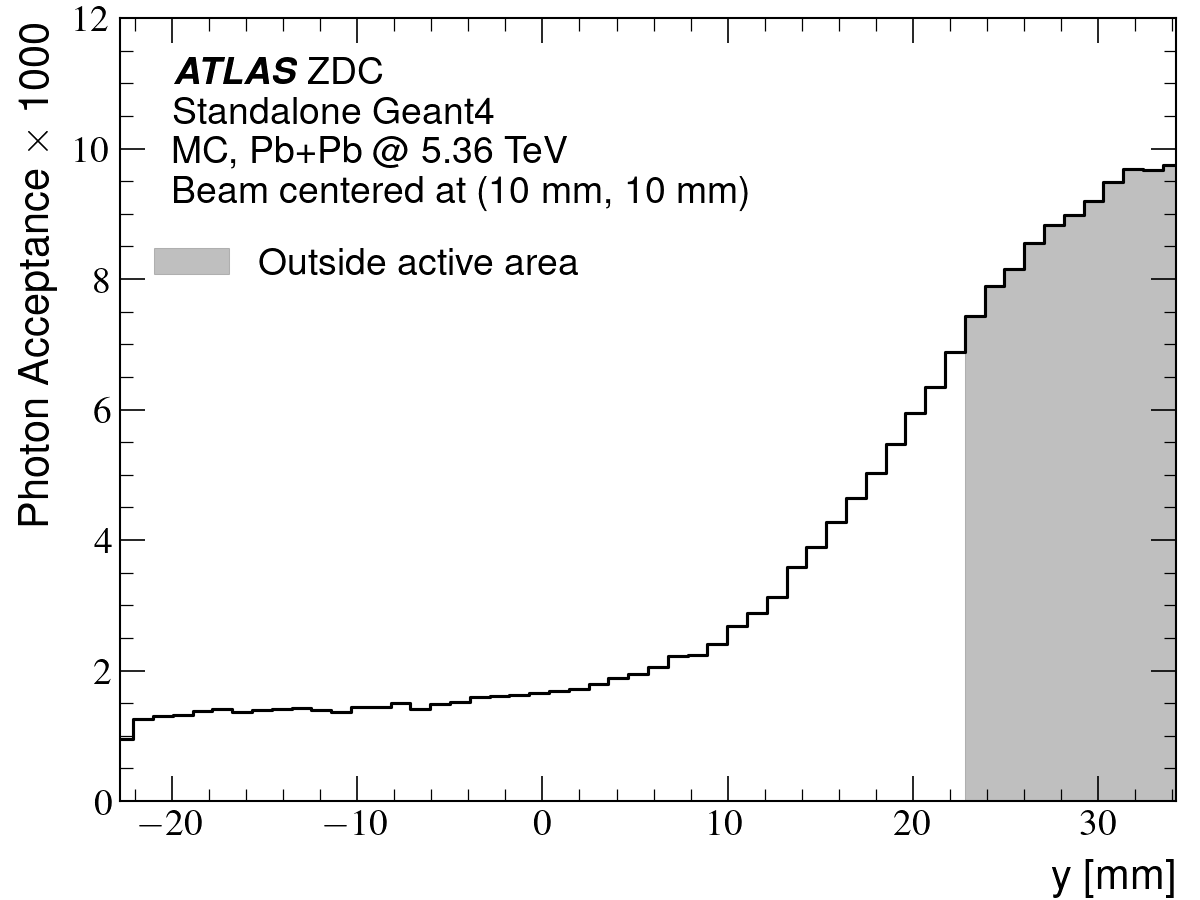}%
    \includegraphics[width=0.49\linewidth]{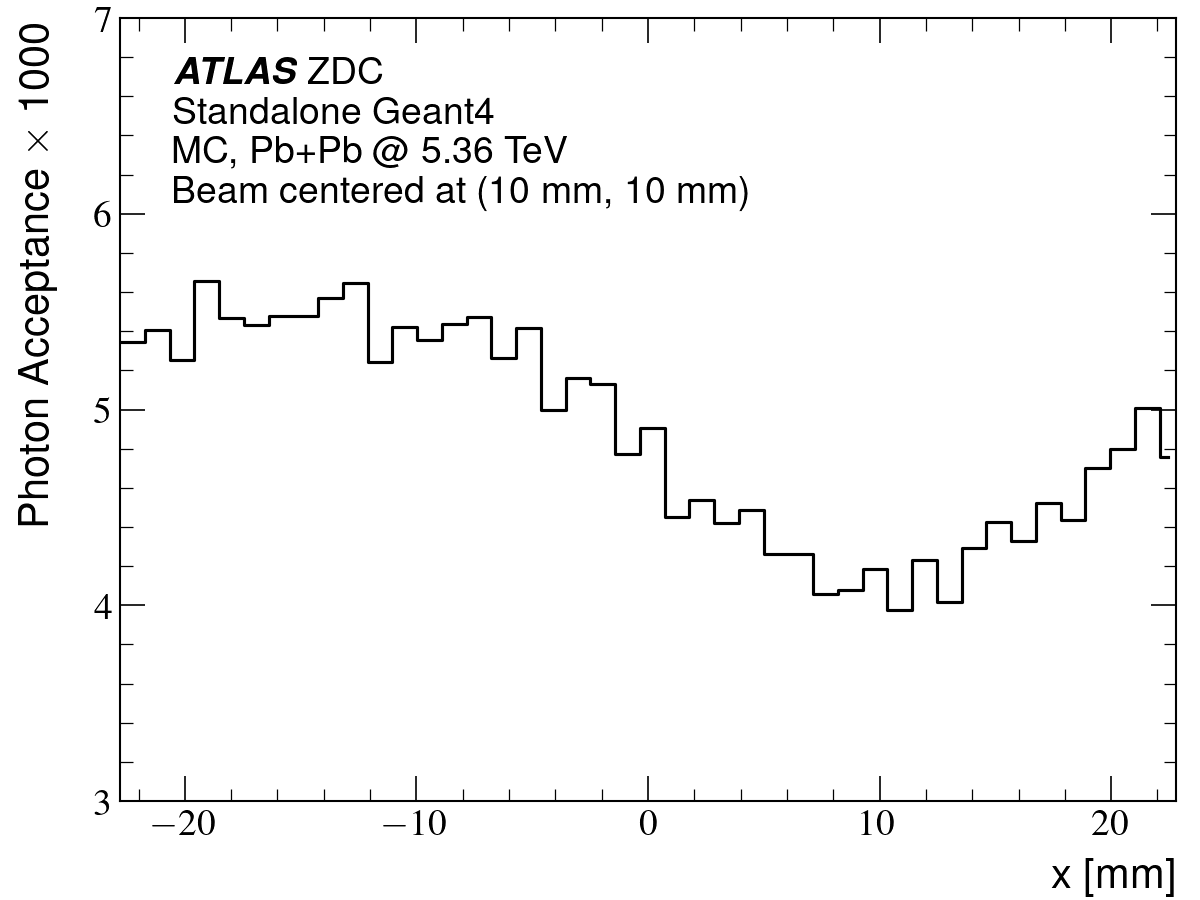}%
    \caption{Demonstration of the $y$ (left panel) and $x$ (right panel) dependence of the light acceptance in the RPD, obtained with a stand-alone \Geant simulation where the beam is off-centered at (10 mm, 10 mm). The shaded area in the $y$-dependence plot distinguish the region above the active area, which extends up to the PMTs. The light acceptance is $\sim$flat for $y$ > 35 mm. }
    \label{fig:rpd_1Dacceptance_offset}
\end{figure}

Interestingly, a structure relative to the position of the beam axis can be noticed also in the $x$-dependence of the light acceptance. In this case, the acceptance increases moving from the region more collimated around the beam direction to the more external part of the detector. The observed increase is of approximately 10\%. To confirm that these structures are related to the beam position, the exercise in Figure~\ref{fig:rpd_1Dacceptance} was repeated after shifting the beam position by +10 mm in both the $x$ and $y$ directions. The results, shown in Figure~\ref{fig:rpd_1Dacceptance_offset}, clearly demonstrate that the described structures track the beam movement. The $y$ dependence of the acceptance seems to stabilize around a constant value even if the beam is move upwards, while for the $x$ direction a change in the relative significance of the deep in correspondence of the beam axis (from 10 to 15\%) can be noticed. This change may be related to the different acceptance of the ZDC for neutrons offset by 10 mm in $x$, as well as the different shower shape illuminating the RPD active area in this case. 

%% file: summary.tex
This paper has described upgrades to the ATLAS Zero Degree Calorimeters performed prior to the start of Run~3 operation of the LHC.
These upgrades include: replacement of the
quartz radiator rods with H$_2$-doped fused silica rods;
replacement of the cables that transmit calorimeter PMT signals to
the ATLAS USA15 cavern with low-dispersion \aircore cables;
upgrade of the front-end electronics using LUCROD boards
originally designed for the LUCID2 detector;  implementation of
fully-digital triggering within the LUCRODs; implementation of a new LED calibration system; and  the addition of new reaction plane detectors 
designed to image the multi-neutron showers in the ZDC to measure the
azimuthal orientation of the directed-flow event plane. This paper 
describes the configurations of the upgraded ZDC detectors for the 2023 and 2024 \PbPb runs. It provides technical details on both the calorimeters and the RPDs, which employ a novel technology, as well as on the electronics used for signal readout, the digital trigger system, calibration methods, and results from Monte Carlo simulations of the detector response.

Run 3 of the LHC is expected to continue through the summer of 2026, ending with a final heavy-ion run.  This marks a major transition for both the accelerator and the experiments, as the machine enters its third long shutdown in preparation for Run~4.   
During this period, extensive upgrades will be made across the accelerator complex as the LHC evolves into the High-Luminosity LHC (HL-LHC). This transformation includes a complete rebuild of the interaction regions, including the neutral absorber housing the ZDCs.
The upgraded absorber will require active cooling to manage the heat load delivered by the forward-going neutral particles, and it has been designed specifically to have slots for the BRAN and the ZDC.  
However, the new absorbers, called TAXNs \cite{SantosDiaz:2022hxl}, have two major differences from the TANs described in this paper: they  are closer to the ATLAS interaction point by about 13~m, and the separation of the outgoing beam pipes only provides 5~cm transverse space for a detector.  
A new detector, the HL-ZDC \cite{Longo:2023ubo}, has been proposed to operate in the TAXNs for LHC Run~4 and beyond; the same detector will be utilized by both the ATLAS and CMS experiments.  
Like the current ATLAS ZDC, the HL-ZDC is a combined EM/hadronic calorimeter with tungsten absorbers and radiation-hard fused silica radiators (planned to be the same as with the ZDC). The HL-ZDC will offset the smaller transverse size and reduced lateral shower containment with increased longitudinal depth.  
The HL-ZDCs also incorporate an RPD with a nearly-identical design as that described in this paper.
These components will be integrated into a single module with a minimal number of multi-core cable connectors, reducing the time and complexity (and thus radiation exposure to workers) of the installation procedure. 
The Cherenkov light from particle-induced showers will be transported vertically, and brought via air light-guides to a set of PMTs, providing 9-fold longitudinal segmentation: three 28~mm PMTs in the EM section, and six 53~mm PMTs in the hadronic section. The HL-ZDC signals will be routed to the ATLAS electronics room in the cavern by air-core cables for the energy measurements, and fast CK50 cables for the RPD. On-detector calibration will be provided by a specially-designed LED system. 
The existing ATLAS ZDC is, thus, providing critical tests of, and important operational experience with, nearly every component of the HL-ZDC well before construction begins during LS3.

%% file: acknowledgements.tex
We thank CERN for the very successful operation of the LHC and its injectors, as well as the support staff including the transport and radiation protection teams who support the installation and de-installation of the ATLAS ZDCs. We also thank the CERN
test beam coordination team at the SPS North Area for the beam time allocated for the project and their support through test beam operations. We gratefully acknowledge the support of the ATLAS collaboration, ATLAS Technical Coordination, and the ATLAS operations team all of whom contributed to the successful upgrade and operation of the ZDCs. We also thank the ATLAS online and offline software teams, without whose efforts this work could not have been accomplished.  We extend our gratitude to the technical staff of the Nuclear Physics Laboratory at the University of Illinois for their continuous technical support and to Sezione di Bologna of INFN for their crucial contributions to the adaptation of the LUCRODs for use in the ZDC.
We also thank Francesco Cerutti and Daniel Prelipcean, from the CERN FLUKA team, for their simulations and their input about the dose accumulated on the ZDC during Run 3.

This work was supported by grants from the US Department of Energy (DE-FG02-86ER40281 and DE-SC0012704), the US National Science Foundation (PHY 2111046, PHY 2110772), the U.S.-Israel Binational Science Foundation (2023767), the Israel Science Foundation (1804/23) and the Italian Ministry of University and Research (Rita Levi Montalcini program). We would like to acknowledge the generous funding support from the Grainger College of Engineering at the University of Illinois, the DaRin Butz Foundation, and the Henry Luce Foundation through the Clare Boothe Luce Undergraduate Research Awards. We also thank the support of the Illinois Scholars Undergraduate Research Program.

%% file: optimization.tex
\section{One-neutron optimization procedure}
An absolute calibration of the ZDC energy scale can be obtained through the use of the prominent single-neutron contribution to the ZDC energy spectrum that results from electromagnetic excitation and decay of the giant dipole resonance.
The calibration is based on the use of single neutrons, which are emitted in photonuclear interactions with fixed mean energy.
The method chosen to calibrate the ZDC energy with single neutrons was developed in Run 2, and was used for the ZDC calibration in a wide range
of ATLAS publications, but has not been documented systematically before this work.
The assumption is that the shower energy is distributed among the four modules, such that the best estimator is a linear combination of the module signal amplitudes, with weights selected to minimize the width 
of the single-neutron peak. 
\begin{equation}
    E = \sum_{i} w_i M_i
\end{equation}
It relies on the assumptions that the fluctuations around the peak are mostly symmetric, and that the contributions from multiple neutrons,
particularly downward fluctuations of the two-neutron energy, can be neglected.
The weights can be determined by minimizing a $\chi^2$ function, with a constraint implemented via a Lagrange multiplier. 
\begin{equation}
\chi^2 = \langle E^2 \rangle - E^2_n + \lambda(\langle E \rangle^2-E_n) 
\end{equation}
where the first two moments of the energy distribution can be written as
\begin{equation}
    \langle E \rangle = \frac{1}{N_{evt}}\sum_{evt}\sum_{i}w_iM_{i,evt} = \sum_{i}w_{i}\langle M_i \rangle
\end{equation}
and
\begin{equation}
    \langle E^2 \rangle = \frac{1}{N_{evt}} \sum_{evt} \sum_{i,j} w_i w_j M_{i,evt} M_{j,evt} = \sum_i\langle M_i M_j \rangle
\end{equation}
for a sample events pre-selected to be within a reasonable range from the one-neutron peak.
The derivatives of the $\chi^2$ function for the minimization are straightforward to calculate:
\begin{equation}
    \frac{d\chi^2}{dw_i} = 2 \sum_j w_j \langle M_i M_j \rangle + \lambda \langle M_i \rangle = 0
\end{equation}
and
\begin{equation}
    \frac{d\chi^2}{d\lambda} = \sum_i w_i \langle M_i \rangle - E_n = 0
\end{equation}
and these can be recast in matrix form as
\begin{equation}
    \left[
    \begin{array}{ccccc}
    2\langle M_0 M_0 \rangle & 2\langle M_0 M_1 \rangle & 2\langle M_0 M_2 \rangle & 2\langle M_0 M_3 \rangle & \langle M_0 \rangle \\
    2\langle M_1 M_0 \rangle & 2\langle M_1 M_1 \rangle & 2\langle M_1 M_2 \rangle & 2\langle M_1 M_3 \rangle & \langle M_1 \rangle \\
    2\langle M_2 M_0 \rangle & 2\langle M_2 M_1 \rangle & 2\langle M_2 M_2 \rangle & 2\langle M_2 M_3 \rangle & \langle M_2 \rangle \\
    2\langle M_3 M_0 \rangle & 2\langle M_3 M_1 \rangle & 2\langle M_3 M_2 \rangle & 2\langle M_3 M_3 \rangle & \langle M_3 \rangle \\
    \langle M_0 \rangle & \langle M_2 \rangle & \langle M_2 \rangle & \langle M_3 \rangle & 0 \rangle 
    \end{array}
    \right]
    \left[     
    \begin{array}{c}
    w_0 \\ w_1 \\ w_2 \\ w_3 \\ \lambda 
    \end{array}
    \right] =
    \left[     
    \begin{array}{c}
    0 \\ 0 \\ 0 \\ 0 \\ E_n
    \end{array}
    \right] 
\end{equation}
which is solved numerically in ROOT using {\tt LUDecomp::Solve} to give the set of linear weights $w_i$ which optimize the one-neutron peak.

This approach is of course very generic, and does not try and obey any further constraints, e.g. the fraction of per-module shower energies expected in simulations.  As such, it does not guarantee a correct measurement of the actual energy measurements in each modules, but just the combination of the module amplitdues which sum to the most precise estimate of the deposited energy.